\documentclass[11pt]{article}
\pdfoutput=1
\usepackage{amsmath,amssymb,amsfonts,dsfont}
\usepackage[citebordercolor={.8 .8 1},urlbordercolor ={.8 .8 1},pdfstartview=FitV]{hyperref}

\newcommand{\hT}{\mathcal{L}}
\newcommand{\hW}{\mathcal{W}}

\numberwithin{equation}{section}

\begin{document}
\begin{center}

\vspace{1cm} { \Large {\bf Boundary Conditions and Partition Functions in Higher Spin AdS$_3$/CFT$_{2}$}}

\vspace{1.1cm}
Jan de Boer and Juan I. Jottar

\vspace{0.7cm}

{\it Institute for Theoretical Physics, University of Amsterdam,\\
Science Park 904, Postbus 94485, 1090 GL Amsterdam, The Netherlands}

{\tt J.deBoer@uva.nl, J.I.Jottar@uva.nl} \\

\vspace{1.5cm}

\end{center}

\begin{abstract}
\noindent  We discuss alternative definitions of the semiclassical partition function in two-dimensional CFTs with higher spin symmetry, in the presence of sources for the higher spin currents. Theories of this type can often be described via Hamiltonian reduction of current algebras, and a holographic description in terms of three-dimensional Chern-Simons theory with generalized AdS boundary conditions becomes available. By studying the CFT Ward identities in the presence of non-trivial sources, we determine the appropriate choice of boundary terms and boundary conditions in Chern-Simons theory for the various types of partition functions considered. In particular, we compare the Chern-Simons description of deformations of the field theory Hamiltonian versus those encoding deformations of the CFT action. Our analysis clarifies various issues and confusions that have permeated the literature on this subject.

\end{abstract}

\pagebreak

\setcounter{page}{1}
\setcounter{equation}{0}
\tableofcontents


\section{Introduction}\label{sec: Intro}
The study of higher spin theories in anti-de Sitter (AdS) space has been recently revitalized, partly because they provide an example of holographic duality in which the field theory is essentially non-interacting, and one often has good analytic control over both local and non-local observables. At least in principle, this feature allows for a very precise holographic dictionary to be established and tested: roughly speaking, the higher spin symmetries emerge in a regime in which one can compute reliably in both the bulk and the boundary sides of the correspondence. A very interesting example of these dualities is the conjecture \cite{Klebanov:2002ja} of Klebanov and Polyakov relating three-dimensional critical $O(N)$ vector models and the Fradkin-Vasiliev higher spin theories in AdS$_{4}$ \cite{Fradkin:1986qy,Fradkin:1987ks,Vasiliev:1995dn}, for which robust evidence has been provided recently (see \cite{Giombi:2012ms} and references therein). 

Another setup where both sides of the duality are amenable to study is that of AdS$_{3}$/CFT$_{2}\,$: here the boundary theories correspond to two-dimensional CFTs with extended current algebras, and the gauge sector of the three-dimensional bulk gravitational theory can be formulated as a Chern-Simons gauge theory. Indeed, starting with the proposal of Gaberdiel and Gopakumar \cite{Gaberdiel:2010pz,Gaberdiel:2012uj} relating the three-dimensional interacting higher spin theories \cite{Prokushkin:1998bq,Prokushkin:1998vn} to a family of minimal model coset CFTs with $\mathcal{W}$-symmetry,\footnote{See \cite{Bouwknegt:1992wg} for a comprehensive review of $\mathcal{W}$-symmetry in CFT.} several results have been obtained that show agreement between quantities computed in CFT and from the bulk duals. These include the spectrum \cite{Castro:2011iw,Perlmutter:2012ds,Campoleoni:2013lma,Campoleoni:2013iha,Gaberdiel:2013cca}, partition functions \cite{Gaberdiel:2011zw,Kraus:2011ds,Gaberdiel:2012yb}, scalar correlators \cite{Hijano:2013fja,Gaberdiel:2013jca}, and entanglement entropies \cite{deBoer:2013vca,Ammon:2013hba,Datta:2014uxa}, to name a few. While the full realization of the duality also involves matter fields in the bulk, which couple to operators other than conserved currents, the pure higher spin sector of the correspondence already provides an interesting arena where universal aspects of the duality can be explored. 

In the present article we will focus on the sector of the latter dualities describing the CFT's conserved currents, where the corresponding symmetries emerge via Hamiltonian reduction of current algebras and admit a simple holographic description in terms of two copies of Chern-Simons theory. Our main goal will be to clarify the interpretation of different boundary conditions in Chern-Simons theory from the point of view of the dual CFT, in the presence of sources for the conserved currents furnishing the extended (possibly higher spin) symmetries. In particular, we will argue that certain boundary conditions correspond to a deformation of the CFT Hamiltonian, while others correspond to deformations of the CFT action. 

More precisely, given a CFT with Hamiltonian $H_{\text{CFT}}$ and action $S_{\text{CFT}}\,$, we can distinguish at least four types of deformations depending on whether they are chiral or non-chiral and whether they are defined as modifications of $S_{\text{CFT}}$ or $H_{\text{CFT}}$ :
\begin{eqnarray}
S
&=&
S_{\text{CFT}} + \int d^{2}z\sum_{s}\mu_{s}\mathcal{W}_{s}
\\
S
&=&
S_{\text{CFT}} + \int d^{2}z\sum_{s}\mu_{s}\mathcal{W}_{s} + \int d^{2}z\sum_{s}\bar{\mu}_{s}\overline{\mathcal{W}}_{s} +  \cdots
\label{non-chiral action def}
\\
H
&=&
H_{\text{CFT}} + \oint d\sigma \sum_{s}\mu_{s}\mathcal{W}_{s}
\label{chiral Hamiltonian def}
\\
H
&=&
H_{\text{CFT}} + \oint d\sigma \sum_{s}\mu_{s}\mathcal{W}_{s} +\oint d\sigma \sum_{s}\bar{\mu}_{s}\overline{\mathcal{W}}_{s}\,.
\end{eqnarray}

\noindent Here $\mathcal{W}_{s}$ and $\overline{\mathcal{W}}_{s}$ are a set of currents of weight $(s,0)$ and $(0,s)$, respectively, obeying appropriate Poisson or Dirac bracket chiral algebras which will typically be non-linear extensions of the Virasoro algebra, and $\sigma$ denotes a compact coordinate on the cylinder. The deformation parameters $\mu_{s}$ and $\bar{\mu}_{s}$ can be thought of as chemical potentials or background gauge fields: provided they transform suitably, the partition functions defined from the above Hamiltonians/actions will be invariant under the symmetry algebra furnished by the currents. The dots in \eqref{non-chiral action def} denote the fact that, in the presence of deformations of both chiralities, the corresponding action requires terms to all orders in the chemical potentials in order to realize the symmetry. On the other hand, as we will discuss in detail in due course, at the level of the Hamiltonian the linear couplings suffice, even when both chiralities are present, because $\mathcal{W}_s$ and $\overline{\mathcal{W}}_s$ Poisson-commute.

The program that we will follow can be summarized quite simply. The fact that the partition functions associated with the various types of deformations above enjoy a symmetry will as usual result in Ward identities for the one-point functions of the currents in the presence of sources. The precise form of these Ward identities will depend on the particular type of deformation under consideration, but in all cases one can encode them as a flatness condition on suitable $2d$ gauge connections in ``Drinfeld-Sokolov form" \cite{Drinfeld:1984qv}. If we now regard the CFT as being defined on the boundary of a $3d$ manifold, these $2d$ gauge connections become boundary conditions for $3d$ Chern-Simons gauge fields, with the flatness conditions enforced by the Chern-Simons equations of motion. From a practical point of view, the advantage of this formulation is that one can now use Chern-Simons theory to derive a number of universal results for the boundary theories quite efficiently, including thermodynamic quantities such as entropy and free energy, and even non-local observables such as entanglement and R\'enyi entropies which are usually quite difficult to obtain using solely CFT techniques. For example, formulae for the thermal entropy in the presence of higher spin chemical potentials written entirely in terms of the Chern-Simons connections were derived in \cite{deBoer:2013gz,Kraus:2013esi}, and two proposals for higher spin entanglement entropy in terms of Wilson lines in Chern-Simons theory were put forward in \cite{deBoer:2013vca} and \cite{Ammon:2013hba}.

It is worth mentioning that the logic behind the holographic formulation of the current sector of these theories predates the advent of the AdS/CFT correspondence, and can be seen as a special case of the usual connection between Chern-Simons theory and Wess-Zumino-Witten (WZW) models. In fact, the different types of deformations we discuss as well as their associated symmetries were studied more than two decades ago in the context of gauging of $\mathcal{W}$-algebras and the so-called $\mathcal{W}$-gravity. Similarly, the connection between deformations of the Hamiltonian and chiral deformations of the CFT action was discussed in \cite{Dijkgraaf:1996iy} from a field-theoretical perspective. Our main goal will be to derive the implications of these results for the Ward identities and their connection to Chern-Simons theory, in the hope that these considerations will help to bridge the gap between the existing literature and the recent discussions in the context of higher spin AdS$_{3}$/CFT$_{2}\,$.

 Importantly, in order to derive the Chern-Simons formulation one does not use holography or the existence of a holographic dual of the starting CFT. Our analysis, however, is only valid at the classical level (which in the dual CFT corresponds to a limit where $c\rightarrow \infty$, with $c$ the central charge), and uses no properties of the CFT except that it possesses particular symmetries. It is only when studying subleading corrections to various quantities that one would need to have a more detailed knowledge of the matter content of the field theory, which in the bulk corresponds to specific couplings of matter fields to Chern-Simons theory. In the latter situation the details of the full-fledged holographic correspondence become important.

While the problem at hand may appear to be of a fairly technical nature, it is conceivable that the techniques developed in the context of the higher spin AdS$_{3}$/CFT$_{2}$ duality may find an application to realistic systems. In fact, Hamiltonians of the form \eqref{chiral Hamiltonian def} feature prominently in the study of the dynamics of one-dimensional integrable condensed matter systems following a quantum quench, where they are referred to as a ``generalized Gibbs ensemble" or GGE (see e.g. \cite{Caux:2012nk} and references therein). Similarly, the large-$N$ limit of certain coset CFTs proposed to describe strange metals in one spatial dimension has been related to higher spin theories on AdS$_{3}$ \cite{Gopakumar:2012gd}. Furthermore, even though most of the results that we will derive are strictly speaking applicable in the large central charge regime, one may hope that some of the conclusions and lessons from the holographic analysis will retain their validity in other corners of parameter space, which would make these results appealing to a wider community. In fact, some of the predictions for entanglement entropy in the presence of sources derived in \cite{deBoer:2013vca} using a novel holographic proposal have been recently argued to apply beyond the large central charge limit from a purely CFT perspective \cite{Datta:2014ska,Datta:2014uxa}, with the first perturbative correction in the higher spin sources being moreover universal.

The rest of the article is organized as follows. In section \ref{sec: canonical} we consider Hamiltonian deformations of the CFT and rewrite the canonical partition function as a path integral in first order form, and exploit this representation to derive the Ward identities obeyed by the one-point function of currents in the presence of sources. Although we employ a free boson realization to perform the calculation, we will find that the resulting Ward identities take a generic form, independent of the specific realization and particular symmetry algebra. Moreover, we will discover that these Ward identities have a slightly different structure from the ones usually discussed in the literature. In section \ref{sec: holo partition function} we consider deformations of the CFT action instead, and exhibit the form of the corresponding Ward identities. In section \ref{sec: Holo} we determine the precise ``Drinfeld-Sokolov pair"  that allows to rewrite the Ward identities for Hamiltonian and action deformations as the flatness condition on $2d$ connections. Using holography, we then extend them into $3d$ flat connections with suitable boundary conditions, and use the associated variational principle to derive expressions for the free energy and entropy, for example. We also revisit and discuss a few results that have generated some confusion in the recent literature, and point out a useful relation obeyed by flat connections in Drinfeld-Sokolov form. We conclude in section \ref{sec: Disc}. Useful formulas and examples that complement the discussion are collected in the appendices.

\section{Hamiltonian deformations and the canonical partition function}\label{sec: canonical}
The basic object of interest is the canonical torus partition function
\begin{equation}\label{canonical partition function as trace}
Z_{\text{can}}\left[\tau,\alpha_{s},\bar{\alpha}_{s}\right]
=
\text{Tr}_{\mathcal{H}}\, \exp 2\pi i\left[\tau\left( L_0-\frac{c}{24}\right) - \bar{\tau}\left( \bar{L}_0-\frac{c}{24}\right)
 + \sum_{s}\left(\alpha_s W_s^{(0)}-\bar{\alpha}_{s}\overline{W}_s^{(0)}\right)\right]
\end{equation}

\noindent where the trace is assumed to be taken over the Hilbert space $\mathcal{H}$ of the CFT, $W_{0}^{(s)}$ and $\overline{W}_0^{(s)}$ denote the zero modes of conserved currents of weight $(s,0)$ and $(0,s)$, respectively, and $\alpha_s\,$, $\bar{\alpha}_{s}$ the corresponding sources. In our conventions the torus has volume $\text{Vol}(T^2)=4\pi^{2}\text{Im}(\tau)\,$ with $\tau = \tau_{1} + i\beta/(2\pi)\,$, where $\beta$ is the inverse temperature. The sum over $s$ runs over the particular spectrum of operators present in the theory, which depends on the symmetry algebra in question.\footnote{In the holographic realization that we will study in section \ref{sec: Holo}, the spectrum is fixed by the choice of gauge algebra $\mathfrak{g}\oplus \mathfrak{g}$ for the bulk Chern-Simons theory, plus a choice of embedding of the $sl(2,\mathds{R})$ factor corresponding to the gravitational (spin-2) degrees of freedom into $\mathfrak{g}\,$.} Before proceeding further it is convenient to clarify our terminology: in agreement with common usage in the literature, we will refer to the CFT operators of conformal dimension greater than two as ``higher spin operators", to the symmetries they generate as ``higher spin symmetries", and to their sources as ``higher spin sources/chemical potentials". Therefore, in our CFT discussion we will often use the terms conformal dimension and spin interchangeably.

One notices that
\begin{align}
2\pi i\tau\left( L_0-\frac{c}{24}\right) - 2\pi i\bar{\tau}\left( \bar{L}_0-\frac{c}{24}\right)
={}&
 -\beta H +2\pi i\tau_{1}J\,,
\end{align}

\noindent  where $H = L_0+\bar{L}_0 -\frac{c}{12}$ is the Hamiltonian and $J = L_0 - \bar{L}_0$ the angular momentum, with $L_0$, $\bar{L}_0$ the Virasoro generators on the cylinder.\footnote{As usual, one thinks of the torus as a cylinder of finite length with the ends identified up to a twist.} Defining the chemical potentials 
\begin{equation}\label{chemical potentials def}
\Omega \equiv  \frac{i\tau_{1}}{\beta}\,,\qquad \mu_{s} \equiv \frac{i\alpha_s}{\beta}\,,\qquad \bar{\mu}_{s}\equiv -\frac{i \bar{\alpha}_s}{\beta}
\end{equation}

\noindent we see that the partition function describes a theory with density operator
\begin{equation}
\hat{\rho} = \frac{e^{-\beta H_{\mu}}}{Z_{\text{can}}\left[\beta,\Omega,\mu_{s},\bar{\mu}_{s}\right]}\,,
\end{equation}

\noindent where the deformed Hamiltonian $H_{\mu}$ is given by
\begin{equation}\label{deformed Hamiltonian}
H_{\mu} \equiv H  -2\pi\Omega J- 2\pi\sum_{s}\left(\mu_s W_s^{(0)}+\bar{\mu}_{s}\overline{W}_s^{(0)} \right).
\end{equation}

\subsection{Partition function in first order form}\label{sec: partition function}
We will now assume the theory possesses a Lagrangian representation. Denoting the set of fields collectively by $\phi$, and their (Euclidean) conjugate momenta by $P\,$, the partition function can be written in a path integral representation as
\begin{equation}\label{2d first order path integral 1}
Z_{\text{can}}\left[\beta,\Omega,\mu_{s},\bar{\mu}_{s}\right]
=
\int\mathcal{D}\phi\,\mathcal{D}P\, e^{\tilde{I}^{(E)}\left(P,\phi\right)}\,,
\end{equation}

\noindent where the Hamiltonian form of the action is
\begin{equation}
\tilde{I}^{(E)}(P,\phi)
=
\int_{0}^{\beta} dt_{E}\int_{0}^{2\pi}d\sigma \left[-P\dot{\phi}-\mathcal{H} +\Omega\mathcal{J}
+ \sum_{s}\left(\mu_{s}\mathcal{W}_{s}+\bar{\mu}_{s} \overline{\mathcal{W}}_{s}\right)\right]
\end{equation}

\noindent with $\dot{\phi} = \partial_{t_{E}}\phi$ and
\begin{equation}
\oint d\sigma\,\mathcal{H} = H\,,\quad \oint\frac{d\sigma}{2\pi}\mathcal{J} = J\,,\quad \oint \frac{d\sigma}{2\pi}\mathcal{W}_{s} = W_{0}^{(s)}\,,\quad \oint \frac{d\sigma}{2\pi}\overline{\mathcal{W}}_{s} = \overline{W}_{0}^{(s)}\,.
\end{equation}

A few comments are in order here. First, we notice that it is the rescaled sources $\mu =i\beta^{-1}\alpha\,$, namely the chemical potentials, that enter in the action. This is the usual result in finite-temperature field theory, and can be established by carefully discretizing the operator trace (see \cite{Kapusta:1981aa,Landsman:1986uw} for example). Secondly, in the reasoning above the potential $\Omega$ for angular momentum was treated in the same footing as the other deformations. We can instead ``geometrize" this potential by introducing a twist in the boundary conditions. Doing so the partition function becomes 
 \begin{align}\label{2d first order path integral 2}
Z_{\text{can}}\left[\beta,\Omega,\mu_{s},\bar{\mu}_{s}\right]
={}&
 \int\mathcal{D}\phi\,\mathcal{D}P\, e^{I^{(E)}\left(P,\phi\right)}
\\
\text{with}\qquad I^{(E)}\left(P,\phi \right)
 ={}&
\int_{T^{2}}d^{2}z \left[-P\dot{\phi}-\mathcal{H} +\sum_{s}\Bigl(\mu_{s}\mathcal{W}_{s}(P,\phi)+\bar{\mu}_{s} \overline{\mathcal{W}}_{s}(P,\phi)\Bigr)\right]
\end{align}

\noindent where $d^{2}z$ is the standard measure on the Euclidean plane (we are assuming a flat torus) and the path integral is performed with boundary conditions
\begin{equation}
\phi(z) = \phi(z+2\pi) = \phi(z + 2\pi \tau).
\end{equation}

\noindent Notice that while we have been working with constant $\mu_s$, $\bar{\mu}_s$ up to now, we are free to make $\mu_s$ and $\bar{\mu}_s$ time- and space-dependent in this path integral representation of $Z_{\text{can}}\,$, as long as we specialize to constant $\mu_s$, $\bar{\mu}_s$ when we want to compute $Z_{\text{can}}\,$. 

A point that will be crucial for the considerations to follow is that in general the currents $\mathcal{W}_{s}$ corresponding to higher spin operators are at least cubic in momenta. Therefore, if we transition to a Lagrangian path integral description by integrating out the momenta (i.e. Legendre-transforming) we find that the resulting action is non-linear in the sources, and in fact it will generically involve mixing between the two chiral sectors. What this means is that the canonical partition function is in general quite different from a simple second order version of the path integral with linear couplings, which we denote by $Z_{\text{Lag,naive}}\,$:
\begin{align}\label{2d first order path integral 3}
Z_{\text{Lag,naive}}\left[\beta,\Omega,\mu_{s},\bar{\mu}_{s}\right]
={}&
 \int\mathcal{D}\phi\, e^{-S_0(\phi)}e^{-\int_{T^{2}}d^{2}z \sum_{s}\left(\mu_{s}\mathcal{W}_{s}(\phi)+\bar{\mu}_{s} \overline{\mathcal{W}}_{s}(\phi)\right)}
\end{align}

\noindent where $S_0$ is the Lagrangian action in the absence of deformations. Fortunately, as we will discuss in detail in the rest of this section, for the purpose of deriving the Ward identities obeyed by the partition function $Z_{\text{can}}$ it will suffice to stay within the first order form of the action, where the deformations appear only linearly and the two chiral sectors do not mix. 

It is important to emphasize that the action deformed by linear couplings which enters the path integral \eqref{2d first order path integral 3} is not invariant under the higher spin symmetries furnished by the currents when both chiral sectors are deformed simultaneously, even if one allows the sources to transform. When both chiralities are present an invariant action involves corrections to all orders in the sources \cite{Hull:1989wu,Hull:1990pg,Schoutens:1990ja}, and we will return to this point in section \ref{subsec: non-chiral defs}. In general, this means that the naive partition function $Z_{\text{Lag,naive}}$ does not obey the usual Ward identities when both $\mu_{s}$ and $\bar{\mu}_{s}$ are switched on. The fact that $Z_{\text{can}}$ and $Z_{\text{Lag,naive}}$ are different objects is in fact true even for deformations involving ``lower spin" currents (relevant operators), and has important consequences for modular invariance, for example. To illustrate this point, in appendix \ref{app: U1} we review an example involving $U(1)$ currents in a free compact boson realization. 

Our next goal is to derive the Ward identities obeyed by the canonical partition function $Z_{\text{can}}\,$. For the sake of concreteness, we will often resort to a theory with $\mathcal{W}_{3}$ symmetry deformed by sources for the stress tensor and weight-3 currents as our basic example. Even though we will use a simple boson realization to derive these identities, we will find that the result is completely fixed by the symmetry algebra and does not rely on details of the explicit realization. By the same token, our conclusions will be general enough to later allow us to make a connection with flat connections in three dimensions and to find the appropriate boundary conditions these should obey in order to reproduce the canonical computations (c.f. section \ref{sec: Holo}). We will first work in Lorentzian signature, where the discussion of symmetries, conserved charges and Ward identities is more transparent. When discussing the Lorentzian theory on the cylinder we will often refer to the chemical potentials $\mu$, $\bar{\mu}$ as the sources. On the other hand, once we transition to the finite temperature theory defined on the torus we will reserve the term sources to denote the $\alpha =- i\beta \mu\,$, $\bar{\alpha} = i\beta \bar{\mu}\,$.

\subsection{A $\mathcal{W}_{3}$ theory in Hamiltonian form}
Free field realizations of the $\mathcal{W}$-current algebras were originally discussed in \cite{Zamolodchikov:1985wn,Fateev:1987vh}. Here we will follow the Hamiltonian approach employed  in \cite{Mikovic:1991rf}, which will prove very advantageous. Consider then a theory of $n$ real bosons $X^{i}$  ($i=1,\ldots,n$) on the cylinder with coordinates $(t,\sigma)$ (where $\sigma \simeq \sigma+ 2\pi$). We denote the canonical momentum conjugate to $X^{i}$ by $P_{i}\,$, with equal-time Poisson brackets
\begin{align}
\Bigl\{P_{i}(\sigma,t),X^{j}(\sigma',t)\Bigr\} &= \delta^{j}_{\hphantom{j}i}\,\delta(\sigma-\sigma')\,,
\end{align}

\noindent and raise and lower Latin indices with the flat metric $\delta_{ij}\,$.\footnote{If so desired, it is possible to introduce a non-trivial metric on the target space \cite{Hull:1989wu,Hull:1990pg}.} Define now
\begin{equation}
\Pi_{\pm}^{i} = \frac{1}{\sqrt{2}}\left(P^{i} \pm \partial_{\sigma} X^{i}\right),
\end{equation}

\noindent which satisfy 
\begin{align}\label{pi pi brackets}
\left\{\Pi_{\pm}^{i}(\sigma,t),\Pi_{\mp}^{j}\left(\sigma',t\right)\right\}&=0
\\
\left\{\Pi_{\pm}^{i}(\sigma,t),\Pi_{\pm}^{j}\left(\sigma',t\right)\right\} &= \mp \delta^{ij}\partial_{\sigma}\delta(\sigma-\sigma')\,.
\label{pi pi brackets 2}
\end{align}

\noindent One then constructs the generators 
\begin{equation}\label{stress tensor}
W^{(s)}_{\pm} = \frac{1}{s}d_{i_{1}\ldots i_{s}}\Pi_{\pm}^{i_{1}}\ldots \Pi^{i_{s}}_{\pm}\,,
\end{equation}

\noindent with $s=2,3,\ldots\, N$, where the $d_{i_{1}\ldots i_{s}}$ are constant symmetric tensors of rank $s\,$. The basic Poisson brackets \eqref{pi pi brackets}-\eqref{pi pi brackets 2} imply that these generators fulfill two decoupled copies of the $\mathcal{W}_{N}$ algebra (with no central extensions) provided the coefficients $d_{i_{1}\ldots i_{s}}$ satisfy certain algebraic relations that guarantee the closure of the algebra  \cite{Hull:1990pg}. For example, defining $T_\pm = W^{(2)}_{\pm}$ and $W_{\pm} = W^{(3)}_{\pm}\,$, in the $\mathcal{W}_{3}$ case one finds\footnote{Since we are working with equal-time Poisson brackets, in order to simplify the notation we will often suppress the explicit time dependence of the currents and other quantities.}
\begin{eqnarray}\label{TT bracket non extended}
\Bigl\{T_{\pm}(\sigma),T_{\pm}(\sigma')\Bigr\} 
&=&
\mp\Bigl[2T_{\pm}(\sigma) \partial_{\sigma}\delta\left(\sigma -\sigma'\right) + \delta\left(\sigma-\sigma'\right)\partial_{\sigma}T_{\pm}(\sigma)\Bigr]
\\
\Bigl\{T_{\pm}(\sigma),W_{\pm}(\sigma')\Bigr\} 
&=&
\mp\Bigl[3W_{\pm}(\sigma)\partial_{\sigma}\delta\left(\sigma-\sigma'\right)+2\delta\left(\sigma-\sigma'\right)\partial_{\sigma}W_{\pm}(\sigma)\Bigr]
\\
\Bigl\{W_{\pm}(\sigma),W_{\pm}(\sigma')\Bigr\} 
&=&
 \mp 4\kappa\Bigl[T_{\pm}(\sigma)^{2}\partial_{\sigma}\delta\left(\sigma-\sigma'\right) + \delta(\sigma-\sigma')T_{\pm}(\sigma)\partial_{\sigma} T_{\pm}(\sigma)\Bigr]
\end{eqnarray}

\noindent provided \cite{Hull:1989wu,Romans:1990ag}
\begin{equation}\label{d tensor}
d_{ij} = \delta_{ij}\,,\qquad 
d_{(ijk}d^{k}_{\hphantom{k}m)n} = \kappa\, \delta_{(ij}\delta_{m)n}\,.
\end{equation}

\noindent Below we will discuss how to generalize this construction to allow for a semiclassical central charge $c\,$, in terms of which $\kappa = -16/c\,$. The condition on $d_{ijk}$ guarantees that the spin-4 term in the r.h.s. of the $\left\{W,W\right\} $ bracket is proportional to $T^{2}\,$, closing the algebra of the stress tensor $T$ and the dimension-3 current $W$, albeit non-linearly. We stress that, since we are using Poisson brackets and working at the semiclassical level, we have considered the product of currents such as $T^{2}\,$ without worrying about operator ordering issues.

Before integrating over momenta, the partition function for the deformed theory involves the first-order action
\begin{equation}
I = \int d\sigma  dt \left[P_{i}\dot{X}^i -\frac{1}{2}\left(P^{i}P_{i} + \partial_{\sigma}X^{i}\partial_{\sigma}X_{i}\right)- \mu_{2}^{+}T_{+} - \mu_{2}^{-}T_{-} - \mu_{3}^{+}W_{+}-\mu_{3}^{-}W_{-}\right]
\end{equation}

\noindent whose symmetries we want to study. The dot notation indicates time derivatives as usual. A convenient feature of the first order formalism is that the $W^{(s)}_{+}$ and $W^{(s)}_{-}$ generators Poisson-commute, so the separation of left- and right-movers is exact. To avoid unnecessary clutter we will often work exclusively with the $+$ sector and drop the subindex to simplify the notation, i.e. we use $T \equiv T_{+}\,$, $W \equiv W_+$ and so forth when there is no risk for confusion. Naturally, all the conclusions apply to the other chiral sector as well. 

The key point we want to stress is that integrating out the momenta one obtains the second order form of the action, which is non-linear in the sources and mixes left- and right-movers in a non-trivial way. A related observation is that, in the absence of deformations (i.e. $\mu_{2}^{\pm}=\mu_{3}^{\pm}=0$) the equation of motion for $P_{i}$ implies $P_{i} = \partial_{t}X^{i}\,$, so that $\Pi_{\pm}^{i} = \partial_{\pm}X^{i}$ in the undeformed theory. The undeformed currents are then schematically of the form $W^{(s)}_{\pm} \sim \left(\partial_{\pm} X\right)^{s}$ and obviously chiral. On the other hand, when the chemical potentials are switched on the $P_{i}$ acquire explicit dependence on them to all orders, and so do the currents themselves. For the purpose of studying the symmetries of the partition function and the associated Ward identities it will be very advantageous to stay within the first order formulation, because the sources enter linearly and the chiral sectors remain factorized.

\subsection{Adding central extensions}
We will now extend the Hamiltonian analysis of \cite{Mikovic:1991rf} to include classical central extensions. This can be achieved by adding improvement terms to the generators, often times called ``background charges" in the literature, along the lines of \cite{Fateev:1987vh,Hull:1989wu,Romans:1990ag}:
\begin{align}\label{improved T}
T ={}&
 \frac{1}{2}\delta_{ij}\Pi^{i} \Pi^{j} + a_{i}\partial_{\sigma} \Pi^{i}
\\
W ={}&
  \frac{1}{3}d_{ijk}\Pi^{i} \Pi^{j} \Pi^{k} + e_{ij}\partial_{\sigma} \Pi^{i} \Pi^{j} + f_{i}\partial_{\sigma}^{2} \Pi^{i}\,,
  \label{improved W}
\end{align}

\noindent where the $a_{i}$, $e_{ij}$ and $f_{i}$ are constant coefficients. With these additions, the $\mathcal{W}_{3}$ Poisson algebra becomes
\begin{eqnarray}\label{TT bracket extended}
\Bigl\{T(\sigma),T(\sigma')\Bigr\} 
&=&
-\left[2T(\sigma) \partial_{\sigma}\delta\left(\sigma -\sigma'\right) + \delta\left(\sigma-\sigma'\right)\partial_{\sigma}T(\sigma)+ \frac{c}{12}\partial_{\sigma}^{3}\delta\left(\sigma-\sigma'\right)\right] 
\\
\label{TW bracket extended}
\Bigl\{T(\sigma),W(\sigma')\Bigr\} 
&=&
-\Bigl[3W(\sigma)\partial_{\sigma}\delta\left(\sigma-\sigma'\right)+2\delta\left(\sigma-\sigma'\right)\partial_{\sigma}W(\sigma)\Bigr]
\\
\label{WW bracket extended}
\Bigl\{W(\sigma),W(\sigma')\Bigr\} 
&=&
 \frac{64}{c}\Bigl[T^{2}(\sigma)\partial_{\sigma}\delta(\sigma-\sigma')+\delta(\sigma-\sigma')T(\sigma)\partial_{\sigma}T(\sigma)\Bigr]
\nonumber\\
&&
+ \, 3\,\partial_{\sigma}\delta(\sigma-\sigma')\partial^{2}_{\sigma}T(\sigma)
+ 5\,\partial_{\sigma}^{2}\delta(\sigma-\sigma')\partial_{\sigma} T(\sigma)
\\
&&
+\, \frac{2}{3}\delta(\sigma-\sigma')\partial_{\sigma}^{3}T(\sigma)+\frac{10}{3}\partial^{3}_{\sigma}\delta(\sigma-\sigma')T(\sigma)+\frac{c}{36}\partial^{5}_{\sigma}\delta(\sigma-\sigma')
\nonumber
\end{eqnarray}

\noindent provided the various coefficients satisfy \eqref{a squared is central charge b}-\eqref{last coeff constraint} (in particular $a_{i}a^{i} = -\frac{c}{12} \,$, where $c$ is the semiclassical central charge), and similarly in the other chiral sector. A feature that distinguishes the non-linear Poisson algebras such as \eqref{TT bracket extended}-\eqref{WW bracket extended} from their linear counterparts is that, upon normal-ordering the products of currents, the Jacobi identities (associativity) will imply that the structure constants in the quantum version of the algebra acquire $\mathcal{O}(1/c)$ corrections (see e.g. \cite{Romans:1990ag}). It is in this sense that the non-linear Poisson bracket algebra is a ``large-$c$" version of the full quantum algebra.

With the Poisson algebra at our disposal, we can compute the transformation of the currents under the various symmetries. Defining the integrated spin-2 and spin-3 charges 
\begin{align}
Q^{(2)} ={}&
 \int d\sigma'\, \epsilon\left(\sigma'\right)T\left(\sigma'\right)
\\
Q^{(3)} ={}&
 \int d\sigma'\, \chi\left(\sigma'\right)W\left(\sigma'\right)\,,
\end{align}

\noindent under an infinitesimal spin-2 transformation one finds
\begin{align}\label{spin 2 transf of currents}
\delta_{\epsilon} T ={}& \left\{Q^{(2)},T\right\}
=
\epsilon\,\partial_{\sigma}T+
2T\,\partial_{\sigma}\epsilon
+\frac{c}{12}\partial^{3}_{\sigma}\epsilon
\\
\delta_{\epsilon} W ={}& \left\{Q^{(2)},W\right\}
=
 \epsilon\,\partial_{\sigma}W+
3W\partial_{\sigma}\epsilon
\end{align}

\noindent (with similar expressions in the other chiral sector) and we recognize the transformation of the stress tensor and a weight-3 primary operator under diffeomorphisms of the form $x^+\rightarrow x^+ + \epsilon(\sigma)$.
Similarly, under the spin-3 symmetry one finds
\begin{align}
\delta_{\chi} T={}& \left\{Q^{(3)},T\right\}
=
 2\chi\partial_{\sigma}W+
3W\partial_{\sigma}\chi
\\
\delta_{\chi} W={}& \left\{Q^{(3)},W\right\}
=
-\Biggl[
\frac{64}{c}\Bigl( \chi \,T\,\partial_{\sigma} T + T^{2}\,\partial_{\sigma}\chi\Bigr) + \frac{c}{36}\partial^{5}_{\sigma}\chi
\nonumber\\
&\qquad\qquad\quad\qquad + \frac{1}{3}\Bigl(2\,\chi\,\partial^{3}_{\sigma}T+9\,\partial_{\sigma}\chi\,\partial^{2}_{\sigma}T+15\,\partial_{\sigma}^{2}\chi\,\partial_{\sigma}T+10\,
T\,\partial^{3}_{\sigma}\chi\Bigr)
\Biggr].
\label{spin 3 transf of currents}
\end{align}

\subsection{Symmetries of the action}
Let us now discuss the symmetries of the action. To this end  it is useful to think of the sources as gauge fields, i.e. Lagrange multipliers imposing constraints that generate the $\mathcal{W}$ algebra or any other symmetry in question. We emphasize however that the sources are background fields which are not being integrated over in the path integral. We will denote the currents generating the symmetry of interest by a vector $\vec{J}$ with components $J_{\alpha}\,$, and the corresponding Lagrange multipliers by a vector $\vec{\mu}$ with components $\mu^{\alpha}\,$. In our $\mathcal{W}_{3}$ example we will have $\vec{J} = \{T_+,T_-,W_+,W_-\}$ and $\vec{\mu} =\{\mu_{2}^{+},\mu_{2}^{-},\mu_{3}^{+},\mu_{3}^{-}\}\,$. The action we consider is then of the generic form
\begin{equation}\label{gauge action}
I = \int d\sigma dt\left(P_i \dot{X}^i - H_0 - \mu^{\alpha}J_{\alpha}\right)
\end{equation}

\noindent where $H_0$ denotes the undeformed Hamiltonian. We will study the symmetries of the associated partition function using the improved generators, i.e. when the algebra acquires semiclassical central extensions:
\begin{align}\label{extended constraint algebra}
\Bigl\{J_{\alpha}(\sigma),J_{\beta}(\sigma')\Bigr\}  ={}&
 \int dx \,f_{\alpha\beta}^{\hphantom{\alpha\beta}\gamma}(\sigma,\sigma',x)J_{\gamma}(x) +  c_{\alpha\beta}(\sigma,\sigma')\,,
\end{align}

\noindent where as before we have suppressed the explicit time dependence of the currents for the sake of notational simplicity. The functions $c_{\alpha\beta}$ are proportional to the semiclassical central charge $c\,$, but do not depend on the phase space variables. 

 Before moving forward, we can take two steps that will simplify the task of finding the Ward identities obeyed by the currents in the presence of sources. First, we note that the undeformed Hamiltonian is simply $H_0 = T_+ + T_-\,$.\footnote{More precisely, in the presence of improvement terms we have $H_0 = \frac{1}{2}\left(P^{i}P_{i} + \partial_{\sigma}X^{i}\partial_{\sigma}X_{i}\right) =T_{+} + T_{-}- a^{+}_{j}\partial_{\sigma}\Pi^{j}_{+}  - a_{j}^{-}\partial_{\sigma}\Pi^{j}_{-}$, but the total $\sigma$-derivatives do not contribute to the Hamiltonian $\int d\sigma\, H_0\,$.} It is then possible to eliminate $H_0$ from the action by shifting the spin-2 chemical potentials as $\mu_{2}^{\pm} \rightarrow \nu_{2}^{\pm}-1$, while keeping the higher spin chemical potentials the same. Consequently, for practical purposes we will define a new vector $\vec{\nu}$ with components  $\{\nu_{2}^+,\nu_{2}^{-},\mu_{3}^+,\mu_{3}^{-},\ldots\}$ and drop $H_{0}\,$. Even though the shift in the spin-2 deformation could be thought of as a ``gauge choice", the undeformed theory has generically a non-zero Hamiltonian $H_{0}\,$, so we must remember to translate our results back to the $\mu^{\alpha}$ at the end if we are to interpret the sources strictly as deformations of the original theory. Secondly, we will define for convenience 
an auxiliary action $I_{c}$ which includes an ``identity gauge field" $\nu_{c}$  which can be thought of as coupling to an extra Abelian generator \cite{Hull:1990mu}:
\begin{equation}\label{def auxiliary action}
I_{c} = \int d\sigma dt\left(P_i \dot{X}^i  - \nu^{\alpha}J_{\alpha} - \nu_{c}\cdot 1\right).
\end{equation}

\noindent The role of this additional Lagrange multiplier, which is purely a bookkeeping device, is to cancel contributions to the variation of the action coming from central extensions. Naturally, at the end of the day we will set $\nu_{c}=0$ in order to obtain the Ward identities obeyed by the original partition function. 

We are now in position to discuss the symmetries of the action in the presence of deformations. It is straightforward to check that under an infinitesimal transformation of the fields and sources of the form
\begin{eqnarray}\label{inf transf 1}
\delta P_{i}(\sigma)
 &=&
 \int d\sigma'\, \epsilon^{\alpha}(\sigma')\Bigl\{J_{\alpha}(\sigma'),P_{i}(\sigma)\Bigr\}
\\
\delta X^{i}(\sigma)
 &=&
 \int d\sigma'\, \epsilon^{\alpha}(\sigma')\Bigl\{J_{\alpha}(\sigma'),X^i(\sigma)\Bigr\}
\\
\delta \nu^{\alpha}(\sigma)
&=&
\dot{\epsilon}^{\alpha}(\sigma)  -  \int d\sigma' dx\, \nu^{\beta}(x)\epsilon^{\gamma}(\sigma')f_{\gamma\beta}^{\hphantom{\gamma\beta}\alpha}(\sigma',x,\sigma)
 \label{inf transf 3}
\\
\delta \nu_{c}(\sigma) 
&=&
 \dot{\zeta}(\sigma)  -  \nu^{\beta}(\sigma) \int d\sigma' \,\epsilon^{\gamma}(\sigma')c_{\gamma\beta}(\sigma',\sigma) \,,
  \label{inf transf 4}
\end{eqnarray}

\noindent the auxiliary action \eqref{def auxiliary action} changes by a boundary term:
\begin{equation}\label{total derivative}
\delta I_{c} = \int dtd\sigma \,\partial_{t}\left(P_{i}\delta X^{i}-\epsilon^{\alpha}J_{\alpha}-\zeta\right).
\end{equation}

Let us now specialize these expressions to our $\mathcal{W}_{3}$ example. Using \eqref{extended struct const 1}-\eqref{central charge matrix 2} we can obtain the explicit transformation of the sources from \eqref{inf transf 3} (with $\epsilon^{\alpha} = \{\epsilon, \chi,\ldots\}$)
\begin{align}\label{delta sources 1b}
\delta \nu_{2}
 ={}&
 \partial_{t}\epsilon  - \nu_{2}\partial_{\sigma}\epsilon + \epsilon\partial_{\sigma}\nu_{2} -\frac{32}{c}\, T\bigl(\chi \partial_{\sigma}\mu_{3} - \mu_{3}\partial_{\sigma}\chi\bigr)
 \nonumber\\
 & 
 + \frac{2}{3}\mu_{3}\partial^{3}_{\sigma}\chi - \frac{2}{3}\chi\partial^{3}_{\sigma}\mu_{3} - \partial_{\sigma}\mu_{3}\partial^{2}_{\sigma}\chi + \partial_{\sigma}\chi \partial^{2}_{\sigma}\mu_{3}
 \\
 \delta\mu_{3} ={}&
 \partial_{t}\chi -\nu_{2}\partial_{\sigma}\chi + 2\chi\partial_{\sigma}\nu_{2}+\epsilon\partial_{\sigma}\mu_{3}-2\mu_{3}\partial_{\sigma}\epsilon
 \label{delta sources 4b}
\end{align}

\noindent where we used the shorthand $\nu_{2} \equiv \nu_{2}^{+}$ and $\mu_{3} \equiv \mu_{3}^{+}$, with similar expressions for the sources in the other chiral sector. Shifting back to the original spin-2 source $\mu_{2} = \nu_{2}-1$ we obtain the desired transformation rules:
\begin{align}\label{delta sources 1}
\delta \mu_{2}
 ={}&
 \partial_{-}\epsilon - \mu_{2}\partial_{\sigma}\epsilon + \epsilon\partial_{\sigma}\mu_{2} -\frac{32}{c}\, T\bigl(\chi \partial_{\sigma}\mu_{3} - \mu_{3}\partial_{\sigma}\chi\bigr)
  \nonumber\\
 & 
 + \frac{2}{3}\mu_{3}\partial^{3}_{\sigma}\chi - \frac{2}{3}\chi\partial^{3}_{\sigma}\mu_{3} - \partial_{\sigma}\mu_{3}\partial^{2}_{\sigma}\chi + \partial_{\sigma}\chi \partial^{2}_{\sigma}\mu_{3}
 \\
 \delta\mu_{3} ={}&
 \partial_{-}\chi -\mu_{2}\partial_{\sigma}\chi + 2\chi\partial_{\sigma}\mu_{2}+\epsilon\partial_{\sigma}\mu_{3}-2\mu_{3}\partial_{\sigma}\epsilon\,.
 \label{delta sources 4}
\end{align}

\noindent  Notice the appearance of the chiral derivative defined as $\partial_{-} = \partial_{t}-\partial_{\sigma}$ ($\partial_{+} = \partial_{t} + \partial_{\sigma}$). We then see that the only effect of the undeformed Hamiltonian $H_{0}$ is to turn the time derivatives in \eqref{inf transf 3} into chiral derivatives. 

It is worth emphasizing that the theory and in particular the partition function are defined at fixed values of the sources. The fact that one needs to transform the $\mu^{\alpha}$ in order to realize the symmetry, therefore moving in the space of theories, shows that generically these deformations will explicitly break the original conformal as well as higher spin and Lorentz 
symmetries.\footnote{The theory naively has new higher spin and Lorentz symmetries which one obtains by (i) performing
a higher spin transformation that puts all sources equal to zero, (ii) performing a higher spin transformation in the undeformed theory 
and (iii) performing the inverse higher spin transformation that puts all sources back to their original value. As we will discuss in section \ref{sec: Disc}, it is not entirely clear whether this is a proper symmetry of the deformed theory.}

\subsection{Ward identities}
Having derived the transformation of the sources, the basic result \eqref{total derivative} showing the invariance of the action under the combined transformation of background sources and fundamental fields will imply a Ward identity for the currents. From the point of view of the path integral, changing the fields $X^{i}$ and momenta $P_{i}$ is a just a change of integration variables. Hence, the symmetry \eqref{inf transf 1}-\eqref{inf transf 4} implies
\begin{equation}
\left\langle \int d\sigma dt \left( \frac{\delta I_{c}}{\delta \nu^{\alpha}}\delta \nu^{\alpha}+ \frac{\delta I_{c}}{\delta \nu_{c}}\delta \nu_{c}\right)\right\rangle \simeq 0\,,
\end{equation}

\noindent where $\simeq$ denotes equivalence up to surface terms (the integral of total time derivatives). Setting $\nu_{c} = 0$ in order to recover the Ward identity obeyed by the original partition function we obtain
\begin{equation}
\int d\sigma dt \left(-J_{\alpha}\delta \nu^{\alpha} +\int d\sigma'\,\nu^{\beta}(\sigma)  \,\epsilon^{\gamma}(\sigma')c_{\gamma\beta}(\sigma',\sigma)\right) \simeq 0\,,
\end{equation}

\noindent where the $J_{\alpha}$ are interpreted as the one-point function of the currents in the presence of external sources. Plugging the explicit form \eqref{inf transf 3}  of the variations $\delta \nu^{\alpha}$ and integrating by parts we find the identity
\begin{equation}\label{master Ward identity}
\partial_{t}J_{\alpha}(\sigma)  + \int d\sigma' dx\, \nu^{\beta}(x)f_{\alpha\beta}^{\hphantom{\alpha\beta}\gamma}(\sigma,x,\sigma')J_{\gamma}(\sigma')  +\int d\sigma' \, \nu^{\beta}(\sigma') c_{\alpha\beta}(\sigma,\sigma') =0\,.
\end{equation}

Note that defining the extended Hamiltonian
\begin{equation}
H_{\nu} \equiv \int d\sigma \,\nu^{\alpha}(\sigma)J_{\alpha}(\sigma)\,,
\end{equation}

\noindent the above Ward identity takes a very compact form:
\begin{equation}\label{master Ward identity 2}
\partial_{t}J_{\alpha}(\sigma)   = \Bigl\{H_{\nu},J_{\alpha}(\sigma)\Bigr\}.
\end{equation}

\noindent In other words, in Hamiltonian language the Ward identities are just the equations of motion of the currents, with the time evolution generated by the deformed Hamiltonian $H_{\nu}\,$. 

We emphasize that the source vector $\nu^{\alpha}$ in \eqref{master Ward identity} and \eqref{master Ward identity 2} contains the shifted spin-2 deformation $\nu_{2} = \mu_{2}+1$ that allowed us to absorb the undeformed Hamiltonian $H_{0}\,$. In the specific $\mathcal{W}_{3}$ example, using \eqref{extended struct const 1}-\eqref{central charge matrix 2} it is easy to see that \eqref{master Ward identity} yields, after shifting back to $\mu_{2}\,$,
\begin{align}\label{T Ward identity}
\partial_{-} T 
={}&
 \mu_{2}\partial_{\sigma}T + 2T\partial_{\sigma}\mu_{2}+\frac{c}{12}\partial^{3}_{\sigma}\mu_{2} + 3W\partial_{\sigma}\mu_{3} + 2\mu_{3}\partial_{\sigma}W 
\\
\label{W Ward identity}
\partial_{-} W 
={}&
\mu_{2}\partial_{\sigma} W + 3W\partial_{\sigma}\mu_{2} -\frac{64}{c}\left(T^{2}\partial_{\sigma}\mu_{3} +\mu_{3}T\partial_{\sigma}T\right) 
\nonumber\\
&
-\frac{10}{3}T\partial^{3}_{\sigma}\mu_{3} -5\partial_{\sigma}T\partial^{2}_{\sigma}\mu_{3} -3\partial^{2}_{\sigma}T\partial_{\sigma}\mu_{3}  -\frac{2}{3}\mu_{3}\partial^{3}_{\sigma}T - \frac{c}{36}\partial^{5}_{\sigma}\mu_{3}\,,
\end{align}

\noindent  where $\partial_{-} = \partial_{t} - \partial_{\sigma}\,$. Just as before, shifting back to $\mu_{2}$ produced an extra term that combined with the time derivatives in \eqref{master Ward identity} to turn them into chiral derivatives. This was to be expected, because $\partial_{-}T = 0$ and $\partial_{-}W=0$ are the Ward identities in the free theory (i.e. when $\mu_{2}=\mu_{3}=0$). More generally, if in a slight abuse of notation we let $J_{\alpha}$, $f_{\alpha\beta}^{\hphantom{\alpha\beta}\gamma}$ and $c_{\alpha\beta}$ denote the currents, structure constants and central extensions on a single chiral copy of the algebra, our results for the Ward identity in terms of the original sources $\mu^{\beta}$ becomes
\begin{equation}\label{master Ward identity mu}
\partial_{-}J_{\alpha}(\sigma)  + \int d\sigma' dx\, \mu^{\beta}(x)f_{\alpha\beta}^{\hphantom{\alpha\beta}\gamma}(\sigma,x,\sigma')J_{\gamma}(\sigma')  +\int d\sigma' \, \mu^{\beta}(\sigma') c_{\alpha\beta}(\sigma,\sigma') =0\,,
\end{equation}

\noindent with a similar expression in the other chiral sector ($\partial_{+}\bar{J}_{\alpha}(\sigma) + \ldots=0$). 

Even though the Ward identities were derived using an explicit realization in terms of scalars, it is clear from \eqref{master Ward identity} and \eqref{master Ward identity mu} that the end result is completely fixed by the symmetry algebra and therefore independent of the particular realization we have chosen. In other words, \eqref{T Ward identity}-\eqref{W Ward identity} are the semiclassical (large-$c$) Ward identities associated to the canonical partition function in any theory with $\mathcal{W}_{3}$ symmetry, in the presence of sources. It is also clear from the derivation that \eqref{master Ward identity mu} extends to any other 
closed symmetry algebra.

It is somewhat peculiar that the right-hand side of the Ward identities involves $\sigma$-derivatives as opposed to $x^{+}$-derivatives,
which to our knowledge has not been emphasized in the literature before. As we have seen this is an automatic consequence of
our canonical treatment, with the Hamiltonian as the starting point. In section \ref{sec: bcs} we will show that these Ward identities can be written as the flatness condition on $sl(N,\mathds{R})\oplus sl(N,\mathds{R})$ gauge fields with appropriate boundary conditions.

\section{Action deformations and the holomorphic partition function}\label{sec: holo partition function}
By now we have established the structure of the Ward identities corresponding to a deformation of the CFT Hamiltonian by higher spin currents. Our next task is to consider a different partition function obtained by deforming the CFT action. Many of the technical aspects in the analysis below are analogous to the canonical case discussed in depth in the previous section, so in what follows we will omit unessential details for the sake of brevity.

\subsection{Chiral deformations}
We will begin by studying the symmetries of the partition function and the associated Ward identities in the presence of chiral deformations. To this end we will again resort to the free boson realization, and consider an action of the form
\begin{equation}\label{chiral def action}
S = \int d^{2}x\left(\frac{1}{2}\partial_{+}X^{i} \partial_{-}X_{i} - \lambda^{\alpha}G_{\alpha} \right)
\end{equation}

\noindent where the vector $G = \{\hT,\hW\}$ contains the currents\footnote{As the notation indicates, these chiral currents are different from their canonical counterparts \eqref{stress tensor}, and only agree with them in the absence of sources.}
\begin{equation}\label{holomorphic currents}
\hT = \frac{1}{2}\partial_+ X^{i}\partial_{+}X^{i}\,,\qquad \hW = \frac{1}{3}d_{ijk}\partial_{+}X^{i}\partial_{+}X^{j}\partial_{+}X^{k}
\end{equation}

\noindent and $\lambda = \{\lambda_{2},\lambda_{3}\}$ the corresponding sources. Following the Noether procedure, it was established  long ago that this linear coupling is in fact enough for the action with chiral deformations to enjoy a gauge invariance \cite{Hull:1989wu,Hull:1990pg}, akin to a chiral half of \eqref{inf transf 1}-\eqref{inf transf 3}. A particularly transparent way of understanding this result, which also allows to make direct contact with the calculations in section \ref{sec: canonical}, is to realize that the above action is amenable to study in a Hamiltonian formalism with the light-cone direction $x^{-}$ thought of as ``time" \cite{Witten:1983ar,Hull:1990pg,Mikovic:1990ha}, and where the undeformed Hamiltonian is identically zero, $H_0=0\,$.

The key observation in \cite{Mikovic:1990ha} is that, after taking into account the presence of the second class constraint $(1/2)\partial_+ X_{i} - P_i = 0\,$, the basic Dirac bracket $\{,\}_{D}$ reads
\begin{equation}
\Bigl\{\partial_{+} X_{i}(x^{+},x^{-}),\partial_{+}X_{j}(y^{+},x^{-})\Bigr\}_{D} = \delta_{ij}\, \partial_{+}\delta\left(x^{+} - y^{+}\right).
\end{equation}

\noindent Given the form of this bracket and the currents \eqref{holomorphic currents} (compare with the basic canonical bracket \eqref{pi pi brackets 2} and currents \eqref{stress tensor}), from the reasoning in the previous section it is clear that the holomorphic currents enjoy the Dirac bracket algebra 
\begin{eqnarray}\label{holo TT bracket non extended}
\Bigl\{\hT(x^{+}),\hT(y^{+})\Bigr\}_{D} 
&=&
-\Bigl[2\hT(x^{+}) \partial_{+}\delta\left(x^{+}-y^{+}\right) + \delta\left(x^{+}-y^{+}\right)\partial_{+}\hT(x^{+})\Bigr]
\\
\Bigl\{\hT(x^{+}),\hW(y^{+})\Bigr\}_{D} 
&=&
-\Bigl[3\hW(x^{+})\partial_{+}\delta\left(x^{+}-y^{+}\right)+2\delta\left(x^{+}-y^{+}\right)\partial_{+}\hW(x^{+})\Bigr]
\\
\Bigl\{\hW(x^{+}),\hW(y^{+})\Bigr\}_{D}
&=&
 - 4\kappa\Bigl[\hT(x^{+})^{2}\partial_{+}\delta\left(x^{+}-y^{+}\right) + \delta(x^{+}-y^{+})\hT(x^{+})\partial_{+} \hT(x^{+})\Bigr]
\end{eqnarray}

\noindent provided $d_{(ijk}d^{k}_{\hphantom{k}m)n} = \kappa\, \delta_{(ij}\delta_{m)n}$ as before. Classical central extensions can be incorporated exactly as in the canonical analysis by adding improvement terms to the generators, which will now involve terms of higher order in chiral derivatives, e.g. $\hT = (1/2)\partial_+X^i\partial_+X^i +a_{i}\partial_{+}^{2}X^{i}$ (compare with \eqref{improved T}). It is then immediate that the improved generators fulfill one copy of the centrally extended algebra \eqref{TT bracket extended}, \eqref{TW bracket extended}, \eqref{WW bracket extended}, with spatial derivatives $\partial_{\sigma}$ replaced by chiral derivatives $\partial_+\,$, provided the coefficients of the improvement terms obey the constraints \eqref{a squared is central charge b}-\eqref{last coeff constraint}. In simple terms, all the calculations performed in the canonical formulation carry over to the chiral deformation case provided one replaces $\Pi^{i}_{+} \to \partial_{+} X^{i}$ and $\partial_{\sigma}\to \partial_{+}\,$. 

Parameterizing the extended Dirac brackets of the currents as (in order to simplify the notation we omit the explicit $x^-$ dependence below)
 \begin{align}\label{extended holomorphic bracket}
\Bigl\{G_{\alpha}(x^{+}),G_{\beta}(y^{+})\Bigr\}_{D}  ={}&
 \int dz^{+} \,f_{\alpha\beta}^{\hphantom{\alpha\beta}\gamma}(x^{+},y^{+},z^{+})G_{\gamma}(z^{+}) +  c_{\alpha\beta}(x^{+},y^{+})\,,
\end{align}

\noindent it was shown in \cite{Mikovic:1990ha} that under the infinitesimal transformation 
\begin{align}
\delta X^{i} ={}&
 \sum_{n\geq 1}(-1)^{n-1}\partial_+^{n-1}\left(\epsilon^{\alpha}\frac{\partial G_{\alpha}}{\partial\left(\partial^{n}_{+}X_{i}\right)}\right)
\\
\delta \lambda^{\alpha} ={}&
\partial_{-}\epsilon^{\alpha}
-\int dy^{+}dz^{+}\, \lambda^{\beta}(y^+,x^-)\epsilon^\gamma(z^+,x^-)f_{\gamma\beta}^{\hphantom{\gamma\beta}\alpha}(z^{+},y^{+},x^{+})
\end{align}

\noindent the action \eqref{chiral def action} transforms as\footnote{Just as in the case of a Hamiltonian deformation, one can alternatively introduce an extra Abelian generator whose transformation is such that the modified action is invariant.}
\begin{equation}\label{transf of chiral action}
\delta S \simeq -\int dx^{-} dx^+dy^{+}\,\lambda^{\beta}(x^{+},x^{-})  \,\epsilon^{\gamma}(y^{+},x^{-})c_{\gamma\beta}(y^{+},x^{+})\,,
\end{equation}

\noindent where as before $\simeq$ denotes equivalence up to surface terms. Repeating the manipulations that lead to \eqref{master Ward identity}, in this case we find the Ward identities 
\begin{equation}\label{master Ward identity chiral}
\partial_{-}G_{\alpha}(x^+)  + \int dz^+dy^+\, \lambda^{\beta}(y^{+})f_{\alpha\beta}^{\hphantom{\alpha\beta}\gamma}(x^+,y^+,z^+)G_{\gamma}(z^+)  +\int dz^+ \, \lambda^{\beta}(z^+) c_{\alpha\beta}(x^+,z^+) =0\,.
\end{equation}

\noindent Using the fact that the structure constants $f_{\alpha\beta}^{\hphantom{\alpha\beta}\gamma}$ and central extensions $c_{\alpha\beta}$ now involve $\partial_+$ derivatives (as opposed to $\partial_{\sigma}$ derivatives), in the $\mathcal{W}_{3}$ example we find
\begin{align}
\label{chiral T Ward identity}
\partial_{-} \hT ={}&
 \mu_{2}\partial_{+}\hT + 2\hT\partial_{+}\mu_{2}+\frac{c}{12}\partial^{3}_{+}\mu_{2} + 3\hW \partial_{+}\mu_{3} + 2\mu_{3}\partial_{+}\hW 
\\
\label{chiral W Ward identity}
\partial_{-} \hW ={}&
\mu_{2}\partial_{+} \hW + 3\hW\partial_{+}\mu_{2} -\frac{64}{c}\left(\hT^{2}\partial_{+}\mu_{3} +\mu_{3}\hT\partial_{+}\hT\right) 
\nonumber\\
&
-\frac{10}{3}\hT\partial^{3}_{+}\mu_{3} -5\partial_{+}\hT\partial^{2}_{+}\mu_{3} -3\partial^{2}_{+}\hT\partial_{+}\mu_{3}  -\frac{2}{3}\mu_{3}\partial^{3}_{+}\hT - \frac{c}{36}\partial^{5}_{+}\mu_{3}\,.
\end{align}

 Just as for the algebra itself, the Ward identities associated with a chiral deformation of the field theory action have the same form as those associated with a chiral deformation of the Hamiltonian, but with spatial derivatives $\partial_{\sigma}$ replaced by light-cone derivatives $\partial_{+}\,$. As shown in e.g. \cite{Gutperle:2011kf}, these Ward identities also follow by computing the one point of $\hT$ and $\hW$ in the presence of the insertion $e^{\int d^{2}z\,\lambda^{\alpha}G_{\alpha}}$ by expanding the exponential and using the OPE of the holomorphic currents. In this sense, \eqref{chiral T Ward identity}-\eqref{chiral W Ward identity} could be said to be the ``usual" Ward identities.  As discussed in section \ref{sec: Holo} and appendix \ref{app: azbar2}, these Ward identities (in fact two chiral copies of them) can be rewritten as the flatness condition on $sl(N)\oplus sl(N)$ gauge fields with appropriate boundary conditions.

\subsection{The coupling to the modular parameter and the notion of energy}\label{subsec: tau coupling}
Consider adding to the Euclidean free boson action a chiral stress tensor deformation and a chiral deformation by a weight-$s$ current:
\begin{equation}\label{chiral spin 2 deformation}
S = \int d^{2}z\left(\frac{1}{2}\partial_{z}X^{i}\partial_{\bar{z}}X^{i} -\mu_{2}\mathcal{L}- \mu_{s} \mathcal{W}_{s}\right)
\end{equation}

\noindent where
\begin{equation}\label{chiral currents}
\mathcal{L}=\frac{1}{2}\partial_{z} X^{i}\partial_{z} X^{i}\,,\qquad \mathcal{W}_{s} = \frac{1}{s}d_{i_{1}\ldots i_{s}}\partial_{z} X^{i_1}\ldots \partial_{z} X^{i_s}\,.
\end{equation}

\noindent For simplicity we have not added improvement terms to the generators, which as we have seen allow to incorporate semiclassical central charges. We will moreover restrict to odd $s$, in which case the closure of the Dirac bracket algebra (obtained interpreting $\bar{z}$ as time) requires \cite{Hull:1990pg}
\begin{equation}\label{d tensor identity}
d_{i\, (i_2\ldots i_s}d^{i}_{\hphantom{i}j_2\ldots )j_s} = \frac{\kappa}{2^{s-1}} \delta_{(i_{2}i_{3}}\ldots\delta_{j_{s-1})j_s}\,.
\end{equation}

\noindent Allowing $\mu_{2}$ and $\mu_{s}$ to have spacetime dependence, the above action is invariant under the following infinitesimal transformation of fields and sources:
\begin{align}
\delta X^{i}
 ={}& \epsilon_{2} \frac{\delta \mathcal{L}}{\delta\left(\partial_{z} X^{i}\right)} + \epsilon_{s} \frac{\delta \mathcal{W}_{s}}{\delta\left(\partial_{z} X^{i}\right)} 
 \\
  ={}&
\epsilon_{2}\,\partial_{z}X^{i}+ \epsilon_{s}\, d^{i}_{\hphantom{i}i_2\ldots i_s}\partial_{z}X^{i_2}\ldots \partial_{z}X^{i_s} 
\\
\delta \mu_{2} ={}&
 \partial_{\bar{z}}\epsilon_{2} - \mu_{2}\partial_{z}\epsilon_{2} + \epsilon_{2} \partial_{z}\mu_{2}
+ \frac{\kappa}{2}\mathcal{L}^{s-2}\bigl(\epsilon_{s} \partial_{z} \mu_{s} - \mu_{s}\partial_{z}\epsilon_{s}\bigr)
  \\
 \delta \mu_{s} ={}&
\partial_{\bar{z}}\epsilon_{s} +\left(s-1\right)\epsilon_{s}\partial_{z} \mu_{2} - \mu_{2}\partial_{z}\epsilon_{s} - \left(s-1\right)\mu_{s}\partial_{z}\epsilon_{2} + \epsilon_{2}\partial_{z}\mu_{s}
\end{align}

\noindent with associated Ward identities
\begin{align}\label{spin s Ward identity}
\partial_{\bar{z}}\mathcal{L}
&=
\mu_{2}\partial_{z}\mathcal{L} + 2\mathcal{L}\partial_{z}\mu_{2}
+
(s-1)\mu_{s}\partial_{z}\mathcal{W}_{s} + s\mathcal{W}_{s}\partial_{z}\mu_{s}
\\
\partial_{\bar{z}}\mathcal{W}_{s}
&=
 \mu_{2}\partial_{z}\mathcal{W}_{s} + s\mathcal{W}_{s}\partial_{z} \mu_{2}
 + \kappa\left(\mathcal{L}^{s-1}\partial_{z}\mu_{s} + \frac{s-1}{2}\mu_{s}\mathcal{L}^{s-2}\partial_{z}\mathcal{L}\right).
\end{align}

In order to discuss a thermal partition function, we now take $\mu_{2}$ and $\mu_{s}$ to be constant chemical potentials and put the theory on a torus with modular parameter $\tau\,$, with $2\pi\text{Im}(\tau) = \beta$ as before. In the canonical formulation of section \ref{sec: canonical}, the modular parameter $\tau$ of the torus couples by definition to the Virasoro zero modes. We would now like to understand what are the quantities that couple to $\tau$ and $\bar{\tau}\,$ in the presence of chiral deformations of the action. This is an important question, as these couplings define for example the quantity that is conjugate to the inverse temperature $\beta$, namely the energy of the system. The original torus has metric and identifications given by
\begin{equation}
ds^{2} =dzd\bar{z} \,,\qquad \text{with}\qquad z \simeq z + 2\pi \simeq z + 2\pi \tau\,,
\end{equation}

\noindent and volume $\text{Vol}(T^{2}) = 4\pi^{2}\text{Im}(\tau)\,$. Since the periodicity of the coordinates depends on $\tau\,$, care must be exercised when taking variations with respect to the modular parameter. A convenient way of dealing with this problem consists in passing first to coordinates $(w,\bar{w})$ of fixed periodicity, e.g. \cite{Kraus:2006wn}
\begin{equation}
z = \frac{1-i\tau}{2}w + \frac{1+i\tau}{2}\bar{w}\,,\qquad \bar{z} = \frac{1-i\bar{\tau}}{2}w + \frac{1 + i\bar{\tau}}{2}\bar{w}\,,
\end{equation}

\noindent which implies
\begin{equation}
w \simeq w + 2\pi \simeq w + 2\pi i\,.
\end{equation}

\noindent One then takes variations of the action in the $(w,\bar{w})$ coordinates, and transforms back to $(z,\bar{z})$ at the end. In this way one obtains for example
\begin{align}\label{variation derivatives 1}
 \delta_{\tau,\bar{\tau}} \left(\partial_{z}X^{i}\right) 
&=
 i\frac{\delta \tau\,\partial_{z}X^{i} +\delta \bar{\tau}\, \partial_{\bar{z}}X^{i}}{2\text{Im}(\tau)}
\\
\delta_{\tau,\bar{\tau}} \left(\partial_{\bar{z}}X^{i}\right) 
&=
-i\frac{ \delta \tau\,\partial_{z}X^{i} +\delta \bar{\tau}\, \partial_{\bar{z}}X^{i}}{2\text{Im}(\tau)}\,.
\label{variation derivatives 2}
\end{align}

Denoting the free (undeformed) boson action by $S_0$ and taking the variation as indicated yields the expected result\footnote{Notice one keeps the invariant measure $\frac{d^{2}z}{\text{Im}(\tau)}=d^{2}w$ fixed in this variation.}
%
\begin{equation}
\delta S_{0} = \int \frac{d^{2}z}{2i\text{Im}(\tau)}\left(\mathcal{L}\delta \tau - \overline{\mathcal{L}}\delta\bar{\tau}\right)
\end{equation}

\noindent with $\mathcal{L}$ as in \eqref{chiral currents} and
\begin{equation}
\overline{\mathcal{L}} = \frac{1}{2}\partial_{\bar{z}} X^{i}\partial_{\bar{z}} X^{i}\,.
\end{equation}

Extending the above computation to include the effects of the chiral deformations requires some caution, as we first have to 
define what exactly are the independent thermodynamic variables that we are going to use. It might be tempting to use $\mu_s$
and $\bar{\mu}_{s}\,$, besides $\tau,\bar{\tau}$, as independent variables, but we will find it more natural and convenient
to use $\text{Im}(\tau)\mu_s\,$, $\text{Im}(\bar{\tau})\bar{\mu}_s\,$, $\tau$ and $\bar{\tau}$ as independent variables.
We make this choice because (i) a similar choice was made in eq. (\ref{chemical potentials def}), (ii) when taking $\mu_s$ and 
$\mathcal{W}_s$ constant the integral $\int d^{2}z\, \mu_s \mathcal{W}_s$ reduces to $4\pi^2 {\rm Im}(\tau)\mu_s \mathcal{W}_s\,$,
and (iii) this is also the standard procedure in thermal field theory in the presence of chemical potentials \cite{Landsman:1986uw}. An additional independent reason supporting this choice of thermal sources, motivated from holographic considerations, will be given in section \ref{subsec: other holomorphic bcs}. In the present context this means that we must take the $\tau$-variation of the action with 
\begin{equation}
\delta \bigl(\text{Im}(\tau)\mu\bigr) = 0\,.
\end{equation}

\noindent This illustrates a subtle yet crucial point: in the presence of deformations by conserved currents, the precise definition of the sources affects the definition of the energy and other thermodynamic quantities of interest.

 Taking into account the contribution of the chiral deformations and performing the variation of \eqref{chiral spin 2 deformation} as described one obtains
\begin{align}\label{tau variation of the action}
\delta S
={}&
 \int \frac{d^{2}z}{2i\text{Im}(\tau)}\Bigl(\mathcal{L} + 2\mu_{2}\mathcal{L} + s\mu_{s}\mathcal{W}_{s}\Bigr)\delta \tau
 \nonumber
 \\
 &
- \int \frac{d^{2}z}{2i\text{Im}(\tau)}\Bigl(\overline{\mathcal{L}}-\mu_{2}\partial_{z}X^{i}\partial_{\bar{z}}X^{i} - \mu_{s}d_{i_{1}\ldots i_{s}} \partial_{z}X^{i_{1}}\ldots \partial_{z}X^{i_{s-1}}\partial_{\bar{z}}X^{i_{s}}\Bigr) \delta\bar{\tau}\,.
\end{align}

\noindent We would now like to rewrite this variation entirely in terms of the generators themselves. To this end we can use the reparametrization freedom of the path integral and consider a (non-local) field redefinition such that
\begin{equation}
\delta \left(\partial_{z}X^{i}\right) = \gamma_{2}\partial_{z}X^{i} + \gamma_{s}d^{i}_{\hphantom{i} i_{2}\ldots i_{s}}\partial_{z}X^{i_{2}}\ldots \partial_{z}X^{i_{s}}\,.
\end{equation}

\noindent The variation of the free action will then cancel the offending terms in the second line of \eqref{tau variation of the action} provided we set
\begin{equation}
\gamma_{2} =  -\frac{\mu_{2}}{2i\text{Im}(\tau)}\delta\bar{\tau}\,\quad\text{and}\quad \gamma_{s} =  -\frac{\mu_{s}}{2i\text{Im}(\tau)}\delta\bar{\tau}\,.
\end{equation}

\noindent Taking into account the new terms generated by the variation of the $(\mu_{2}\mathcal{L} + \mu_{s}\mathcal{W}_{s})$ piece, the final result for the combined variation of the complex structure plus field redefinition is
\begin{align}\label{tau variation of the action 2}
\delta S
={}&
 \int \frac{d^{2}z}{2i\text{Im}(\tau)}\left(E\delta \tau - \overline{E}\delta\bar{\tau}\right)
\end{align}

\noindent where we defined the ``energies" $E$ and $\overline{E}$ as
\begin{equation}\label{modified left and right moving energies}
E = \mathcal{L} + 2\mu_{2}\mathcal{L} + s\mu_{s}\mathcal{W}_{s}\,,\qquad \overline{E} = \overline{\mathcal{L}}-2\mu_{2}^{2}\mathcal{L}-\kappa \mu_{s}^{2}\mathcal{L}^{s-1}\,.
\end{equation}

The above simple-minded calculation glossed over many details: it did not take central terms into account, it applies to a single higher spin deformation only, and the field redefinition we have performed is non-local. A rigorous calculation should involve e.g. treating the kernel of the derivative operator (in particular zero modes) carefully. Barring these technical complications, the naive calculation exemplifies some facts that should remain true once these subtleties are taken into account. In particular, it shows that even for a chiral deformation the notion of energy on the opposite chiral sector is modified. In fact, the generalization of \eqref{modified left and right moving energies} was obtained in \cite{deBoer:2013gz} using Chern-Simons theory.\footnote{The calculation performed in \cite{deBoer:2013gz} moreover involved non-chiral deformations, but reduces to the above results once the chemical potentials in the barred sector are switched off.} We now see that the mixing of chiralities has a very simple origin in field theory: it arises due to the mixing of left- and right-movers in \eqref{variation derivatives 1} and \eqref{variation derivatives 2}. We will return to this result and its interpretation in section \ref{subsec: holomorphic part fn} and appendix \ref{app: azbar2}.

\subsection{Non-chiral deformations}\label{subsec: non-chiral defs}
Having studied chiral deformations of the CFT action, a natural question is whether one can simultaneously turn on sources for both left- and right-moving chiral algebras in such a way that the Ward identities consist of two copies of \eqref{master Ward identity chiral} (with $\partial_+$ and $\partial_{-}$ interchanged). As we have anticipated, a naive second order path integral with linear couplings, i.e.
\begin{align}\label{2d first order path integral 4}
Z_{\text{Lag,naive}}\left[\beta,\mu_{s},\bar{\mu}_{s}\right]
={}&
 \int\mathcal{D}\phi\, e^{-S_0(\phi)}e^{-\int_{T^{2}}d^{2}z \sum_{s}\left(\mu_{s}\mathcal{W}_{s}(\phi)+\bar{\mu}_{s} \overline{\mathcal{W}}_{s}(\phi)\right)}
\end{align}

\noindent would not lead to the desired Ward identities. A simple way to appreciate the problems associated with this definition is to notice that in order to derive the desired Ward identities one would need to assume that the chiral sectors are decoupled, whereas in practice the OPE between e.g. $\mathcal{W}_{s}$ and $\overline{\mathcal{W}}_{s}$ involves contact terms. In terms of free bosons $X^{i}\,$, these contact terms arise for example from  $\partial X^{i}(z,\bar{z})\bar{\partial} X^{j}(w,\bar{w}) \sim \delta^{ij}\delta^{(2)}(z-w,\bar{z}-\bar{w})\,$. Though contact terms are perhaps often associated to quantum effects, 
it is straightforward to see that here $\overline{\mathcal{W}}_s$ transforms non-trivially under a higher spin transformation generated by $\mathcal{W}_s$ already at the classical level, thereby spoiling the derivation of the Ward identities.

In certain cases one can indeed write down a partition function whose symmetries result in two copies of the chiral Ward identities, at the expense of introducing auxiliary fields \cite{Schoutens:1990ja}. As anticipated,  integrating out the auxiliary fields results in an action involving infinitely many higher order terms in $\mu_{s}$ and $\bar{\mu}_{s}$, which would be the Lagrangian version of the theory with well-separated left- and right-movers. Even though we will not discuss the auxiliary field formalism in detail, in appendix \ref{app: stress tensor defs} we review an example involving non-chiral stress tensor deformations that illustrates various general features of the construction.  We emphasize that the difficulties associated with non-chiral deformations do not arise in the holographic formulation using Chern-Simons theory. In particular, it was already shown in \cite{Gutperle:2011kf}  that two copies of the chiral Ward identities arise as the flatness condition on $sl(N)\oplus sl(N)$ gauge fields with appropriate boundary conditions (c.f. section \ref{subsec: holomorphic part fn}). The difficult only emerges when one tries to associate a deformed CFT path integral to the bulk theory with these boundary conditions.

In our considerations above, the non-decoupling of the chiral sectors in the path integral \eqref{2d first order path integral 4} was easy to see because it involved classical field variations only. One could however contemplate other definitions of the path integral, for example using conformal perturbation theory, where contact terms play no role since one regularizes the integrated correlators by excising small disks around each operator insertion. At first sight, this prescription then leads to a factorization of the chiral and non-chiral deformations, since they only interact through contact terms in correlation functions of the form $\langle \mathcal{W}\overline{\mathcal{W}}\rangle$, and therefore also to the correct separate Ward identities.\footnote{We thank Per Kraus for bringing this scenario to our attention.} It is somewhat puzzling that conformal perturbation theory naively yields an answer which differs from that obtained using classical field variations, especially since the disagreement is already there at the classical level and has nothing to do with quantum issues. One possibility is that the treatment using conformal perturbation theory becomes subtle when going to higher orders, since one needs to separate the chiral and anti-chiral insertions from each other, and that this induces some mixing. Alternatively, we are dealing with two different prescriptions which simply differ by finite local counterterms. It would be interesting to investigate this issue further.

\section{Holography} \label{sec: Holo}
The semiclassical analysis in sections \ref{sec: canonical} and \ref{sec: holo partition function} culminating in the Ward identities \eqref{master Ward identity mu} and \eqref{master Ward identity chiral}, respectively, was purely field-theoretical and did not presume the existence of a holographic description. In particular, for any theory with $\mathcal{W}_{3}$ symmetry we showed that the symmetries of the canonical partition function in the presence of sources result in equations \eqref{T Ward identity}-\eqref{W Ward identity} for the one-point function of the stress tensor and the dimension $3$ operator, in the semiclassical limit. For chiral deformations of the field theory action, the analogous results \eqref{chiral T Ward identity}-\eqref{chiral W Ward identity} hold. Solely a consequence of symmetries, these results are generally valid in the large central charge limit, and in particular independent of the existence of a holographic realization. 

As we have mentioned, the $\mathcal{W}_{N}$ algebras are an example of symmetries that emerge via the so-called Hamiltonian or Drinfeld-Sokolov \cite{Drinfeld:1984qv} reduction of current algebras, and such theories can be described in terms of Chern-Simons theories on a three-dimensional manifold with boundary (see the recent \cite{Gaberdiel:2010pz,Gaberdiel:2010ar,Campoleoni:2010zq,Henneaux:2010xg,Campoleoni:2011hg,Gaberdiel:2011wb,Gutperle:2011kf,Ammon:2011nk,Gaberdiel:2012yb,
Kraus:2011ds}, and \cite{Verlinde:1989ua,Bais:1990bs,Bilal:1991cf,deBoer:1991jc,DeBoer:1992vm,deBoer:1993iz,deBoer:1998ip} for earlier work). The pure Chern-Simons sector is in fact a consistent truncation of the full interacting Prokushkin-Vasiliev theory \cite{Prokushkin:1998bq,Prokushkin:1998vn}, where the matter sector decouples. When the connections are valued in $sl(N,\mathds{R})\oplus sl(N,\mathds{R})$ the dual CFT possesses $\mathcal{W}_{N}$ symmetry \cite{Campoleoni:2010zq,Gaberdiel:2010ar}. Replacing the gauge algebra by two copies of the infinite-dimensional $\text{hs}[\lambda]$ algebra, the resulting theory enjoys $\mathcal{W}_{\infty}[\lambda]$ symmetry \cite{Henneaux:2010xg,Gaberdiel:2011wb}. For a succinct overview of the basic facts concerning the formulation of three-dimensional gravitational theories in Chern-Simons language, as applied to higher spin AdS$_{3}$/CFT$_{2}\,$, we refer the reader to \cite{deBoer:2013gz,deBoer:2013vca}. Full details can be found in the comprehensive reviews \cite{Campoleoni:2011hg,Gaberdiel:2012uj,Ammon:2012wc}.

We do not need the full machinery referred to in the previous paragraph to describe the connection to Chern-Simons
theory, however. Consider Chern-Simons theory, augmented with a suitable boundary term and boundary conditions, on a three-manifold 
with boundary. The Chern-Simons gauge fields will depend in some particular way on the sources $\mu_s$
and $\bar{\mu}_s\,$, and also on the expectation values (EVs) $\langle \mathcal{W}_s\rangle_{\mu}$ and $\langle \overline{\mathcal{W}}_s \rangle_{\mu}$
in the deformed CFT. If the variation of the Chern-Simons action with boundary term and boundary conditions takes the schematic form
\begin{equation} \label{rough form}
\delta S \sim \int_M \text{Tr}\left[\delta A \wedge F\right] + \int_{\partial M}d^{2}x \left(\text{EVs}\right)\delta(\text{sources}) 
\end{equation}

\noindent and $F=0$ restricted to the boundary agrees with the Ward identities of the dual deformed CFT, then the on-shell value 
of the Chern-Simons action plus boundary term yields a functional which will automatically solve the Ward identities. 
Interestingly, we obtain a solution for each choice of three-manifold $M$ with the same boundary $\partial M$. What this analysis
does not tell us is which $M$ to pick, whether to sum over all possible $M$, and whether all solutions of the Ward identities
can be obtained in this way. In the remainder, we will assume the latter to be true, and in order to select a three-manifold
we will pick the dominant saddle point suggested by AdS/CFT in the case where the sources are turned off. We expect this to remain
the dominant saddle for sufficiently small values of the sources, but an analysis of exactly which saddle dominates for which
values of the sources is beyond the scope of the present paper. 

Note that (\ref{rough form}) requires one to identify precisely what sources one chooses, and different choices of sources
or thermodynamic variables will correspond to different boundary terms and boundary conditions, as 
recently discussed in \cite{deBoer:2013gz}. A class of boundary conditions was studied in  \cite{Banados:2012ue,Perez:2012cf,Perez:2013xi,Henneaux:2013dra,Compere:2013nba,Compere:2013gja,Bunster:2014mua} that lead to the so-called ``canonical thermodynamics", consistent with canonical definitions of conserved charges and thermodynamics in gravitational theories, with a perturbative application of Wald-like formulae for the entropy and energy \cite{Campoleoni:2012hp}, and with the thermal limit of entanglement entropy calculations in higher spin theories \cite{deBoer:2013vca,Ammon:2013hba}. A feature of the canonical approach is that, once sources for the currents are switched on, quantities such as the energy and higher spin charges, for example, acquire an explicit dependence on the chemical potentials and differ from their undeformed counterparts. On the other hand, alternative ``holomorphic" boundary conditions were employed in \cite{Gutperle:2011kf,Ammon:2011nk,Kraus:2011ds,Kraus:2013esi} which yielded results consistent with various independent CFT calculations \cite{Gaberdiel:2012yb,Gaberdiel:2013jca,Hijano:2013fja}. The question that concerns us here is what is the precise interpretation of these boundary conditions in terms of the dual field theory. 

The picture we want to put forward is that while canonical boundary conditions are associated with deformations of the CFT Hamiltonian, of the type studied in section \ref{sec: canonical}, the holomorphic ones are related to deformations of the CFT action as described in section \ref{sec: holo partition function}.\footnote{See \cite{Dijkgraaf:1996iy} for a detailed discussion of the relation between chiral deformations of the action and Hamiltonian in CFT.}

\subsection{Canonical boundary conditions and variational principle} \label{sec: bcs}
We will begin our discussion from the perspective of holography by deriving the boundary conditions that realize the canonical structure discussed in section \ref{sec: canonical}. Taking the $\mathcal{W}_{3}$ case as our guiding example, we then ask what are the boundary conditions in Chern-Simons theory that are consistent with the symmetry transformations \eqref{spin 2 transf of currents}-\eqref{spin 3 transf of currents} of the currents, and the Ward identities \eqref{T Ward identity}-\eqref{W Ward identity} (or, equivalently, the transformation \eqref{delta sources 1}-\eqref{delta sources 4} of the sources). The first part of the question, concerning the transformation of the charges, was already answered in \cite{Campoleoni:2010zq}: focusing on the unbarred sector for simplicity, one starts by gauging away the dependence of the connection on the bulk radial coordinate $\rho$ and works with the reduced or ``two-dimensional" connection $a$ defined through 
\begin{equation}
A = b^{-1}(\rho)a(t,\sigma)b(\rho) + b^{-1}(\rho)db(\rho)\,.
\end{equation}

\noindent To obtain the right Ward identities, one needs to choose the asymptotic boundary conditions 
to be of Drinfeld-Sokolov form, i.e.\footnote{We adopt the same convention as \cite{Campoleoni:2010zq} for the $sl(3,\mathds{R})$ generators $L_{1}, L_{0},L_{-1}$ and $W_{j}$ ($j=-2,-1,\ldots,2$), but rescale the currents by a factor of $2\pi\,$.}
\begin{equation}\label{a sigma DS}
a_{\sigma} = L_1 + Q 
\end{equation}

\noindent where $Q$ a highest weight matrix ($\left[Q,L_{-1}\right]=0$) whose entries contain the stress tensor and the higher spin currents. For example, in the spin-3 case we write the spatial component of the $sl(3,\mathds{R})$ connection as
\begin{equation}\label{a sigma DS spin 3}
a_\sigma = L_1 + \frac{T}{k}L_{-1} - \frac{W}{4k}W_{-2}
\end{equation}

\noindent where $k=\frac{\ell}{4G_{3}}\,$. By definition, the asymptotic symmetry algebra is generated by the gauge transformations that respect these boundary conditions, and it corresponds to the (infinite) global symmetries of the dual CFT. Perfoming an infinitesimal gauge transformation with parameter $\lambda$ as $\delta a = d\lambda + \left[a,\lambda\right]$, one finds that \eqref{a sigma DS spin 3} is preserved if $\lambda$ takes the form \cite{Campoleoni:2010zq} 
\begin{equation}\label{gauge parameter 1}
\lambda = \sum_{i=-1}^{1}\epsilon^{i}L_i + \sum_{m=-2}^{2}\chi^{m}W_{m}
\end{equation}

\noindent with the parameters fixed in terms of $\epsilon^{1} \equiv \epsilon$ and $\chi^{2} \equiv \chi\,$ by \eqref{gauge param 1}-\eqref{gauge param 2}. Under such transformations, the change in the currents is precisely given by \eqref{spin 2 transf of currents}-\eqref{spin 3 transf of currents}.

The remaining question is how to incorporate the sources $\mu_{2}$, $\mu_{3}$ in the connection. The guiding principle is that the asymptotic equations of motion, namely the flatness condition on the reduced connection $a(t,\sigma)\,$, should reproduce the Ward identities \eqref{T Ward identity}-\eqref{W Ward identity}. Having fixed the form of $a_{\sigma}$, the complete $sl(3,\mathds{R})$ flat connection is found to be
\begin{eqnarray}
a_\sigma 
&=& L_1 + \frac{T}{k}L_{-1} - \frac{W}{4k}W_{-2}
\\
a_{t}  
&=&
a_{\sigma}+ \mu_{2}L_{1} + \mu_{3}W_{2} - \partial_{\sigma}\mu_{2}\, L_0-\partial_{\sigma}\mu_{3}\, W_1  + \left(\frac{1}{2}\partial^{2}_{\sigma}\mu_{3} + \frac{2T}{k}\mu_{3}\right)W_0
\nonumber\\
&& + \left(\frac{1}{2}\partial_{\sigma}^{2}\mu_{2} + \frac{2W}{k}\mu_{3} + \frac{T}{k}\mu_{2}\right)L_{-1}+ \left(-\frac{1}{6}\partial^{3}_{\sigma}\mu_{3} - \frac{5}{3k}T\partial_{\sigma}\mu_{3} - \frac{2}{3k}\mu_{3}\partial_{\sigma}T\right)W_{-1} 
\label{general at}
\\
&&
+ \bigg(\frac{1}{24}\partial^{4}_{\sigma}\mu_{3} + \frac{2}{3k}T\partial^{2}_{\sigma}\mu_{3} + \frac{7}{12k}\partial_{\sigma}T\partial_{\sigma}\mu_{3}
+\left(\frac{T^{2}}{k^{2}} +\frac{1}{6k}\partial^{2}_{\sigma}T\right)\mu_{3} - \frac{1}{4k}\mu_{2}W\biggr)W_{-2}\,.
\nonumber
\end{eqnarray}

The general structure at play is more clearly appreciated in terms of a ``Drinfeld-Sokolov pair", consisting of one component of the connection carrying the currents as highest weights, and a conjugate component carrying the corresponding sources as lowest weights. This is conveniently summarized as\footnote{In the $N=3$ case, these boundary conditions have been recently advocated in \cite{Henneaux:2013dra,Bunster:2014mua} from a purely bulk perspective. Here we have arrived at them following a different route, using as guiding principle the $(1+1)$-dimensional field theory Ward identities in the presence of Hamiltonian deformations.}
\begin{align}\label{DS sigma t}
a_{\sigma} ={}&
L_1 + Q
\\
a_{t} - a_{\sigma} ={}&
 M + \ldots
 \label{DS sigma t 2}
\end{align}

\noindent where as before $Q$ is linear in the currents and satisfies $\left[Q,L_{-1}\right]=0$, $M$ is a matrix linear in the sources which satisfies $\left[M,L_{1}\right]=0\,$, and the dots stand for higher weight terms completely fixed by the equations of motion once a suitable normalization of the sources is chosen (see \cite{deBoer:2013gz,Compere:2013nba} and appendix \ref{app: azbar2} for details). In particular, in the above example we have
\begin{equation}
M =  \mu_{2}L_{1} + \mu_{3}W_{2}\,.
\end{equation}

As a further consistency check, acting on \eqref{general at} with the gauge parameter \eqref{gauge parameter 1} (subject to \eqref{gauge param 1}-\eqref{gauge param 2}) one easily verifies that the change in the lowest weights of the connection is
\begin{equation}
\delta\left(a_{t} - a_{\sigma}\right) = \delta\mu_{2}L_{1} + \delta\mu_{3}W_{2} + \text{(higher weights)}\,,
\end{equation}

\noindent with $\delta \mu_{2}$ and $\delta \mu_{3}$ given precisely by \eqref{delta sources 1}-\eqref{delta sources 4}. We have then shown that the equations of motion of Chern-Simons theory with boundary conditions \eqref{DS sigma t}-\eqref{DS sigma t 2} (and a Dirichlet variational principle for the sources) agree with the Ward identities we obtained from the canonical partition function in field theory. To make sure that the partition functions also agree, all that remains is to find an appropriate boundary term which is compatible with the Dirichlet boundary conditions on the sources, and which will guarantee that the charges are indeed coupled in the right way to the currents. We will turn back to these boundary terms momentarily.

In order to facilitate comparison with the recent literature, we note that the above boundary conditions written in light-cone coordinates $x^{\pm} = t\pm \sigma$ read
\begin{align}\label{light cone DS}
a_+ - a_{-} = L_1 + Q
\\
2a_{-} = M + \ldots
\label{light cone DS 2}
\end{align}

\noindent Recalling that the $a_{-}$ component is zero for undeformed solutions (such as pure AdS), we see that incorporating the sources in $a_{-}$ we can readily interpret them as deformations of the original theory. 

Turning back to the boundary terms, the necessary techniques to find these were introduced in \cite{deBoer:2013gz}, generalizing the results of \cite{Banados:2012ue}. The Lorentzian Chern-Simons action on a three-dimensional manifold $M$ reads\footnote{All traces in this section are taken in the fundamental representation.}
\begin{align}\label{CS action}
I_{CS} =
 \frac{k_{cs}}{4\pi}\int_{M} \mbox{Tr}\Bigl[CS(A) - CS(\bar{A})\Bigr]
\end{align}

\noindent where \cite{Castro:2011iw}
\begin{equation}
k_{cs} = \frac{k}{2\text{Tr}\left[L_0L_0\right]} 
\end{equation}

\noindent in order to match with the normalization of the Einstein-Hilbert action in the pure gravity case. With this normalization, the central charge in the dual CFT is given by $c=12k_{cs}\text{Tr}\left[L_0L_0\right]\,$. The total action will be of the form
\begin{equation}\label{full action}
I = I_{CS} + I_{B}\,,
\end{equation}

\noindent where $I_{B}$ is the required boundary term. In terms of the $\rho$-independent connections $a$, $\bar{a}$, the variation of the bulk action $I_{CS}$, evaluated on-shell, is easily seen to be
\begin{align}\label{CS action variation}
\left.\delta I_{CS}\right|_{os}  
={}&
- \frac{k_{cs}}{4\pi}\int_{\partial M}\text{Tr}\Bigl[a\wedge \delta a - \bar{a}\wedge \delta \bar{a}\Bigr]
\\
 ={}&
 - \frac{k_{cs}}{2\pi}\int_{\partial M}d^{2}x\,\text{Tr}\Bigl[a_{+} \delta a_{-} - a_{-}\delta a_{+} -\bar{a}_{+} \delta \bar{a}_{-} + \bar{a}_{-}\delta \bar{a}_{+} \Bigr],
\end{align}

\noindent where $d^{2}x\equiv (1/2)dx^{-}\wedge dx^{+} = dt\,d\sigma\,$. The necessary boundary term is
\begin{align}\label{Lorentzian boundary term}
I_{B}
=
-\frac{k_{cs}}{2\pi} \int_{\partial M} d^{2}x\, \text{Tr}\Bigl[\left(a_{+}-a_{-} -2 L_1\right)a_{-}\Bigr] 
-\frac{k_{cs}}{2\pi}\int_{\partial M} d^{2}x\,  \text{Tr}\Bigl[\left(\bar{a}_{-}-\bar{a}_{+}+2 L_{-1}\right)\bar{a}_{+} \Bigr],
\end{align}

\noindent and the variation of the full action $I$, evaluated on-shell, is then
\begin{align}
\left.\delta I\right|_{os}
={}&
 - \frac{k_{cs}}{2\pi}\int_{\partial M}d^{2}x\,\text{Tr}\Bigl[\left(a_{+}-a_{-}-L_1\right)\delta \left(2a_{-}\right) + \left(\bar{a}_{-}-\bar{a}_{+}+ L_{-1}\right)\delta\left(2 \bar{a}_{+}\right)\Bigr].
\end{align}

\noindent This confirms that the boundary term above is well-suited to the Dirichlet problem (fixed sources).

\subsection{Canonical thermodynamics revisited} \label{sec: Thermo}
We will now describe the boundary conditions in the Euclidean formulation of Chern-Simons theory, and derive general expressions for the free energy and entropy in the dual theory. In order to introduce temperature, the Euclidean time direction is compactified and the topology of the three-dimensional manifold $M$ becomes that of a solid torus. Complex coordinates $(z,\bar{z})$ are introduced by analytically continuing the light-cone directions as $x^{+} \to z\,$, $x^{-}\to -\bar{z}\,$, with identifications $z \simeq z + 2\pi \simeq z + 2\pi \tau $, where $\tau$ is the modular parameter of the boundary two-torus. In the semiclassical limit (large temperature and central charges), the CFT partition function is obtained from the saddle point approximation of the Euclidean on-shell action:
\begin{equation}
\ln Z = -I^{(E)}_{os} =\left. -\left(I^{(E)}_{CS} + I^{(E)}_{B}\right)\right|_{os}\,,
\end{equation}

\noindent where
\begin{equation}\label{Euclidean Chern-Simons action}
I^{(E)}_{CS} =  \frac{ik_{cs}}{4\pi}\int_{M} \mbox{Tr}\Bigl[CS(A) - CS(\bar{A})\Bigr]
\end{equation}

\noindent and $I_{B}^{(E)}$ denotes the Euclidean continuation of the boundary term \eqref{Lorentzian boundary term},
\begin{equation}\label{Euclidean boundary term}
I^{(E)}_{B} =- \frac{k_{cs}}{2\pi} \int_{\partial M} d^{2}z\, \text{Tr}\Bigl[\left(a_{z}+a_{\bar{z}} -2 L_1\right)a_{\bar{z}}\Bigr] 
-\frac{k_{cs}}{2\pi}\int_{\partial M} d^{2}z\, \text{Tr}\Bigl[\left(\bar{a}_{\bar{z}} + \bar{a}_{z} -2 L_{-1}\right)\bar{a}_{z} \Bigr].
\end{equation}

Mirroring the field theory discussion in section \ref{subsec: tau coupling}, when computing the variation of the Chern-Simons action one should acknowledge that the modular parameter of the torus is varying. As before, a convenient way of dealing with this fact is to compute the variation in coordinates with fixed-periodicity (where $\tau$ appears in the connection itself), and change back to the $z$ coordinates at the end. Following the steps detailed in \cite{deBoer:2013gz}, in the present case we find that the variation of the full action, evaluated on-shell, is given by
\begin{align}\label{Euclidean variation 1}
\delta I^{(E)}_{os} =-2\pi i k_{cs}\int_{\partial M}\frac{d^{2}z}{4\pi^{2}\text{Im}(\tau)}\text{Tr}&\left[ \frac{1}{2}\left(a_{z} + a_{\bar{z}}\right)^{2}\delta\tau +\left(a_{z} + a_{\bar{z}} - L_1\right)\delta \left((\bar{\tau}-\tau)a_{\bar{z}}\right)\right.
\nonumber\\
&\left. 
-\frac{1}{2}\left(\bar{a}_{z} + \bar{a}_{\bar{z}}\right)^2\delta\bar{\tau} + \left(\bar{a}_{z} + \bar{a}_{\bar{z}} - L_{-1}\right)\delta \left((\bar{\tau}-\tau)\bar{a}_{z}\right)\right].
\end{align}

\noindent First, we notice that the quantities conjugate to $\tau$ and $\bar{\tau}$, namely the left- and right-moving energies $T$ and $\overline{T}$, are given by
\begin{equation}\label{canonical stress tensor zero modes}
T= -\frac{k_{cs}}{2}\text{Tr}\left[\left(a_{z} + a_{\bar{z}}\right)^{2}\right]\,,
\qquad
\overline{T} = -\frac{k_{cs}}{2}\text{Tr}\left[\left(\bar{a}_{z} + \bar{a}_{\bar{z}}\right)^2\right].
\end{equation}

\noindent In particular one notices that the mixing of chiralities encountered in \ref{subsec: tau coupling} from the field theory perspective, and in \cite{deBoer:2013gz} from the Chern-Simons perspective, does not arise when using canonical boundary conditions. Secondly, we see that the quantities coupling to the higher spin currents are $\left(\bar{\tau} -\tau\right) a_{\bar{z}}$ and $\left(\bar{\tau}-\tau\right)\bar{a}_{z}\,$, so the Euclidean version of the boundary conditions \eqref{light cone DS}-\eqref{light cone DS 2} is
\begin{align}\label{Euclidean DS 1}
a_{z} + a_{\bar{z}}={}&
 L_{1} + Q&
\bar{a}_{z} + \bar{a}_{\bar{z}}={}& L_{-1} - \overline{Q}
 \\
\left(\bar{\tau} -\tau\right) a_{\bar{z}} ={}& M +\ldots
&
\left(\bar{\tau} -\tau\right) \bar{a}_{z} ={}& \overline{M} +\ldots
\label{Euclidean DS}
\end{align}

\noindent with the difference that the matrix $M$ does not contain the spin-2 source anymore, because the latter has been incorporated as the modular parameter of the torus. Equation \eqref{Euclidean DS} makes manifest the fact that the sources get rescaled by the temperature when transitioning to the Euclidean formalism. In other words, the matrix elements of $a_{\bar{z}}$ and $\bar{a}_{z}$ contain the chemical potentials $\mu$ (i.e. the deformation parameters in the Lorentzian description), while the matrices $M$ and $\overline{M}$ contain the actual sources $\alpha \simeq \text{Im}(\tau)\mu\,$.\footnote{A slightly different definition of the sources was employed in \cite{Compere:2013nba}, as lowest weights in $\tau a_{z} +\bar{\tau}a_{\bar{z}}\,$. One notes however that $\tau a_{z} +\bar{\tau}a_{\bar{z}} = (\bar{\tau}-\tau)a_{\bar{z}} + \tau\left(a_{z}+a_{\bar{z}}\right)$, and since $a_{z}+a_{\bar{z}}$ is a highest weight matrix, this implies that the lowest weights in  $\tau a_{z} +\bar{\tau}a_{\bar{z}}$ and $(\bar{\tau} -\tau)a_{\bar{z}}$ are in fact the same. By the same token, in the Lorentzian theory the sources can be said to be the lowest weights in $a_{t} -a_{\sigma}$ or equivalently in $a_{t}$, because $a_{\sigma}$ is a highest weight matrix. The difference between these two approaches amounts simply to a shift in the definition of the spin-2 source.} This agrees with our field theory discussion in section \ref{sec: partition function} (c.f. \eqref{chemical potentials def}).

 As explained in \cite{deBoer:2013gz}, for the theory determined by choosing the principal embedding of $sl(2)$ into $sl(N)$, resulting in $\mathcal{W}_{N}$ as the asymptotic symmetry algebra, the normalization of the currents and sources can be chosen such that 
\begin{align}\label{normalization 1}
 -k_{cs}\left(\bar{\tau} -\tau\right)\text{Tr}\bigl[Q a_{\bar{z}}\bigr] ={}&
  \sum_{s=3}^{N}\alpha_{s}\mathcal{W}_{s}
\\
-k_{cs}\left(\bar{\tau} -\tau\right)\text{Tr}\bigl[L_1 a_{\bar{z}}\bigr] ={}&
 \sum_{s=3}^{N}(s-1)\alpha_{s}\mathcal{W}_{s}
 \label{normalization 2}
\end{align}

\noindent Similar expressions hold in the other chiral sector. To adapt the above formulae to non-principal embeddings one simply replaces $s$ by the conformal weight of the operator, with the sum running over the appropriate spectrum. The above formulae rely solely on the lowest/highest weight structure of the solutions, and therefore are valid even for non-constant connections. See appendix \ref{app: azbar2} for a general derivation.

So far we have been discussing the variation of the on-shell value of the Chern-Simons action, for which the choice of three-manifold
was irrelevant. To find the actual value of the Chern-Simons action, we however need to pick a three-manifold $M$. In the absence
of sources the dominant saddle point in the high-temperature regime is the one where the Euclidean time-circle is smoothly contractible
in the interior. We will therefore pick this particular three-manifold $M$, as we expect this to still be the dominant saddle point
for sufficiently small values of the sources. 
In this particular case, where moreover the connections are constant, we can explicitly evaluate the on-shell action and therefore the partition function $Z$ and the free energy $F$ as $-\beta F=\ln Z = -\left.I^{(E)}\right|_{os}\,$,\footnote{See \cite{Banados:2012ue} for a discussion of the subtleties associated with the evaluation of the bulk piece.} obtaining
\begin{equation}\label{canonical free energy}
\ln Z_{\text{can}}
 =
  -2\pi i k_{cs}\text{Tr}\left[\frac{1}{2}\left(a_{z} + a_{\bar{z}}\right)^{2}\tau +\left(\bar{\tau}-\tau\right)L_1 a_{\bar{z}}-\frac{1}{2}\left(\bar{a}_{z} + \bar{a}_{\bar{z}}\right)^2\bar{\tau} +  \left(\bar{\tau}-\tau\right)L_{-1} \bar{a}_{z} \right].
\end{equation}

\noindent As usual, the free energy is a function of the temperature and chemical potentials. A standard Legendre transform produces the entropy, a function of the charges. The term implementing the Legendre transformation can be read off from \eqref{Euclidean variation 1}, and the thermal entropy is then
\begin{align}
S_{\text{can}}
 =
  \ln Z_{\text{can}} - 2\pi ik_{cs}\text{Tr}&\biggl[
  \frac{1}{2}\left(a_{z} + a_{\bar{z}}\right)^{2}\tau -\frac{1}{2}\left(\bar{a}_{z} + \bar{a}_{\bar{z}}\right)^2\bar{\tau}
\nonumber\\
& \quad
 + \left(a_{z} + a_{\bar{z}} - L_1\right)\left(\bar{\tau}-\tau\right)a_{\bar{z}}
   +\left(\bar{a}_{z} + \bar{a}_{\bar{z}} - L_{-1}\right) \left(\bar{\tau}-\tau\right)\bar{a}_{z}
  \biggr]
\end{align}

\noindent which after using \eqref{canonical free energy} yields
\begin{equation}\label{canonical entropy}
S_{\text{can}} = -2i\pi k_{cs}\text{Tr}\Bigl[\left(a_{z}+a_{\bar{z}}\right)\left(\tau a_{z} +\bar{\tau}a_{\bar{z}}\right)-\left(\bar{a}_{z}+\bar{a}_{\bar{z}}\right)\left(\tau\bar{a}_{z} + \bar{\tau}\bar{a}_{\bar{z}}\right)\Bigr].
\end{equation}

\noindent This formula for the entropy was first derived in \cite{deBoer:2013gz}. In the particular case of the $\mathcal{W}_{3}$ theory  ($N=3$) in the principal embedding, it agrees with a result derived in the metric formulation \cite{Perez:2013xi}, as well as the perturbative application of Wald's entropy formula \cite{Campoleoni:2012hp}. We emphasize however that equations \eqref{canonical free energy} and \eqref{canonical entropy} are valid for any $N$, and any choice of embedding.  Moreover, they are valid for the $\text{hs}[\lambda]$ theory as well, provided the trace is interpreted accordingly (see section 4 of \cite{Compere:2013nba} for a complete discussion of this case). The above form of the entropy has been also recovered as the thermal limit of entanglement entropy proposals for higher spin theories \cite{deBoer:2013vca,Ammon:2013hba}.

It is important to mention that the charges and their conjugate sources have to be related in a particular way for the first law of thermodynamics to hold. In the Chern-Simons formulation this requirement has been encoded in an elegant way in terms of holonomies of the connection \cite{Gutperle:2011kf}. In a few words, one demands that the connection has trivial holonomy around the thermal cycle of the boundary torus, that becomes contractible in the bulk. This is the Chern-Simons analogue of the familiar statement for Euclidean black holes that the thermal circle should be smoothly contractible. Using these holonomy conditions, it was also shown in \cite{deBoer:2013gz} that the above formula for the entropy can be written very compactly as
\begin{equation}\label{our entropy formula}
S_{\text{can}} = 2\pi k_{cs}\text{Tr}\Bigl[\left(\lambda - \overline{\lambda}\right)L_0\Bigr],
\end{equation}

\noindent where $\lambda$ and $\overline{\lambda}$ are diagonal matrices containing the eigenvalues of the component of the connection along the non-contractible cycle of the boundary torus, i.e.
\begin{equation}
\lambda \equiv \text{Eigen}\left(a_{z}+a_{\bar{z}}\right)\,,\qquad \overline{\lambda} \equiv \text{Eigen}\left(\bar{a}_{z}+\bar{a}_{\bar{z}}\right)\,.
\end{equation}

 Given the boundary conditions \eqref{Euclidean DS 1}-\eqref{Euclidean DS}, it is evident from \eqref{our entropy formula} that the entropy is a function of the charges. The particular combination of zero modes implied by \eqref{our entropy formula} can be then viewed as the generalization of the Cardy formula for higher spin theories. As a side remark we note that in the principally-embedded $sl(N,\mathds{R})\oplus sl(N,\mathds{R})$ theory, the above expression for the entropy can be also written in a representation-independent way as \cite{deBoer:2013vca}
\begin{equation}\label{our Cardy formula v2}
S_{\text{can}}=  2\pi k_{cs}\left\langle \vec{\lambda}-\vec{\overline{\lambda}}\,\,,\,\vec{\rho}\, \right\rangle 
\end{equation}

\noindent where $\vec{\lambda}$, $\vec{\overline{\lambda}}$ are the weight vectors dual to $\lambda$ and $\overline{\lambda}$ (which belong to the Cartan subalgebra), $\vec{\rho}$ denotes the Weyl vector of $sl(N)\,$ (which is dual to $L_0$), and the brackets denote the usual inner product induced by the Killing form.

\subsection{Holomorphic boundary conditions and thermodynamics}\label{subsec: holomorphic part fn}
The holomorphic partition functions we discussed in this paper correspond to deformations of the Lagrangian instead of the Hamiltonian.
The analysis proceeds exactly as for the canonical case, the main difference being that the boundary conditions become
\begin{align}\label{Euclidean DS 2}
a_{z} ={}&
 L_{1} + Q&
  \bar{a}_{\bar{z}}={}& L_{-1} - \overline{Q}
 \\
\left(\bar{\tau} -\tau\right) a_{\bar{z}} ={}& M +\ldots
&
\left(\bar{\tau} -\tau\right) \bar{a}_{z} ={}& \overline{M} +\ldots
\label{Euclidean DS}
\end{align}

\noindent instead of \eqref{Euclidean DS 1}, and similarly in Lorentzian signature (see  \cite{deBoer:2013gz})  where they result in two copies of the Ward identities \eqref{master Ward identity chiral} \cite{Gutperle:2011kf}. Accordingly, instead of \eqref{Euclidean boundary term} the appropriate boundary term now reads
\begin{equation}\label{Euclidean boundary term 2}
I^{(E)}_{B}
 =
 - \frac{k_{cs}}{2\pi} \int_{\partial M} d^{2}z\, \text{Tr}\Bigl[\left(a_{z} -2 L_1\right)a_{\bar{z}}\Bigr] 
-\frac{k_{cs}}{2\pi}\int_{\partial M} d^{2}z\, \text{Tr}\Bigl[\left(\bar{a}_{\bar{z}}  -2 L_{-1}\right)\bar{a}_{z} \Bigr].
\end{equation}

\noindent As discussed in \cite{deBoer:2013gz}, the corresponding free energy is
\begin{align}\label{holomorphic free energy}
-\beta F_{\text{holo}} =
 \ln Z_{\text{holo}}
  ={}&
   -2\pi i k_{cs}\text{Tr}\biggl[\tau\left(\frac{a_{z}^{2}}{2} + a_{z}a_{\bar{z}} -\frac{\bar{a}_{z}^{2}}{2}\right) -\bar{\tau}\left( \frac{\bar{a}_{\bar{z}}^{2}}{2} + \bar{a}_{\bar{z}}\bar{a}_{z} -\frac{a_{\bar{z}}^{2}}{2}\right)
   \nonumber\\
   {}&\hphantom{-2\pi i k_{cs}\text{Tr}\Bigl[ a}
     + \left(\bar{\tau}-\tau\right)\left(L_1 a_{\bar{z}}+L_{-1} \,\bar{a}_{z}\right)\biggr].
\end{align}

There are several marked differences with the canonical case. For example, under variations of the complex structure one finds \cite{deBoer:2013gz} 
\begin{equation}\label{complex structure variation}
\delta_{\tau,\bar{\tau}} \ln Z_{\text{holo}}
 =
  -2\pi i\int \frac{d^{2}z}{4\pi^{2}\text{Im}(\tau)}\left(E\delta \tau -\overline{E}\delta \bar{\tau}\right)
\end{equation} 

\noindent with 
\begin{equation}\label{modified energies}
E
=
-\frac{k_{cs}}{2}\text{Tr}\Bigl[a_{z}^{2} + 2a_{z}a_{\bar{z}} - \bar{a}_{z}^{2}\Bigr]\,,
\qquad
\overline{E}
=
-\frac{k_{cs}}{2}\text{Tr}\Bigl[\bar{a}_{\bar{z}}^{2} + 2\bar{a}_{\bar{z}}\bar{a}_{z} - a_{\bar{z}}^{2}\Bigr].
\end{equation}

\noindent We then see that the operator that couples to $\tau$ is now much more complicated, and involves a mixture of left and right movers.  The content of \eqref{modified energies} is that the energy of the system is manifestly modified by the higher spin sources, and in particular it does no longer correspond to the zero modes of the stress tensor as defined in the undeformed theory. We presented a qualitative field theory explanation of this mixture in section \ref{subsec: tau coupling}, and it would be very interesting to derive the form of this operator directly from the path integral. A hint as to how this might come about comes from a property of Drinfeld-Sokolov connections that we discuss below.

In order to characterize the operators $E$ and $\overline{E}\,$, we note that the Drinfeld-Sokolov form of the connection with holomorphic boundary conditions is easily seen to imply (c.f. appendix \ref{app: azbar2})
\begin{align}
-\frac{k_{cs}}{2}\text{Tr}\left[a_{z}^{2}\right] 
&=
 \mathcal{L}
\\
 -k_{cs}\text{Tr}\big[a_{z}a_{\bar{z}}\bigr] 
 &=
   \sum_{s \geq 3}s\mu_{s}\mathcal{W}_{s}\,,
\end{align}

\noindent with $\mathcal{L}$ the stress tensor, and similarly in the other chiral sector.\footnote{Note that we have not included a spin-2 source $\mu_{2}$ in the connection, as we would have done if using coordinates with fixed periodicity, i.e. for a square torus. The full general expressions containing $\mu_{2}$ can be found in appendix \ref{app: azbar2}.} The only quantity left to characterize is then $-k_{cs}\text{Tr}\left[a_{\bar{z}}^{2}\right]\,$. In appendix \ref{app: azbar2} we point out the useful relation
\begin{equation}\label{trazbarsq}
-k_{cs}\text{Tr}\left[a_{\bar{z}}^{2}\right] = \text{Res}_{z\to w}\Bigl[(z-w)\Delta L_{N}(z)\Delta L_{N}(w)\Bigr] + \partial^{2}\left(P_{N}\right),
\end{equation}

\noindent where $\Delta L_{N}\equiv  \sum_{s=3}^{N}\mu_{s}\mathcal{W}_{s}\,$ is the deformation operator, and provide the explicit form of $P_{N}$ for $N=2,3,4$. We notice however that $P_{N}$ does not contribute under the integral sign in \eqref{complex structure variation}. In other words, the contribution of  $-k_{cs}\text{Tr}\left[a_{\bar{z}}^{2}\right]$ to the energies is given precisely by the second order pole in the OPE of the Lagrangian deformation with itself. We emphasize that \eqref{trazbarsq} holds for arbitrary spacetime-dependent sources, and it therefore applies beyond the thermodynamic analysis. We leave a further study of this curious relation to future work.

In order to obtain the entropy we need to perform an appropriate Legendre transform of the free energy, which in this case reads
\begin{align}
S_{\text{holo}} =
\ln  Z_{\text{holo}}  -2\pi i k_{cs}\text{Tr}&\left[ \left(\bar{\tau}-\tau\right)\left(a_{z}-L_1\right)a_{\bar{z}} +\tau\left(\frac{a_{z}^{2}}{2} + a_{z}a_{\bar{z}} -\frac{\bar{a}_{z}^{2}}{2}\right)  
\right.
\nonumber\\
&\quad \left.
- \left(\bar{\tau} - \tau\right)\left(-\bar{a}_{\bar{z}}+L_{-1}\right)\bar{a}_{z}-\bar{\tau}\left( \frac{\bar{a}_{\bar{z}}^{2}}{2} + \bar{a}_{\bar{z}}\bar{a}_{z} -\frac{a_{\bar{z}}^{2}}{2}\right)\right]
\end{align}

\noindent and evaluates to the same expression \eqref{canonical entropy} for the entropy as in the canonical case, namely
\begin{equation}\label{our old entropy formula 3}
S_{\text{holo}} = -2\pi i k_{cs}\,\text{Tr}\Bigl[\left(a_{z}+a_{\bar{z}}\right)\left(\tau a_{z} +\bar{\tau}a_{\bar{z}}\right)-\left(\bar{a}_{z}+\bar{a}_{\bar{z}}\right)\left(\tau\bar{a}_{z} + \bar{\tau}\bar{a}_{\bar{z}}\right)\Bigr].
\end{equation}

\noindent In particular, this result can be written in the same form as \ref{our entropy formula}. However, despite the apparent similarity,
there is an important difference once again: whereas in the canonical case  $(a_z+a_{\bar{z}})$ depends on the charges only and the eigenvalues $\lambda$ and $\bar{\lambda}$ immediately yield an expression for the entropy as a function of the charges, in the holomorphic case $a_z+a_{\bar{z}}$ depends on both the charges and chemical potentials in a complicated way.  Hence, in order to find an expression for the entropy, in the latter case one needs to explicitly solve the monodromy conditions which allow to express the sources in terms of the charges.

\subsection{Other holomorphic boundary conditions}\label{subsec: other holomorphic bcs}

In the context of holography, the first discussion of boundary conditions in the presence of higher spin sources and the associated thermodynamics was given in the original work \cite{Gutperle:2011kf} of Gutperle and Kraus on higher spin black holes. For the bulk theory based on $sl(3,\mathds{R})\oplus sl(3,\mathds{R})$ with principally embedded $sl(2,\mathds{R})$, for example, the boundary conditions advocated therein agree with our holomorphic boundary conditions, and resulted in two copies of the Ward identities \eqref{chiral T Ward identity}-\eqref{chiral W Ward identity}.\footnote{The structure of a general Drinfeld-Sokolov connection obeying these boundary conditions is described in appendix \ref{app: azbar2}, and detailed examples are provided for the theory based on the $sl(N,\mathds{R})\oplus sl(N,\mathds{R})$ algebra for $N=2,3,4\,$.}  

For chiral deformations of this sort, we have given a particular partition function in CFT which corresponds to a deformation of the action by a linear coupling, which indeed reproduces a single chiral copy of these Ward identities. We have also pointed out that when sources for currents of both chiralities are present, the corresponding partition function in CFT that reproduces the Ward identities involves terms to all orders in the sources, including terms that mix both chiralities. This can be understood, for example, using the auxiliary field formalism introduced in \cite{Schoutens:1990ja}, which unfortunately needs to be formulated on a case by case basis.

By considering thermodynamics of Chern-Simons theory on a solid torus, an entropy was found in \cite{Gutperle:2011kf} whose precise form was determined by the first law of thermodynamics based on a definition of the sources $(\alpha,\bar{\alpha})$ that involved rescaling the chemical potentials $(\mu,\bar{\mu})$ by the modular parameter $\tau$ of the boundary torus torus, e.g. $\alpha = \bar{\tau}\mu$ and $\bar{\alpha} = \tau \bar{\mu}\,$. It was then shown in \cite{deBoer:2013gz} that the entropy formula obtained in \cite{Gutperle:2011kf} can be written quite generally as
\begin{equation}\label{holomorphic entropy 4}
S_{\text{G-K}} = -2i\pi k_{cs}\text{Tr}\Bigl[a_{z}\left(\tau a_{z} +\bar{\tau}a_{\bar{z}}\right)-\bar{a}_{\bar{z}}\left(\tau\bar{a}_{z} + \bar{\tau}\bar{a}_{\bar{z}}\right)\Bigr],
\end{equation}

\noindent or, equivalently,
\begin{equation}\label{our entropy formula holomorphic 4}
S_{\text{G-K}} = 2\pi k_{cs}\text{Tr}\bigl[\left(\lambda_{z} - \overline{\lambda}_{z}\right)L_0\bigr],
\end{equation}

\noindent where $\lambda_{z}$ and $\overline{\lambda}_{z}$ are diagonal matrices containing the eigenvalues of the $a_{z}$ and $\bar{a}_{\bar{z}}$ components of the connection. 

It may appear strange that although the Ward identities take the same ``holomorphic'' form in \cite{Gutperle:2011kf} as we obtained
from deformations of the action, the entropy (\ref{our old entropy formula 3}) one obtains from the latter formulation 
is clearly distinct from \eqref{holomorphic entropy 4}. The explanation of this discrepancy lies in the different choices
of sources $\alpha$, $\bar{\alpha}$ in the thermal case: even with identically looking Ward identities (in terms of the chemical potentials $\mu$, $\bar{\mu}$), different choices of sources $\alpha$, $\bar{\alpha}$ can give rise to different notions of free energy and entropy. Moreover, the modular parameter $\tau$ of the torus does not enter the Ward identities directly and always needs a separate treatment. 

These results illustrate a rather subtle point that was alluded to from a field theory perspective in section \ref{subsec: tau coupling}, namely that the precise definition of the sources in the thermal theory affects the notion of energy and other thermodynamic quantities. The bottom line is that an unambiguous definition of the thermal partition function requires to specify not only the boundary conditions on the plane/cylinder, but also the precise scaling of the sources with the complex structure of the torus. In this light it should come as no surprise that the same flat connection can yield two different results \eqref{canonical entropy} and \eqref{holomorphic entropy 4} for the entropy, depending on precisely how the thermal sources are related to the chemical potentials and the temperature. 

We would like to mention, in passing, one more argument in favor of the definition $\alpha \simeq \text{Im}(\tau) \mu$ for the sources and the resulting canonical form \eqref{our old entropy formula 3} of the holomorphic entropy. In the $N=2$ theory, corresponding to pure gravity, there is an independent holographic notion of entropy in the dual CFT in terms of the thermal entropy of black hole solutions. The latter can be of course computed by the Bekenstein-Hawking area law or any other standard method. As shown in \cite{Compere:2013nba}, it is \eqref{our old entropy formula 3} and not \eqref{holomorphic entropy 4} that coincides with the area of the black hole horizon in the $N=2$ theory. Moreover, the canonical entropy was derived in \cite{Compere:2013nba} by adapting the Wald formalism to Chern-Simons theory, and recently rederived in \cite{Hijano:2014sqa} using these techniques.  These results indicate that the definition $\alpha \simeq \text{Im}(\tau) \mu$ and consequently the entropy  \eqref{our old entropy formula 3} are preferred from a bulk perspective. This is reassuring, because it implies that both the canonical boundary conditions studied in sections \ref{sec: bcs}-\ref{sec: Thermo} and the holomorphic boundary conditions studied in \cite{deBoer:2013gz} and reviewed in section \ref{subsec: holomorphic part fn} yield the same expression for the thermal entropy, consistent with the idea that there should be a single notion of thermal entropy in a bulk theory containing gravity.

Finally, to see how the Gutperle-Kraus result fits in our general framework, we would like to present a 
computation in deformed $2d$ CFT which reproduces the appropriate free energy and entropy. To this end, we first recall that one can find the
free energy by e.g. varying the entropy to read off the sources and charges
\begin{align}\label{variation GK entropy}
\delta S_{\text{G-K}} 
={}&
    -2\pi i k_{cs}\text{Tr}\Biggl[ \tau\, \delta\left( \frac{a_{z}^{2}}{2}  \right)-\bar{\tau}\,\delta \left( \frac{\bar{a}_{\bar{z}}^{2}}{2}\right)
    +\bar{\tau}a_{\bar{z}} \,\delta \left(a_{z}-L_1 \right) +\tau\bar{a}_{z}\,\delta\left(-\bar{a}_{\bar{z}}+L_{-1}\right)\Biggr],
\end{align}

\noindent and the free energy is then given by the corresponding Legendre transform \cite{deBoer:2013gz}
\begin{align}\label{GK free energy}
-\beta F_{\text{G-K}} =
 \ln Z_{\text{G-K}}
  ={}&
   -2\pi i k_{cs}\text{Tr}\biggl[\tau\left(\frac{a_{z}^{2}}{2} \right) -\bar{\tau}\left( \frac{\bar{a}_{\bar{z}}^{2}}{2}\right)
     +\left(\bar{\tau} L_1 a_{\bar{z}}-\tau L_{-1} \,\bar{a}_{z}\right)\biggr].
\end{align}

\noindent One can derive from the results in \cite{deBoer:2013gz} that this free energy follows by computing the partition
function on a \textit{square torus} of the following deformed CFT
\begin{equation}
\label{def-theory}
S = S_{\text{CFT}} + c_1 \left( 1 + \frac{i\tau}{2} \right) \int d^2 z\, T_{\text{CFT}} + c_2 \bar{\tau} \int d^2 z \sum_{s} 
\mu_{s} W_{s} +\text{c.c.}
\end{equation}

\noindent with some numerical constants $c_1,c_2$ which we did determine explicitly. Let us reemphasize that this theory lives on a square torus 
of fixed periodicities and that the dependence on $\tau$ is only through the explicit appearance in the action. It remains to be seen whether deformations of the type \eqref{def-theory} have any particularly nice intrinsic properties, or whether they were merely stumbled upon by accident as a by-product of the definitions in \cite{Gutperle:2011kf}.

\subsection{Field redefinitions}\label{subsec: field redefs}
Even though different boundary conditions in Chern-Simons theory describe different partition functions in the two-dimensional boundary theory, it was already pointed out in \cite{deBoer:2013gz} that field redefinitions exist which allow to map between different Drinfeld-Sokolov pairs. This suggests that redefinitions of the sources might be possible which allow to relate partition functions corresponding to, say, a chiral deformation of the Hamiltonian and a chiral deformation of the action. We will now discuss to what extent this is indeed possible. It will in general turn out to be relatively easy to find redefinitions of the charges in such a way
that the entropies transform into each other, but difficult to find redefinitions of the sources to map the free energies into
each other. 

We will first use chiral stress tensor deformations as an example. This is, for the theory defined on a torus $T^{2}$ with modular parameter $\tau\,$, we would like to relate a Hamiltonian deformation of the form
\begin{align}
Z_{\text{can}}\left[\tau,\alpha_{2}\right] &= \text{Tr}_{\mathcal{H}}\,\left[q^{L_0-\frac{c}{24}}\bar{q}^{\bar{L}_0-\frac{c}{24}}\exp\left(2\pi i \alpha_{2}L_0\right)\right]
\end{align}

\noindent with $q = e^{2\pi i\tau}\,$, and an action deformation of the form 
\begin{equation}
Z_{\text{holo}}\left[\tau, \lambda_{2}\right]
=
 \int\mathcal{D}\phi\, e^{-S_0(\phi)}e^{-i\int_{T^{2}}\frac{d^{2}z}{2\pi \text{Im}(\tau)} \lambda_{2}\mathcal{L}}
\end{equation}

\noindent with $\mathcal{L}$ the left-moving stress tensor. 

Given a Drinfeld-Sokolov pair, for constant  sources the flatness of the gauge connection implies that the conjugate components of the gauge field commute. In the particular case of a stress tensor deformation, that can be described by $sl(2,\mathds{R})$ connections, the two components are actually proportional to each other. Denoting the connection describing the Hamiltonian deformation by $a$ and that describing an action deformation by $b$, from our discussion above in the canonical case plus the corresponding results for the holomorphic case (see \cite{deBoer:2013gz}) it follows that
\begin{equation}\label{stress tensor deformations}
a_{\bar{z}} = \frac{\alpha_{2}}{\bar{\tau}-\tau}\left(a_{z}+a_{\bar{z}}\right)\qquad \text{and}\qquad b_{\bar{z}} = \frac{\lambda_{2}}{\bar{\tau}-\tau}b_{z}\,.
\end{equation}

\noindent The precise proportionality coefficient is fixed by relations such as \eqref{normalization 1}-\eqref{normalization 2}. The idea is to now relate the two sets of gauge fields on the torus to each other through gauge transformations.

Recall now that the only gauge-invariant information carried by the connection is contained in the holonomy around cycles, with $a_{z} + a_{\bar{z}}$ being the component of the connection along the non-contractible cycle of the boundary torus, and $\tau a_{z} + \bar{\tau}a_{\bar{z}}$ the component along the thermal cycle, which becomes contractible in the bulk (and similarly for $b$). Hence, the two sets of gauge fields are gauge-equivalent if their spectrum matches (up to conjugation):
\begin{align} \label{equiv-spec}
\text{spec}\bigl(a_{z} + a_{\bar{z}}\bigr)
 &\sim
   \text{spec}\bigl(b_{z} + b_{\bar{z}}\bigr)
\\
\text{spec}\bigl(\tau a_{z} + \bar{\tau}a_{\bar{z}}\bigr)
 &\sim
   \text{spec}\bigl(\tau b_{z} + \bar{\tau}b_{\bar{z}}\bigr)\,.
\end{align}

\noindent Using the on-shell relations \eqref{stress tensor deformations} these conditions become
\begin{align}
\text{spec}\bigl(a_{z} + a_{\bar{z}}\bigr)
 &\sim
   \left(1+\frac{\lambda_{2}}{\bar{\tau}-\tau}\right)\text{spec}\left(b_{z}\right)
\\
\left(\tau + \alpha_{2}\right)\text{spec}\bigl(a_{z} + a_{\bar{z}} \bigr)
 &\sim
   \left(\tau +\frac{\bar{\tau}\lambda_{2}}{\bar{\tau}-\tau}\right) \text{spec}\bigl( b_{z}\bigr)\,,
\end{align}

\noindent implying
\begin{equation}\label{relation can and holo sources}
1 + \frac{i\lambda_{2}}{2\text{Im}(\tau)} = \left(1 - \frac{i\alpha_{2}}{2\text{Im}(\tau)}\right)^{-1}\,.
\end{equation}

\noindent This is precisely the same relation obtained in \cite{Dijkgraaf:1996iy} using field theory techniques, reproduced here with very simple manipulations in terms of flat connections in Chern-Simons theory.

Why did this work? The reason is that (\ref{equiv-spec}) implies that the gauge fields transform under a global gauge transformation, i.e.
\begin{align} \label{trafo}
(a_{z} + a_{\bar{z}}) &= U^{-1}(b_{z} + b_{\bar{z}})U
\\
(\tau a_{z} + \bar{\tau}a_{\bar{z}}) &= U^{-1} (\tau b_{z} + \bar{\tau}b_{\bar{z}})U \, ,
\end{align}

\noindent and therefore any quantity which consists of the trace of the products of gauge fields will be left invariant under this 
transformation. The entropy is of this form in general, and that is why transformations of this type can be used to 
find charge redefinitions which leave the entropy invariant. For the free energy, however, the situation is more complicated.
The on-shell value of the Chern-Simons action is left invariant under the global gauge transformations (\ref{trafo}), but
the boundary terms are not, because these contains terms like $\text{Tr}\left[L_1 a_{\bar{z}}\right]$ which are not invariant 
under (\ref{trafo}), given that $L_1$ is kept fixed. Then why did the computation for chiral stress-tensor deformations work? 
It did because it so happens that the boundary terms vanish for such deformations. 

The general lesson is therefore that although we can relate in a fairly straightforward way 
the different entropies to each other with redefinitions of the charges, the free energies do not share this property. 
Although this does not say that there can not exist redefinitions of the sources which relate the free energies, 
Chern-Simons theory does not appear to provide a natural candidate, except in the case of stress-tensor deformations. 

Let us now comment on an application involving non-chiral deformations. As we have mentioned, using Chern-Simons theory the results \eqref{our entropy formula} and \eqref{our entropy formula holomorphic 4} were derived in \cite{deBoer:2013gz}, the first corresponding to canonical boundary conditions, and the second to holomorphic boundary conditions, a particular choice of stress tensor coupling to $\tau$ and a particular scaling $\alpha = \bar{\tau}\mu$ and $\bar{\alpha} = \tau \bar{\mu}\,$ of the sources with the modular parameter. On the other hand, the components of the connection carrying the charges in either case are $a_{z} + a_{\bar{z}} = L_1 + \tilde{Q}$ and $a_{z} = L_1 + Q\,$, with similar expressions in the barred sector. Since both $Q$ and $\tilde{Q}$ are highest weight matrices which are linear in the corresponding charges, it follows that the matrix $a_{z} + a_{\bar{z}}$ in the canonical description has the same form as function of the tilded charges that $a_{z}$ has as function of the untilded charges in the holomorphic description. From  \eqref{our entropy formula} and \eqref{our entropy formula holomorphic 4} it is then immediate that the functional form of the canonical entropy, as a function of the canonical charges, is exactly the same as the functional form of the Gutperle-Kraus entropy as a function of the holomorphic charges. This agreement was first noticed in \cite{Compere:2013nba}, and while establishing it from a field theory perspective would be presumably quite involved, it emerges in a very transparent way when using the holographic description in terms of Chern-Simons theory. 

To be a bit more explicit about the above  map at the level of free energies, consider the canonical free energy (\ref{canonical free energy}) and the Gutperle-Kraus free energy \eqref{GK free energy}. It is easy to see that if we start with the former, and make
the following change of variable (restricting to the chiral sector for simplicity)
\begin{equation}\label{field redefinition example}
(\bar{\tau}-\tau)a_{\bar{z}} \rightarrow \bar{\tau} b_{\bar{z}}\,,\qquad
a_z+a_{\bar{z}} \rightarrow b_z 
\end{equation}

\noindent we get precisely the Gutperle-Kraus free energy in terms of $b_z,b_{\bar{z}}\,$. In addition, if $(a_z+a_{\bar{z}},a_{\bar{z}})$
was a Drinfeld-Sokolov pair, then so is $(b_z,b_{\bar{z}})$, and the trivial monodromy around the contractible cycle is
preserved because $\tau a_z + \bar{\tau} a_{\bar{z}} = \tau b_z + \bar{\tau} b_{\bar{z}}\,$.  The field redefinition \eqref{field redefinition example} then realizes a map between Hamiltonian deformations, dual to canonical boundary conditions, and deformations of the type \eqref{def-theory}, which are dual to Gutperle-Kraus boundary conditions. 

Given that the Gutperle-Kraus boundary conditions are dual to action deformations, while the canonical boundary conditions are dual to Hamiltonian deformations, it might seem surprising that a detailed agreement was found between the free energies computed from the bulk theory with Gutperle-Kraus boundary conditions \cite{Kraus:2011ds} and a CFT calculation that involved Hamiltonian deformations by zero modes \cite{Gaberdiel:2012yb}. From the map \eqref{field redefinition example} and the above discussion it is clear that the \textit{functional form} of the canonical free energy, as a function of the canonical sources $\alpha\simeq \text{Im}(\tau)\mu$\,, is exactly the same as that of the Gutperle-Kraus free energy as a function of the sources $\alpha_{\text{G-K}} = \bar{\tau}\mu\,$. This explains why the two calculations seemingly agreed, even though they involve two a priori different partition functions. We will further comment on the implications of these findings in section \ref{sec: Disc}.

\subsection{Modular transformations}\label{subsec: modular trafos}

Recall that in $2d$ CFT modular transformations can be understood as a change of coordinates followed by a scale transformation, which are symmetries of the deformed action when the currents and sources transform appropriately \cite{Dijkgraaf:1996iy}. In our example involving stress tensor deformations, this implies in particular that $\lambda_{2}$ above transforms covariantly under modular transformations; from \eqref{relation can and holo sources} it then follows that the canonical source $\alpha_{2}$ transforms in a complicated way. Another way to understand this fact is to notice that the rescaling amounts to a gauge transformation, and that the combined effect of the change of coordinates and gauge transformation preserves the Drinfeld-Sokolov form of the pair $(b_{z}\,,b_{\bar{z}})\,$, but not that of $(a_{z} + a_{\bar{z}}\,,a_{\bar{z}})$. An additional compensating transformation would be necessary to put the gauge field $a$ back into the appropriate Drinfeld-Sokolov form, which explains why the canonical source $\alpha_{2}$ transforms in a complicated way under modular transformations. Whether
such transformations exist when higher sources are turned on is not clear.

It is instructive to describe the Chern-Simons perspective on modular transformations, an issue that was recently investigated in \cite{Li:2013rsa}. Here we will provide a succinct derivation that will once more make it clear why modular transformations are simple for deformed Lagrangians and complicated for deformed Hamiltonians. We will only consider chiral deformations in what follows, but the results can be generalized to the non-chiral case in a straightforward way. 

As we have discussed, there are different possible three-manifolds we can use to evaluate the Chern-Simons action. If the boundary
is a two-torus, there is an entire $SL(2,\mathds{Z})$ family of three-manifolds we can choose, each yielding a different answer
for the on-shell value of the action. To write this answer explicitly, we rewrite (\ref{holomorphic free energy}) for one chiral sector as
\begin{align}\label{holomorphic free energy again}
 \ln Z_{\text{holo}}
  ={}&
   -\pi i k_{cs}\text{Tr}\biggl[(\tau a_{z} + \bar{\tau} a_{\bar{z}})(a_z+a_{\bar{z}}) +
   \left(\bar{\tau}-\tau\right)\bigl(\left(2L_1-a_z\right) a_{\bar{z}}\bigr)  \biggr]
\end{align}

\noindent where the first term is the contribution from the on-shell value of the Chern-Simons action, and the second term is the contribution from the boundary term. For a different three-manifold, labeled by an $SL(2,\mathds{Z})$ matrix
\begin{equation}
R =
 \left(
\begin{array}{cc}
 \alpha & \beta \\
  \gamma & \delta
\end{array}
\right)
\end{equation}

\noindent the partition function becomes\footnote{As we have emphasized the boundary term is the same for any choice of three-manifold, but the on-shell value of the Chern-Simons action depends on how the two-torus is filled, namely the choice of contractible and non-contractible cycles in the bulk.}
\begin{align}\label{holomorphic free energy again}
 \ln Z_{\text{holo}}[R]
  =
   -\pi i k_{cs}\text{Tr}\biggl[
   &\Bigl(\alpha\left( \tau a_{z} + \bar{\tau} a_{\bar{z}} \right) +\beta ( a_z+a_{\bar{z}} ) \Bigr)
\Bigl(\gamma \left( \tau a_{z} + \bar{\tau} a_{\bar{z}} \right) +\delta ( a_z+a_{\bar{z}} ) \Bigl) 
\nonumber\\
&+
   \left(\bar{\tau}-\tau\right)\Bigl((2L_1-a_z) a_{\bar{z}}\Bigr)  \biggr].
\end{align}

\noindent In the first term, we recognize the product of the monodromies of the gauge field along the new a-cycle and b-cycle of the boundary two-torus. 

Suppose that in the above me make the substitution
\begin{equation}\label{modtrafo}
a_z \rightarrow \left(\gamma\tau+\delta\right)^{-1} U b_z U^{-1}\,,\qquad
a_{\bar{z}} \rightarrow \left(\gamma\bar{\tau}+\delta\right)^{-1} U b_{\bar{z}} U^{-1}
\end{equation}

\noindent with
\begin{equation}
U = \exp\Bigl[\ln\left(\gamma\tau + \delta\right) L_0\Bigr].
\end{equation}

\noindent This substitution preserves the Drinfeld-Sokolov form of the gauge field, i.e. if $(a_z,a_{\bar{z}})$
is a Drinfeld-Sokolov pair then so is $(b_z,b_{\bar{z}})$. Morever, by direct calculation, we observe
that after this substitution the partition function takes the original form (\ref{holomorphic free energy})
with $\tau$ replaced by $(\alpha\tau+\beta)/(\gamma\tau+\delta)$. Thus, to summarize, we have shown that
\begin{equation}
 \ln Z_{\text{holo}}[R]\left[\tau;(\gamma\tau+\delta)^{-1} U b_z U^{-1},(\gamma\bar{\tau}+\delta)^{-1} U b_{\bar{z}} U^{-1}\right]
= 
 \ln Z_{\text{holo}}[\mathds{1}]\left[\frac{\alpha\tau+\beta}{\gamma\tau+\delta};b_z,b_{\bar{z}}\right],
\end{equation}

\noindent where $Z_{\text{holo}}[\mathds{1}]$ on the r.h.s. denotes the partition function in the original manifold, labeled by $R =\mathds{1}\,$. This is the Chern-Simons version of modular invariance, and we see that (\ref{modtrafo}) provides the
transformation rules for the sources, in agreement with what one gets directly from the deformed action.

Interestingly, for deformations of the Hamiltonian the above computation does not work due to the different structure of the boundary term, and we have not succeeded in deriving a general transformation rule under modular transformations for the sources in that case. It would be very interesting to explore this issue further.

\section{Discussion} \label{sec: Disc}
Starting from two-dimensional CFTs with a (possibly higher spin) current symmetry algebra, we have reviewed different types of deformations that are possible once sources are switched on. While some of these can be understood as deformations of the CFT Hamiltonian, others are defined as changes directly at the level of the action. Associated with each of these theories there is a notion of partition function that is a function of the background sources, and whose associated Ward identities we have studied. Using the Ward identities as the guiding principle, we have argued that these different theories map to different boundary conditions in a holographic realization in terms of Chern-Simons theory on a three-dimensional manifold with boundary. The issue of boundary conditions in the higher spin AdS$_{3}$/CFT$_{2}$ correspondence has proven to be particularly subtle, and it is therefore worth summarizing how our analysis fits with the recent literature.

In the holographic context, a first set of boundary conditions in the presence of higher spin sources was proposed in \cite{Gutperle:2011kf,Ammon:2011nk}, with the flatness condition on the connection resulting in Ward identities of the form \eqref{master Ward identity chiral}. We have argued that these boundary conditions most naturally correspond to a deformation of the CFT action of the form \eqref{chiral def action} in the chiral case, and to an action involving infinitely many higher order terms in the sources in the non-chiral case. The latter can be rewritten linearly in the sources at the expense of introducing auxiliary fields, but this formulation has to be constructed on a case by case basis. 

For the finite temperature version of the holomorphic theory on the torus, two definitions of the thermal higher spin sources have been proposed. The first alternative was put forward in \cite{Gutperle:2011kf,Ammon:2011nk} and identifies the sources schematically as $\alpha = \bar{\tau}\mu$ and $\bar{\alpha} = \tau \bar{\mu}\,$, where $\mu$ and $\bar{\mu}$ are the chemical potentials. This choice implies in particular that the expression for the energy is the same as in the absence of sources, and leads to an entropy of the form \eqref{holomorphic entropy 4}. This identification of the sources is not what one gets from deformations of the form \eqref{chiral def action}, but maps instead to a peculiar deformation of the form \eqref{def-theory}, with the theory defined on a square torus. An alternative definition was studied in \cite{deBoer:2013gz}, which consists in defining the thermal sources as $\alpha = -i\beta\mu\,$, $\bar{\alpha} = i\beta\bar{\mu}$ with $\beta = 2\pi\text{Im}(\tau)$ the inverse temperature. This case precisely describes deformed actions of the form \eqref{chiral def action} and the expression for the energy is explicitly modified with respect to the undeformed theory, a fact that we have rederived from a field theory perspective in section \ref{subsec: tau coupling}, and one is led in particular to the formula \eqref{our old entropy formula 3} for the entropy \cite{deBoer:2013gz}.

A different set of ``canonical" boundary conditions in the presence of sources was proposed in \cite{Compere:2013gja,Compere:2013nba,Henneaux:2013dra,Bunster:2014mua} from a bulk perspective, with the flatness condition on the connection resulting in Ward identities of the form \eqref{master Ward identity mu}. We have shown that these boundary conditions correspond to deformations of the CFT Hamiltonian of the form \eqref{deformed Hamiltonian}. In our discussion of the finite temperature version of this theory on the torus, we have exploited the holographic description in terms of Chern-Simons theory to provide expressions for the stress tensor \eqref{canonical stress tensor zero modes}, free energy \eqref{canonical free energy}, and entropy \eqref{canonical entropy}, which are written entirely in terms of the gauge connections and are valid in any embedding. It is satisfying to note that, provided the thermal sources are always identified as $\alpha = -i\beta\mu\,$, Chern-Simons theory yields the same functional for the entropy in theories corresponding to Hamiltonian and Lagrangian deformations (c.f. \eqref{canonical entropy} and \eqref{our old entropy formula 3}), consistent with the expectation that there should exist an unambiguous functional that computes the thermal entropy in a bulk theory containing gravity. 

It has been proposed \cite{Henneaux:2013dra,Bunster:2014mua} that the solutions constructed in \cite{Gutperle:2011kf,Ammon:2011nk} that realize the holomorphic $\mathcal{W}_{3}$ boundary conditions are in fact $\mathcal{W}_{3}^{(2)}$ boundary conditions in disguise. This conclusion was arrived at by interpreting the solutions of \cite{Gutperle:2011kf,Ammon:2011nk} in light of a canonical Drinfeld-Sokolov pair of the form \eqref{DS sigma t}-\eqref{DS sigma t 2}. In the original proposal, however, these solutions are interpreted instead in terms of a holomorphic Drinfeld-Sokolov pair of the form $a_{z} = L_1 + Q$, $a_{\bar{z}} = M +\ldots$. For chiral deformations, we have shown that the latter choice realizes a canonical structure where one of the light-cone directions is chosen as the ``time" coordinate \cite{Witten:1983ar}, c.f. the Dirac bracket algebra \eqref{extended holomorphic bracket} obtained by acknowledging the presence of a second class constraint $P_i = (1/2)\partial_+ X_i\,$. It is conceivable that a canonical structure based on a null coordinate could be at odds with a well-posed Cauchy problem in the bulk when sources of both chiralities are switched on, and this issue deserves further scrutiny. On the other hand, we have argued that well-defined partition functions exist in CFT whose Ward identities are indeed those obtained in \cite{Gutperle:2011kf,Ammon:2011nk}, and we expect them to have a dual description in the bulk. Consequently, our point of view is that the $\mathcal{W}_{3}$ boundary conditions proposed in \cite{Gutperle:2011kf,Ammon:2011nk} do indeed give rise to $\mathcal{W}_{3}$ symmetry, and that no conflict arises when they are interpreted in a light-cone framework as in \cite{Witten:1983ar} (or a suitable generalization thereof in the non-chiral case). 

To add to this, we emphasize than just providing a solution of the Chern-Simons field equations, i.e. a flat connection, is not sufficient; 
we also need to specify an a priori choice of boundary conditions, boundary terms, and identification of sources and dual
expectation values, and different choices can provide different interpretations for the same flat connections. For one choice the flat connections in \cite{Gutperle:2011kf,Ammon:2011nk} describe a solution with $\mathcal{W}_{3}$ boundary conditions, and for another 
choice they describe a solution with $\mathcal{W}_{3}^{(2)}$ boundary conditions. Both are valid but inequivalent points of view.

Regarding the matching between bulk and boundary computations, it might appear as somewhat surprising that a chiral half of the partition function (free energy) derived in \cite{Kraus:2011ds} from the bulk theory with Gutperle-Kraus boundary conditions, which as we have seen here correspond to a linear deformation of the CFT action, has been matched by a CFT calculation involving a chiral deformation of the Hamiltonian by zero modes \cite{Gaberdiel:2012yb}. To clarify this issue, in section \ref{subsec: field redefs} we have shown that the \textit{functional form} of the partition function (as a function of the sources) and  of the entropy (as a function of the charges) is the same with canonical or Gutperle-Kraus boundary conditions, even though different definitions of the sources and charges themselves are been used in one version of the theory or the other. As we have discussed in depth, the detailed matching between charges and sources in the bulk and boundary, namely the holographic dictionary, will however change depending on what precise version of the theory we want to describe. As a consequence, one should in principle expect observables such as correlators, which are generically not fixed by symmetry or otherwise, to be sensible to these choices. These subtle differences have indeed been noticed in calculations of thermal correlators of  scalar primaries in CFTs with higher spin symmetry \cite{Gaberdiel:2013jca}, and we expect our analysis to shed light on these issues as well. 

It is perhaps worthwhile to briefly discuss the validity and interpretation of the irrelevant deformations that we considered. 
A priori, theories deformed by irrelevant deformations are ill-defined. In the present case we are deforming by conserved currents,
which might improve the situation. Let us first think what happens when we expand the theories as a power series
in terms of the sources, with each term being an integrated correlation function. These correlation functions are singular when
points coincide and some regularization has to be employed. In standard conformal perturbation theory, one cuts out small
disks around the points and subtracts all singularities that arise when shrinking the disks to zero size. We expect this procedure
to yield finite, well-defined answers, in particular since the conserved currents cannot develop anomalous dimensions.\footnote{One might worry that contact terms produce divergences containing new operators which would need to be added to the theory to make it consistent. For example, the OPE of two spin-three currents contains $T^2$, the square of the energy-momentum tensor, which
does not appear in the deformed theory. We do not see any need, at least classically, to add such deformations to the theory.}
Therefore, the theories we consider may well have well-defined perturbative expansions in $\mu_s$ and $\bar{\mu}_s\,$. These
perturbative expansions presumably have zero radius of convergence, and it is an interesting questions whether one can directly define
the deformed theories non-perturbatively e.g. by choosing suitable complex contours. 

There are two other arguments that these deformed theories make sense. First, one the plane, we can perform a higher spin transformation which puts all $\mu_s=0\,$, mapping the deformed theory to the original, undeformed theory. The latter is clearly well-defined, and so should the former? Perhaps, except
that it is not clear that the required higher spin transformations act in a reasonable way, they could for example map normalizable
field configurations into non-normalizable field configurations. Moreover, on a torus one cannot get rid of the zero modes of
the $\mu_s$ and $\bar{\mu}_s$ in this way and the argument no longer applies. A second argument that these deformed theories
are well-defined is that we can use Chern-Simons theory to compute their partition functions, and the result is a non-pathological
function of $\mu_s$ and $\bar{\mu}_s\,$. Clearly, more work is required before we can make a definite statement about the existence
of CFT's deformed by irrelevant deformations of conserved currents.

We have by no means exhausted the possible deformations of $2d$ conformal field theories, nor have we exhausted the possible list
of boundary conditions in Chern-Simons theory. It would be interesting to examine whether other interesting boundary conditions exist
and if so what their $2d$ CFT interpretation is. Similarly, one could extend our considerations to encompass the non-AdS (non-CFT) higher spin dualities studied in \cite{Afshar:2012nk,Gary:2012ms}. As discussed in section \ref{subsec: modular trafos} and appendix \ref{app: U1}, the different types of partition functions we have studied moreover differ in their modular transformations properties. We have not found a change of variable which directly connects deformations
of the action to deformations of the Hamiltonian, however, and in particular we have not been able to determine the behavior of the 
latter under modular transformations. It is possible that in order to find such a change of variable additional operators need
to be included, such as normal-ordered products of higher spin fields and their derivatives, and it would be interesting to explore whether
such more general deformed theories still admit dual Chern-Simons descriptions. These interesting questions will be discussed elsewhere.

\vskip 1cm
\centerline{\bf Acknowledgments}
\noindent It is a pleasure to thank Per Kraus and Daniel Robbins for enlightening discussions and comments on a draft of this paper. We are also grateful to Marco Baggio, Max Ba\~nados, Alejandra Castro, Geoffrey Comp\`ere, Matthias Gaberdiel, Daniel Grumiller, Diego Hofman, Romuald Janik, Rob Leigh, Wei Li, Eric Perlmutter, Wei Song, Hai-Siong Tan and Erik Verlinde for helpful conversations. J.I.J. is supported by funding from the European Research Council, ERC Grant agreement no.~268088-EMERGRAV. This work is part of the research programme of the Foundation for Fundamental Research on Matter (FOM), which is part of the Netherlands Organization for Scientific Research (NWO).

\appendix

\section{A $U(1)$ example}\label{app: U1}
Here we will briefly review a non-higher spin example from \cite{Kraus:2006nb}, involving deformations by $U(1)$ currents in a compact boson realization. The canonical partition function with sources for left- and right-momenta is 
\begin{equation}\label{canonical boson part function}
Z_{\text{can}}\left[\tau,\alpha_{L},\alpha_{R}\right]= \frac{1}{\left(q\bar{q}\right)^{1/24}}\text{Tr}\left[q^{L_0}\bar{q}^{\bar{L}_0}e^{2\pi i \alpha_L p_L}e^{-2\pi i \alpha_{R}p_{R}}\right]
\end{equation}

\noindent where $q = \exp(2\pi i\tau)$ and
\begin{align}\label{left momentum}
p_{L} ={}& \oint \frac{d\sigma}{2\pi}\bigl(\partial_{\sigma}X - i\partial_{t_{E}}X\bigr)\,,
\\
p_{R} ={}&
 \oint \frac{d\sigma}{2\pi}\bigl(\partial_{\sigma}X + i\partial_{t_{E}}X\bigr)\,.
 \label{right momentum}
\end{align}

\noindent It is tempting to conclude that the path integral representation of this partition function is
\begin{equation}\label{Zhol}
Z_{\text{Lag,naive}}= \int \mathcal{D}X\,e^{-S_{0} + \int_{T^{2}} d^{2}\sigma \sqrt{g} A^{i}\partial_{i}X }\,,
\end{equation}

\noindent where $S_0$ is the free action
\begin{equation}
S_0 = \frac{1}{4\pi}\int_{T^{2}}d^{2}\sigma \sqrt{g}\, g^{ij}\partial_{i} X \partial_{j}X =  \frac{1}{4\pi}\int_{T^{2}}d^{2}\sigma\left[ \left(\partial_{t_{E}} X\right)^2 + \left( \partial_{\sigma}X\right)^2\right],
\end{equation}

\noindent (we consider a flat torus with $ds^{2}(T^{2}) = dzd\bar{z} =dt_{E}^{2}+d\sigma^{2}$) and the background gauge field, whose components are the chemical potentials, given by
\begin{equation}\label{external gauge field}
A_{z} = -i\frac{\alpha_{R}}{2\pi\text{Im}(\tau)} = \mu_{R} \,,\qquad A_{\bar{z}} = i\frac{\alpha_{L}}{2\pi\text{Im}(\tau)} = \mu_{L}\,.
\end{equation}

\noindent  In particular, since modular transformations correspond to a change of coordinates followed by a Weyl rescaling, which are symmetries of the deformed action, $Z_{\text{Lag,naive}}$ is modular invariant:
\begin{equation}\label{Zhol is modular invariant}
Z_{\text{Lag,naive}}\left[\frac{a\tau + b}{c\tau + d},\frac{\alpha_{L}}{c\tau + d},\frac{\alpha_{R}}{c\bar{\tau}+d}\right] = Z_{\text{Lag,naive}}\left[\tau,\alpha_{L},\alpha_{R}\right].
\end{equation}

%
%

On the other hand, following the standard steps to discretize the operator trace, the path integral representation of $Z_{\text{can}}$ is found to be
\begin{equation}
Z_{\text{can}} =
 \int  \mathcal{D}P\, \mathcal{D}X\, e^{\int_{T^{2}}d^{2}\sigma \left[-\frac{1}{2\pi}\left(P\dot{X}- \frac{1}{2} \left(P\right)^{2} +  \frac{1}{2}  \left( \partial_{\sigma}X\right)^2\right) + A_{t_{E}}P + A_{\sigma}\partial_{\sigma} X\right]}
\end{equation}

\noindent with $P$ the momentum conjugate to $X\,$.  Integrating out $P$ one concludes \cite{Kraus:2006nb}
\begin{equation}
Z_{\text{can}}\left[\tau,\alpha_{L},\alpha_{R}\right]=e^{-\frac{\pi\left(\alpha_{L} + \alpha_{R}\right)^2}{\text{Im}(\tau)}} Z_{\text{Lag,naive}}\left[\tau,\alpha_{L},\alpha_{R}\right],
\end{equation}

\noindent which in particular implies (c.f. \eqref{Zhol is modular invariant})
\begin{equation}\label{modular transformation Z}
Z_{\text{can}}\left[\frac{a\tau + b}{c\tau + d},\frac{\alpha_{L}}{c\tau + d},\frac{\alpha_{R}}{c\bar{\tau}+d}\right]= e^{\frac{2\pi i c}{c\tau+d}\alpha_{L}^{2}}e^{-\frac{2\pi i c}{c\bar{\tau}+d}\alpha_{R}^{2}}Z_{\text{can}}\left[\tau,\alpha_{L},\alpha_{R}\right].
\end{equation}

\noindent Therefore, the canonical partition function in the presence of sources is not modular invariant, but rather modular covariant. The bottom line is that, even in simple examples such as a deformation of the Hamiltonian by constant $U(1)$ chemical potentials, it is important to acknowledge that the proper representation of the canonical partition function involves the path integral in first order form, and to exercise care when Legendre-transforming to pass to the Lagrangian version of the theory.

\section{Useful $\mathcal{W}_{3}$ formulae}
As explained in the main text, the improved $\mathcal{W}_{3}$ generators in the bosonic realization are
\begin{align}\label{improved T b}
T ={}&
 \frac{1}{2}\delta_{ij}\Pi^{i} \Pi^{j} + a_{i}\partial_{\sigma} \Pi^{i}
\\
W ={}&
  \frac{1}{3}d_{ijk}\Pi^{i} \Pi^{j} \Pi^{k} + e_{ij}\partial_{\sigma} \Pi^{i} \Pi^{j} + f_{i}\partial_{\sigma}^{2} \Pi^{i}\,.
  \label{improved W b}
\end{align}

\noindent The improved $TT$ bracket takes the usual form \eqref{TT bracket extended} provided
\begin{equation}\label{a squared is central charge b}
a_{i}a^{i} = -\frac{c}{12} \,,
\end{equation}

\noindent where $c$ denotes the \textit{classical} central charge. Similarly, the form of the $TW$ bracket requires
\begin{equation}\label{TW conditions b}
a_{i}f^{i} = 0\,,\qquad f_{i} = a^{j}e_{ji} \,, \qquad 3a^ie_{ij} = a^{i}e_{ji}\,,\qquad e_{(ij)} = d_{ij}^{\hphantom{ij}k}a_{k}\,.
\end{equation}

\noindent Finally, the improved $WW$ bracket \eqref{WW bracket extended} requires \eqref{d tensor} to be satisfied with
\begin{align}
 \kappa = -\frac{16}{c}\,,
\end{align}

\noindent and 
\begin{eqnarray}
 f_{i}f^{i}
 &=&
  -\frac{c}{36}
\\  
   a_{i} &=&
 e_{ji}f^{j}
\\   
 e_{ij}f^{j}
&=&
 \frac{1}{3}e_{ji}f^{j}
\\
d_{ij}^{\hphantom{ij}k}(e_{k\ell} - e_{\ell k}) - 2d_{\ell(j}^{\hphantom{(\ell(j}k}e_{ki)}
&=&
\frac{32}{c}\delta_{ij} a_{\ell}
\\
-2d_{ij}^{\hphantom{ij}k}f_{k} + e^{k}_{\hphantom{k}i}e_{kj}
&=&
 \frac{5}{3}\delta_{ij} 
 \\
e_{k[i}e_{j]}^{\hphantom{j]}k}
&=&
0
\\
e_{(ik}e^{k}_{\hphantom{k}j)}
&=&
\delta_{ij} 
\\
6d_{ij}^{\hphantom{ij}k}f_{k} + e_{ik}e_{j}^{\hphantom{j}k} + \frac{64}{c}a_{i}a_{j} 
&=&
- \delta_{ij}
\label{last coeff constraint}
\end{eqnarray}

\noindent in addition to \eqref{a squared is central charge b} and \eqref{TW conditions b}.  It is worth pointing out that there is some degree of redundancy in these constraints; if so desired, one could choose a minimal set that contains all the information. We emphasize that the above conditions were derived semiclassically, at the level of Poisson brackets, and therefore ignoring operator ordering issues. The resulting expressions can be viewed as the ``large-$c$" version of the full constraints obtained from the quantum $\mathcal{W}_{3}$ algebra, derived in \cite{Romans:1990ag}. An immediate consequence of the conditions on the various coefficients is that at least two scalars are needed in order to support an arbitrary semiclassical central charge.

When the above conditions are satisfied, the improved generators satisfy the Poisson algebra
\begin{align}
\left\{J_{\alpha}(\sigma),J_{\beta}(\sigma')\right\}  ={}&
 \int dx \,f_{\alpha\beta}^{\hphantom{\alpha\beta}\gamma}(\sigma,\sigma',x)J_{\gamma}(x) +  c_{\alpha\beta}(\sigma,\sigma')\,,
\end{align}

\noindent with
\begin{eqnarray}\label{extended struct const 1}
f_{TT}^{\hphantom{TT}T}(\sigma,\sigma',x) &=&
 -\delta\left(\sigma -x\right)  \partial_{x}\delta\left(x -\sigma'\right) + \delta\left(x-\sigma'\right)  \partial_{x}\delta\left(\sigma-x\right) 
 \\
 f_{TW}^{\hphantom{TW}W}(\sigma,\sigma',x) &=&
 -\delta\left(\sigma -x\right)  \partial_{x}\delta\left(x -\sigma'\right) + 2\delta\left(x-\sigma'\right)  \partial_{x}\delta\left(\sigma-x\right) 
 \\
 f_{WT}^{\hphantom{TW}W}(\sigma,\sigma',x) &=&
 \delta\left(\sigma' -x\right)  \partial_{x}\delta\left(x -\sigma\right) -2\delta\left(x-\sigma\right)  \partial_{x}\delta\left(\sigma'-x\right)
 \\
 f_{WW}^{\hphantom{WW}T}(\sigma,\sigma',x) &=&
2\kappa\, T(x)\left[-\delta\left(\sigma -x\right)  \partial_{x}\delta\left(x -\sigma'\right) + \delta\left(x-\sigma'\right)  \partial_{x}\delta\left(\sigma-x\right) \right]
\nonumber\\
&& 
-\frac{2}{3}\partial^{3}_{x}\delta(\sigma-x)\delta(x-\sigma') + \partial^{2}_{x}\delta(\sigma-x)\partial_{x}\delta(x-\sigma')
\nonumber\\
&&
 - \partial_{x}\delta(\sigma-x)\partial^{2}_{x}\delta(x-\sigma') + \frac{2}{3}\delta(\sigma-x)\partial^{3}_{x}\delta(x-\sigma')
\label{extended struct const 4}
\end{eqnarray}

\noindent and
\begin{eqnarray}\label{central charge matrix 1}
c_{TT}(\sigma,\sigma') &=&
 -\frac{c}{12}\partial^{3}_{\sigma}\delta(\sigma-\sigma')
 \\
 c_{WW}(\sigma,\sigma') &=&
 \frac{c}{36}\partial^{5}_{\sigma}\delta(\sigma-\sigma')\,.
 \label{central charge matrix 2}
\end{eqnarray}

Writing the boundary Chern-Simons connection in highest weight gauge,\footnote{We follow the conventions of \cite{Campoleoni:2010zq} up to a rescaling of the currents by a factor of $2\pi\,$.}
\begin{equation}
a_\sigma = L_1 + \frac{T}{k}L_{-1} - \frac{W}{4k}W_{-2}\,,
\end{equation}

\noindent it was found in \cite{Campoleoni:2010zq} that the gauge transformations $\delta a = d\lambda + \left[a,\lambda\right]$ that respect the Drinfeld-Sokolov form of $a_{\sigma}$ are generated by an infinitesimal parameter
\begin{equation}\label{gauge parameter}
\lambda = \sum_{i=-1}^{1}\epsilon^{i}L_i + \sum_{m=-2}^{2}\chi^{m}W_{m}
\end{equation}

\noindent with 
\begin{eqnarray}\label{gauge param 1}
\epsilon^0 &=& -\partial_{\sigma}\epsilon
\\
\epsilon^{-1} &=& \frac{1}{2}\partial^{2}_{\sigma}\epsilon + \frac{T}{k}\epsilon + \frac{2W}{k}\chi
\\
\chi^{1} &=& -\partial_{\sigma}\chi
\\
\chi^{0} &=& \frac{1}{2}\partial^{2}_{\sigma}\chi + \frac{2}{k}\chi T
\\
\chi^{-1} &=& -\frac{1}{6}\partial^{3}_{\sigma}\chi -\frac{5}{3k}T\partial_{\sigma}\chi - \frac{2}{3k}\chi \partial_{\sigma}T
\\
\chi^{-2} &=& \frac{1}{24}\partial^{4}_{\sigma}\chi + \frac{2}{3k}T\partial^{2}_{\sigma}\chi + \frac{7}{12k}\partial_{\sigma}T\partial_{\sigma}\chi
\nonumber\\
&& 
+ \frac{1}{6k}\chi \partial^{2}_{\sigma}T + \frac{1}{k^{2}}\chi T^{2} -\frac{\epsilon}{4k}W
\label{gauge param 2}
\end{eqnarray}

\noindent where $\epsilon^{1} \equiv \epsilon$ and $\chi^{2} \equiv \chi\,$. Under such transformations, the change in the charges is precisely given by \eqref{spin 2 transf of currents}-\eqref{spin 3 transf of currents}.

\section{Non-chiral stress tensor deformations}\label{app: stress tensor defs}
In certain cases it is possible to write down a partition function whose symmetries give rise to two decoupled copies of Ward identities of the type \eqref{chiral T Ward identity}-\eqref{chiral W Ward identity}, at the expense of introducing auxiliary fields \cite{Schoutens:1990ja} (see \cite{Hull:1993kf} for a review). The auxiliary field formalism is non-universal and has to be constructed on a case-by-case basis, but we can illustrate many of its important features by considering a simple example involving stress tensor deformations. Consider then the action for the scalar field theory with both left- and right-moving stress tensor deformations
\begin{equation}
S_{\text{aux}}
= 2\int d^{2}x\left(-\frac{1}{2}\partial_+ X^i \partial_{-}X^i - \Pi^i_+\Pi^i_- + \Pi^i_+\partial_- X^i + \Pi^i_-\partial_+X^i - \mu_{--}T_{++} - \mu_{++}T_{--}\right)
\end{equation}

\noindent where $\Pi^{i}_{\pm}$ denote the auxiliary fields and 
\begin{equation}\label{def stress tensor auxiliary fields}
T_{\pm\pm} \equiv \frac{1}{2}\Pi^{i}_{\pm}\Pi^{i}_{\pm}\,.
\end{equation}

\noindent For the sake of simplicity, we have omitted improvement terms that would generate classical central extensions. When $\mu_{\pm \pm} = 0\,$, $S_{\text{aux}}$ yields the free boson action upon integrating out the auxiliary fields. When deformations are present, the action is invariant under the infinitesimal transformation
\begin{eqnarray}
\delta X^i &=& p_- \Pi^i_+ + p_+ \Pi^i_-
\\
\delta \mu_{\pm\pm} &=& \partial_{\pm}p_{\pm} + p_{\pm}\partial_{\mp}\mu_{\pm\pm} - \mu_{\pm\pm}\partial_{\mp}p_{\pm}
\label{variation sources}
\\
\delta \Pi^i_{\pm} &=&
\partial_{\pm}\left(p_{\mp}\Pi^{i}_{\pm}\right)\,.
\end{eqnarray}

\noindent Since the path integral contains an integration over $X$ and $\Pi$, this symmetry yields the Ward identity
\begin{equation}
\int d^{2}x\left\langle \frac{\delta S_{\text{aux}}}{\delta \mu_{++}}\delta \mu_{++} +  \frac{\delta S_{\text{aux}}}{\delta \mu_{--}}\delta \mu_{--}\right\rangle = 0\,.
\end{equation}

\noindent Plugging the explicit variation \eqref{variation sources} of the sources we obtain
\begin{align}\label{holo spin 2 Ward}
\partial_{-}T_{++} ={}&
 \mu_{--}\partial_+ T_{++} + 2T_{++}\partial_{+} \mu_{--}
\\
\partial_{+}T_{--} ={}&
 \mu_{++}\partial_- T_{--} + 2T_{--}\partial_{-} \mu_{++}
 \label{holo spin 2 Ward 2}
\end{align}

\noindent which are the familiar holomorphic Ward identities (in the absence of central extensions). In the context of holography, these Ward identities (including central extensions) and their supersymmetric extension were derived in \cite{Banados:2004nr} using the Chern-Simons formulation of three-dimensional anti-de Sitter gravity. 

One notices that the equation of motion for the auxiliary fields is
\begin{equation}
\Pi^i_{\pm} = \partial_{\pm}X^{i}-\mu_{\pm\pm}\Pi^{i}_{\mp}\,,
\end{equation}

\noindent which can be solved to give
\begin{equation}\label{solution for Pi}
\Pi^i_{\pm} = \frac{\partial_{\pm}X^{i}-\mu_{\pm\pm}\partial_{\mp}X^{i}}{1 -\mu_{--}\mu_{++}}\,.
\end{equation}

\noindent From \eqref{def stress tensor auxiliary fields} we see that the stress tensor obeying the holomorphic Ward identities is not merely $\sim  \left(\partial_{\pm}X\right)^{2}$, but rather
\begin{equation}\label{Aux T}
T_{\pm\pm} = \frac{1}{2}\left( \frac{\partial_{\pm}X^{i}-\mu_{\pm\pm}\partial_{\mp}X^{i}}{1 -\mu_{--}\mu_{++}}\right)^{2}\,.
\end{equation}

\noindent The fact that the naive free-field expressions for the currents are modified in a source-dependent way in the presence of non-chiral deformations is a general feature of the construction.

Another general feature we have emphasized in the body of the paper is that the process of integrating out the auxiliary fields results in a second order action which contains corrections to all orders in the sources. To illustrate this point we can replace \eqref{solution for Pi} back into the action, obtaining a flat space theory with Lagrangian
\begin{equation}\label{Lag}
\text{Lag} \equiv \frac{1}{(1-\mu_{--}\mu_{++})}\biggl[\left(1+\mu_{--}\mu_{++}\right)\partial_+X^i\partial_{-}X^i - \mu_{++}\partial_-X^i\partial_-X^i - \mu_{--}\partial_+ X^i\partial_+X^i \biggr].
\end{equation}

\noindent The spin-2 symmetries are of course still present: under the transformations (note the infinitesimal parameters $k_{\pm}$ below are different from the $p_{\pm}$ above)
\begin{eqnarray}\label{transf scalar}
\delta X^{i} &=& k^{+}\partial_+ X^{i} + k^{-}\partial_{-}X^{i}
\\
\delta \mu_{++} &=& \partial_{+}\left( k^{-} + \mu_{++}k^{+}\right) + \left( k^{-} + \mu_{++}k^{+}\right)\partial_{-}\mu_{++} - \mu_{++}\partial_{-}\left( k^{-} + \mu_{++}k^{+}\right)
\label{transf sources 1}
\\
\delta \mu_{--} &=& \partial_{-}\left(k^{+} + \mu_{--}k^{-}\right) + \left(k^{+} + \mu_{--}k^{-}\right)\partial_{+}\mu_{--} - \mu_{--}\partial_{+}\left(k^{+} + \mu_{--}k^{-}\right)
\label{transf sources 2}
\end{eqnarray}

\noindent the second order action changes as
\begin{equation}
\delta S = \int d^{2}x\Bigl[\partial_+\left(k^{+}\text{Lag} \right)+\partial_-\left(k^{-}\text{Lag} \right)\Bigr].
\end{equation}

The fact that the action is non-linear in the sources should come as no surprise if we recall that the gauging of spin-2 deformations is equivalent to putting the theory on a curved background metric. Indeed, the second order action involving the Lagrangian  \eqref{Lag} can be written covariantly as
\begin{equation}
S = \frac{1}{2}\int d^{2}x\,\sqrt{-g} g^{\mu\nu}\partial_\mu X^i \partial_{\nu}X^i
\end{equation}

\noindent with metric \cite{Hull:1993kf}\footnote{Notice that our parameterization of the metric differs slightly from that in \cite{Hull:1993kf}.}
\begin{equation}\label{metric}
g_{\mu\nu} = \frac{\Omega}{\left(1-\mu_{--}\mu_{++}\right)}\left(\begin{array}{cc}
2\mu_{++} & 1+\mu_{--}\mu_{++} \\ 
1+\mu_{--}\mu_{++} & 2\mu_{--}
\end{array} \right).
\end{equation}

\noindent As emphasized in \cite{Hull:1993kf}, \eqref{metric} does not correspond to partial gauge fixing: it is a general parameterization of a two-dimensional metric. In our conventions $\sqrt{-g} = \Omega$ is the conformal mode of the metric, which as usual drops from the action because of Weyl invariance. It is straightforward to verify that the transformations of the covariant fields induced by the transformation \eqref{transf scalar}-\eqref{transf sources 2} of the sources are simply
\begin{align}
\delta X^{i} ={}&
 \pounds_{k}X^{i}
\\
\delta g_{\mu\nu} ={}&
 \pounds_{k}g_{\mu\nu} - \left(\nabla_{\rho}k^{\rho}\right) g_{\mu\nu}\,.
\end{align}

\noindent In other words, the symmetry transformations \eqref{transf scalar}-\eqref{transf sources 2}  are a combination of diffeomorphism generated by $k^{\mu}$ plus a Weyl rescaling generated by $-\left(\nabla_{\rho}k^{\rho}\right) \,$.

Note that the components of the covariant stress tensor
\begin{equation}
\hat{T}_{\mu\nu} = \frac{1}{2}\left(\partial_{\mu}X^i\partial_{\nu}X^i  - \frac{g_{\mu\nu}}{2}\partial_{\alpha}X^{i}\partial^{\alpha}X^{i}\right)
\end{equation}

\noindent are given by
\begin{align}
\hat{T}_{++} ={}&
 T_{++} + \mu_{++}^{2}T_{--}
 \label{Thatpp}
\\
\hat{T}_{--} ={}&
 T_{--} + \mu_{--}^{2}T_{++}
  \label{Thatmm}
\\
\hat{T}_{+-} ={}&
 \mu_{++}T_{--} + \mu_{--}T_{++}\,.
\end{align}

\noindent This illustrates yet another subtle point: the definition of the stress tensor depends on what the sources are, namely what is kept fixed in the variation. While the covariant stress tensor couples to the metric $g_{\mu\nu}\,$, the currents $T_{\pm \pm}$ satisfying the usual Ward identities \eqref{holo spin 2 Ward}-\eqref{holo spin 2 Ward 2} couple instead to the sources $\mu_{\pm\pm}\,$.

\section{$\text{Tr}\bigl[a_{\bar{z}}^{2}\bigr]$ and the OPE}\label{app: azbar2}
In what follows we will exemplify various relations satisfied by flat connections in Drinfeld-Sokolov form. A 2$d$ Drinfeld-Sokolov connection consists of a component $a_{J}$ that contains a set of currents as highest weights, and a conjugate component $a_{\mu}$ whose lowest weights are linear in the corresponding sources. The various relations we discuss below rely exclusively on this lowest/highest weight structure, and therefore apply to any choice of boundary conditions. However, for the sake of concreteness we will exemplify them for holomorphic boundary conditions, where the currents sit in $a_{z}$ and the sources in $a_{\bar{z}}\,$ and $(z,\bar{z})$ denote complex coordinates. We will moreover work with the theory defined by the principal $sl(2)$ embedding into $sl(N)\,$, but the expressions adapt straightforwardly to other embeddings as well (see \cite{deBoer:2013gz} for example).

In the principal embedding, the $sl(N)$ generators organize into $N-1$ multiplets with $sl(2)$ spin $s-1$ ($s=2,\ldots,N$), spanned by generators $W^{(s)}_{j}$ with $j=-s+1,\ldots, s-1$. In particular, the $sl(2)$ generators $L_{j}$ ($j=-1,0,1$) correspond to the spin one multiplet $W^{(2)}_{j} = L_{j}\,$. The structure of the general Drinfeld-Sokolov connection is then
\begin{align}\label{DS az component}
a_{z} 
&= L_{1} + \frac{T(z,\bar{z})}{k} L_{-1} + \sum_{s = 3}^{N}\alpha_{s}J_{s}(z,\bar{z})W^{(s)}_{-s +1}
\\
a_{\bar{z}}
&=
\mu_{2}(z,\bar{z}) L_{1} + \sum_{s=3}^{N}\beta_{s}\mu_{s}(z,\bar{z})W^{(s)}_{s-1} + \text{(higher weights)}\,.
\label{DS azbar component}
\end{align}

\noindent Here $k \equiv c/6\,$ and $\alpha_{s}$ and $\beta_{s}$ are normalization constants which will be fixed as indicated below, and the higher weight terms in $a_{\bar{z}}$ are completely determined by solving the flatness conditions. The latter contain $N-1$ additional constraints which amount to the Ward identities obeyed by the currents $J_{s}$ in the presence of sources $\mu_{s}\,$. 

 In order to derive the symmetries associated to the above connection, one notices that the most general gauge transformation  $\delta a = d\Lambda  + \left[a,\Lambda\right]$ that preserves the form of $a_{z}$ contains $N-1$ independent infinitesimal parameters $\epsilon_{2},\ldots ,\epsilon_{N}$. Moreover, given that the flatness condition $F_{z\bar{z}}=0$ and the condition $\delta a_{z} =0$ are essentially the same equation (save for two components that yield the Ward identities in the former case and the transformation of the currents in the latter), it is not hard to see that the matrix parameter $\Lambda$ that generates such a gauge transformation is obtained from $a_{\bar{z}}$ by simply replacing $\mu_{s} \to \epsilon_{s}$ for $s=2,\ldots,N\,$:
\begin{equation}\label{gauge parameter from azbar}
\Lambda = \left. a_{\bar{z}}\right|_{\mu_{s} \to \epsilon_{s}}\,.
\end{equation}

\noindent Under this transformation the stress tensor and higher spin currents $J_{s}$ will transform, so that
\begin{equation}
\delta a_{z} = \frac{\delta T}{k}L_{-1} + \sum_{s = 3}^{N}\alpha_{s}\delta J_{s}(z,\bar{z})W^{(s)}_{-s +1}\,.
\end{equation}

\noindent Comparing these transformations with Noether's theorem
 \begin{equation}\label{Noether theorem}
\delta_{\lambda}\mathcal{O}(w) = \text{Res}_{z\to w}\Bigl[\lambda(z)J(z)\mathcal{O}(w)\Bigr]
\end{equation}

\noindent one reads off the semiclassical (large-$c$) OPEs of the $\mathcal{W}_{N}$ currents. Their normalization $\alpha_{s}$ can be then determined (up to a sign) by fixing the normalization of the OPEs to be
\begin{equation}\label{OPE normalization}
J_{s}(z)J_{s}(w) \sim \frac{c/s}{(z-w)^{2s}} + \ldots
\end{equation} 

\noindent Having determined the normalization of the currents in this way, the normalization $\beta_{s}$ of the sources is fixed by demanding 
\begin{align}\label{trace relation 1}
-k_{cs}\text{Tr}\bigl[\left(a_{z}-L_1\right)a_{\bar{z}}\bigr] 
&=
 \mu_{2}(z,\bar{z})T(z,\bar{z}) + \sum_{s=3}^{N}\mu_{s}(z,\bar{z})J_{s}(z,\bar{z})
\\
-k_{cs}\text{Tr}\big[L_1a_{\bar{z}}\bigr] 
&=
  \mu_{2}(z,\bar{z})T(z,\bar{z}) + \frac{c}{12}\partial^{2}\mu_{2}(z,\bar{z}) + \sum_{s=3}^{N}(s-1)\mu_{s}(z,\bar{z})J_{s}(z,\bar{z})
  \label{trace relation 2}
\end{align}

\noindent where $\partial \equiv\partial_{z}$ ($\bar{\partial} \equiv \partial_{\bar{z}}$), all traces are taken in the fundamental representation, and
\begin{equation}
k_{cs} =\frac{k}{2\text{Tr}\left[L_0L_0\right]} =  \frac{c}{12\text{Tr}\left[L_0L_0\right]}
\end{equation}

\noindent is the Chern-Simons level. The trace relations \eqref{trace relation 1}-\eqref{trace relation 2} follow from properties of the $sl(N)$ algebra and the flatness condition on the Drinfeld-Sokolov pair, and are valid for arbitrary spacetime-dependent sources as we now show. 

Without loss of generality, for the purpose of proving \eqref{trace relation 1}-\eqref{trace relation 2} we choose the normalization of the generators in the principal embedding such that
\begin{align}
\left[L_m,L_n\right] &= (m-n)L_{m+n}
\\
\left[L_m,W^{(s)}_{n}\right] 
&= \left(m(s-1) - n\right)W_{m+n}^{(s)}
\label{LW commutator}
\end{align}

\noindent and the Cartan-Killing form on $sl(N,\mathds{R})$ is then
\begin{align}\label{Cartan Killing}
\text{Tr}\left[W_{m}^{(s)}W_{n}^{(r)}\right] = t^{(s)}_{m}\delta^{r,s}\delta_{m,-n}
\end{align}

\noindent where the explicit form of the coefficients $t_{m}^{(s)}$ can be found in e.g. \cite{Castro:2011iw}. Since highest-weight generators have non-vanishing trace only against lowest-weight generators in the same multiplet, it is immediate from \eqref{DS az component}-\eqref{DS azbar component} that
\begin{align}
-k_{cs}\text{Tr}\bigl[\left(a_{z}-L_1\right)a_{\bar{z}}\bigr] 
={}&
-\frac{\text{Tr}\bigl[L_{-1} L_{1}\bigr] }{2\text{Tr}\left[L_0L_0\right]}\mu_{2} T(z,\bar{z})-k_{cs} \sum_{s = 3}^{N}\alpha_{s}\beta_{s}\mu_{s}(z,\bar{z})J_{s}(z,\bar{z})t^{(s)}_{s-1}\,.
\end{align}

\noindent Our normalization implies $\text{Tr}\left[L_{-1}L_{1}\right]  = -2 \text{Tr}\left[L_{0}L_{0}\right]$, so the last equation will be precisely \eqref{trace relation 1} provided we choose
\begin{equation}\label{normalization sources}
\beta_{s} = -\frac{1}{k_{cs}\alpha_{s}t^{(s)}_{s-1}} =  -\frac{2\text{Tr}\left[L_{0}L_{0}\right]}{k\alpha_{s}t^{(s)}_{s-1}}.
\end{equation}

\noindent Since we are always free to normalize the sources in this way, this proves \eqref{trace relation 1}. 

 In order to prove \eqref{trace relation 2} we will first obtain the useful intermediate results
\begin{align}\label{L0 Abarz trace}
-k_{cs}\text{Tr}\left[L_0a_{\bar{z}}\right] 
={}&
 \frac{k}{2}\partial \mu_{2}
 \\
 \label{temp result 1}
k_{cs}\text{Tr}\Bigl[\bigl[(a_{z} - L_{1}),L_0\bigr]a_{\bar{z}}\Bigr]
 ={}&
\mu_{2}(z,\bar{z})T(z,\bar{z})+\sum_{s = 3}^{N}(s-1)\mu_{s}(z,\bar{z})J_{s}(z,\bar{z}) .
\end{align}

\noindent To this end, consider the flatness condition $F_{z\bar{z}} = \partial a_{\bar{z}} - \bar{\partial}a_{z} + \left[a_{z},a_{\bar{z}}\right]=0$ and its trace against $L_{-1}\,$:
\begin{equation}\label{temp trace 1}
0 = \text{Tr}\left[L_{-1}F_{z\bar{z}}\right] = \partial \text{Tr}\left[L_{-1}a_{\bar{z}}\right] + \text{Tr}\bigl[\left[L_{-1},a_{z}\right]a_{\bar{z}}\bigr],
\end{equation}

\noindent where we used the cyclicity of the trace and $\bar{\partial} \text{Tr}\left[L_{-1}a_{z}\right] =0$, which follows from $\text{Tr}\left[L_{-1}a_{z}\right] = \text{Tr}\left[L_{-1}L_{1}\right] = \text{constant}$. Noticing  $\text{Tr}\left[L_{-1}a_{\bar{z}}\right] = \text{Tr}\left[L_{-1}L_{1}\right]\mu_{2}$ and also $\left[L_{-1},a_{z}\right] = \left[L_{-1},L_{1}\right] =- 2L_0$, which follows from \eqref{DS az component} and \eqref{LW commutator}, \eqref{temp trace 1} becomes
\begin{equation}
0 = \text{Tr}\left[L_{-1}L_{1}\right]  \partial \mu_{2} -2\text{Tr}\left[L_0a_{\bar{z}}\right]\quad \Rightarrow \quad \text{Tr}\left[L_0a_{\bar{z}}\right] = -  \text{Tr}\left[L_{0}L_{0}\right]  \partial \mu_{2}\,.
\end{equation}

\noindent  Multiplying this last equation by $-k_{cs}$ we obtain \eqref{L0 Abarz trace}. 

In order to derive \eqref{temp result 1}, let us define the matrix $Q = a_{z} - L_{1}\,$. It follows that
\begin{align}
\left[Q,L_0\right] ={}&
 \frac{T(z,\bar{z})}{k} \left[ L_{-1},L_{0}\right]+ \sum_{s = 3}^{N}\alpha_{s}J_{s}(z,\bar{z})\left[ W^{(s)}_{-s +1},L_0\right]
\nonumber \\
 ={}&
 -\frac{T(z,\bar{z})}{k} L_{-1} - \sum_{s = 3}^{N}(s-1)\alpha_{s}J_{s}(z,\bar{z}) W^{(s)}_{-s +1}
\end{align}

\noindent and therefore
\begin{align}\label{temp result 2}
k_{cs}\text{Tr}\Bigl[\left[Q,L_0\right]a_{\bar{z}}\Bigr]
 ={}&
 -\frac{\text{Tr}\left[L_{-1}L_{1}\right]}{2\text{Tr}\left[L_0L_0\right]}\mu_{2}(z,\bar{z})T(z,\bar{z}) -k_{cs}\sum_{s = 3}^{N}(s-1)\alpha_{s}\beta_{s}\mu_{s}(z,\bar{z})J_{s}(z,\bar{z}) t^{(s)}_{s-1}
  \nonumber\\
 ={}&
\mu_{2}(z,\bar{z})T(z,\bar{z})+\sum_{s = 3}^{N}(s-1)\mu_{s}(z,\bar{z})J_{s}(z,\bar{z}) 
\end{align}

\noindent where in the last equality we used $\text{Tr}\left[L_{-1}L_{1}\right]  = -2 \text{Tr}\left[L_{0}L_{0}\right]$ and the normalization \eqref{normalization sources}.  

With these results in hand we can now compute
\begin{align}\label{partial result 3}
-k_{cs}\text{Tr}\left[L_{1}a_{\bar{z}}\right] 
={}&
 -k_{cs}\text{Tr}\bigl[\left[L_{1},L_{0}\right]a_{\bar{z}}\bigr]
\nonumber\\
 ={}&
 -k_{cs}\text{Tr}\bigl[\left[a_{z} -Q,L_{0}\right]a_{\bar{z}}\bigr]
 \nonumber\\
 = {}&
  -k_{cs}\text{Tr}\bigl[\left[a_{\bar{z}},a_{z}\right]L_{0}-\left[Q,L_{0}\right]a_{\bar{z}}\bigr]
  \nonumber\\
 = {}&
  -k_{cs}\text{Tr}\bigl[\left(\partial a_{\bar{z}} - \bar{\partial}a_{z}\right)L_{0}-\left[Q,L_{0}\right]a_{\bar{z}}\bigr]
\nonumber\\
 = {}&
  -k_{cs}\partial\text{Tr}\bigl[ a_{\bar{z}}L_{0}\bigr]   +k_{cs}\text{Tr}\Bigl[\left[Q,L_{0}\right]a_{\bar{z}}\Bigr]
\end{align}

\noindent where as before we used the flatness condition and the cyclicity of the trace. Using \eqref{L0 Abarz trace} and \eqref{temp result 1}, equation \eqref{partial result 3} becomes precisely \eqref{trace relation 2}, completing the proof. As it should be clear from the above derivations, it is a straightforward matter to extend these general results to non-principal embeddings. 

Let us continue with our discussion of symmetries. The transformation $\delta \mu_{s}$ of the sources can be read off from the lowest weights of $\delta a_{\bar{z}}\,$ under the same allowed gauge transformation with parameter \eqref{gauge parameter from azbar} we employed above. We also note that, by construction, the Ward identities in the presence of sources are obtained from the variation of the currents by simply replacing the infinitesimal parameters $\epsilon_{s}$ by the sources $\mu_{s}\,$, i.e.
\begin{equation}\label{Ward identities from variations}
\overline{\partial}J_{s} = \left. \delta J_{s}\right|_{\epsilon_{s} \to \mu_{s}}\,.
\end{equation}

Define now the quantity
\begin{equation}
\Delta L_{N}(z,\bar{z})= \mu_{2}(z,\bar{z})T(z,\bar{z}) + \sum_{s=3}^{N}\mu_{s}(z,\bar{z})J_{s}(z,\bar{z})
\end{equation}

\noindent which is the deformation of the CFT Lagrangian in the chiral case. With the above normalization one finds the curious relation
\begin{equation}\label{Trazbar sq}
-k_{cs}\text{Tr}\left[a_{\bar{z}}^{2}\right] = \text{Res}_{z\to w}\Bigl[(z-w)\Delta L_{N}(z)\Delta L_{N}(w)\Bigr] + \partial^{2}\left(P_{N}\right),
\end{equation}

\noindent where $P_{N}$ will be determined below for $N=2,3,4\,$. In other words, up to a total second derivative, the quantity $-k_{cs}\text{Tr}\left[a_{\bar{z}}^{2}\right]$ is the coefficient of the second order pole in the $\Delta L_{N}(z)\Delta L_{N}(w)$ OPE.

\subsection{$N=2$}
We employ the usual two-dimensional representation of $sl(2,\mathds{R})$ in terms of matrices
\begin{equation}\label{2d representation of sl2r}
L_{0}=
\frac{1}{2}\left(\begin{array}{cc}
1 & 0  \\
0 & -1
\end{array}
\right),\qquad
L_{1}=
\left(\begin{array}{cc}
0 & 0  \\
1 & 0
\end{array}
\right),
\qquad
L_{-1}=
\left(\begin{array}{cc}
0 & -1  \\
0& 0
\end{array}
\right).
\end{equation}

\noindent The Drinfeld-Sokolov connection is
\begin{align}
a_{z} &=
 L_{1} + \frac{1}{k}T(z,\bar{z})L_{-1}
 \\
a_{\bar{z}} &=
 \mu_{2}(z,\bar{z}) L_{1} -\partial \mu_{2}L_0 +\left( \frac{1}{k}T\mu_{2}+ \frac{1}{2}\partial^{2}\mu_{2}\right) L_{-1} 
\end{align}

\noindent  and the flatness condition amounts to the Ward identity
\begin{equation}
\bar{\partial}T = \mu_{2}\partial T + 2T\partial \mu_{2} + \frac{k}{2}\partial^{3}\mu_{2}\,.
\end{equation}

The general infinitesimal gauge transformation that preserves the form of $a_{z}$ has parameter
\begin{equation}
\Lambda =
\left. a_{\bar{z}}\right|_{\mu_{2}\to \epsilon} =\epsilon(z,\bar{z})L_{1} -\partial \epsilon\, L_0 +\left(\frac{1}{k}T\epsilon + \frac{1}{2}\partial^{2}\epsilon\right)L_{-1}\,.
\end{equation}

\noindent Under such a transformation, the stress tensor changes as
\begin{equation}\label{delta T spin 2}
\delta T = \epsilon \partial T + 2T \partial \epsilon + \frac{k}{2}\partial^{3}\epsilon\,.
\end{equation}

\noindent Similarly, from $\delta a_{\bar{z}} = \delta \mu_{2} L_1 + (\text{higher weights})$ we read off the transformation of the source
\begin{equation}
\delta \mu_{2} = \bar{\partial}\epsilon - \mu_{2}\partial \epsilon + \epsilon \partial \mu_{2}\,.
\end{equation}

\noindent Comparing the variation \eqref{delta T spin 2} with Noether's theorem $\delta_{\epsilon}T(w) = \text{Res}_{z\to w}\left[\epsilon(z)T(z)T(w)\right]$ we obtain the stress tensor OPE. The standard normalization requires 
\begin{equation}\label{alpha spin 2}
k = \frac{c}{6} = 2\text{Tr}\left[L_0L_0\right]k_{cs}=k_{cs}
\end{equation}

\noindent and we find
\begin{equation}\label{TT OPE}
T(z)T(w) \sim \frac{c/2}{(z-w)^{4}} + \frac{2T(w)}{(z-w)^{2}} + \frac{\partial T(w)}{z-w}
\end{equation}

\noindent as expected. 

With the normalization \eqref{alpha spin 2} the Drinfeld-Sokolov flat connection satisfies
%
\begin{equation}
-k_{cs}\text{Tr}\left[a_{\bar{z}}^{2}\right] = \mu_{2}^{2}(2T) + \frac{c}{4}\mu_{2}\partial^{2}\mu_{2} - \partial^{2}\left(\frac{c}{24}\mu_{2}^2\right).
\end{equation}

\noindent  In the $N=2$ case we have $\Delta L_{2} = \mu_{2}T$ and
\begin{align}
\text{Res}_{z\to w}\Bigl[(z-w)\Delta L_{2}(z)\Delta L_{2}(w)\Bigr]
&= \mu_{2}^{2}(2T) + \frac{c}{4}\mu_{2}\partial^{2}\mu_{2} \,.
\end{align}

\noindent Therefore, $-k_{cs}\text{Tr}\left[a_{\bar{z}}^{2}\right]$ is indeed of the form \eqref{Trazbar sq} with $P_{2} = -\frac{c}{24}\mu_{2}^2\,$.

\subsection{$N=3$}
Our convention for the $sl(3,\mathds{R})$ generators in the principal embedding is

\begin{align*}
L_{0} &=
\left(\begin{array}{ccc}
1 & 0 & 0 \\
0 & 0 & 0 \\
0& 0 & -1
\end{array}
\right),&
 L_{1} &=
\left(\begin{array}{ccc}
0 & 0 & 0 \\
1 & 0 & 0 \\
0& 1 & 0
\end{array}
\right),&
L_{-1} &=
-2\left(\begin{array}{ccc}
0 & 1 & 0 \\
0& 0 & 1\\
0 & 0 & 0
\end{array}
\right),
\\
W_{2} &=
 2
\left(
\begin{array}{ccc}
0& 0 & 0 \\
0 & 0 & 0\\
1& 0 & 0
\end{array}
\right),&
W_{1} &=
\left(
\begin{array}{ccc}
0& 0 & 0 \\
1 & 0 & 0\\
0& -1 & 0
\end{array}
\right),&
W_{-2} &=
8
\left(
\begin{array}{ccc}
0& 0 & 1 \\
0 & 0 & 0\\
0& 0 & 0
\end{array}
\right),&
\nonumber\\
W_{-1} &=
2
\left(
\begin{array}{ccc}
0& -1 & 0 \\
0 & 0 & 1\\
0& 0 & 0
\end{array}
\right),&
W_{0} &= \frac{2}{3}
\left(
\begin{array}{ccc}
1& 0 & 0 \\
0 & -2 & 0\\
0& 0 & 1
\end{array}
\right).
\end{align*}

\noindent The Drinfeld-Sokolov connection is of the form
\begin{align}\label{az DS spin 3}
a_{z} &=
 L_{1} + \frac{1}{k}T(z,\bar{z})L_{-1}+ \frac{1}{k\, \beta}W(z,\bar{z})W_{-2}
 \\
a_{\bar{z}} &=
 \mu_{2}(z,\bar{z}) L_{1} -\frac{\beta}{4}\mu_{3}(z,\bar{z})W_{2} +\sum_{j=-1}^{0}f_{j}(z,\bar{z})L_{j} + \sum_{m=-2}^{1}g_{m}(z,\bar{z})W_{m}\,.
\label{azbar DS spin 3}
\end{align}

\noindent Solving the flatness condition yields
\begin{align}
f_{0} 
&=  -\partial \mu_{2}
\\
f_{-1}
&=
\frac{1}{k}T\mu_{2}+ \frac{2}{k}\mu_{3}W+\frac{1}{2}\partial^{2}\mu_{2}
\\
g_{1}
&=
\frac{\beta}{4}\partial\mu_{3}
\\
g_{0}
&=
-\frac{\beta}{2}\left(\frac{1}{k}\mu_{3}T+\frac{1}{4}\partial^{2}\mu_{3}\right)
\\
g_{-1}
&=
\frac{\beta}{12k}\left(2\mu_{3}\partial T + 5T\partial \mu_{3} + \frac{k}{2}\partial^{3}\mu_{3}\right)
\\
g_{-2}
&=
-\frac{\beta}{48k}\left(\frac{12}{k}\mu_{3}T^{2} - \frac{48}{\beta^{2}}\mu_{2}W + 7\partial T\partial \mu_{3} + 2\mu_{3}\partial^{2}T + 8T\partial^{2}\mu_{3} + \frac{k}{2}\partial^{4}\mu_{3}\right)
\end{align}

\noindent plus two additional  constraints that correspond to the Ward identities
\begin{align}
\overline{\partial}T &=
 \mu_{2}\partial T + 2T\partial \mu_{2} + 2\mu_{3}\partial W + 3W\partial \mu_{3} + \frac{k}{2}\partial^{3}\mu_{2}
\\
\overline{\partial}W &=
\mu_{2}\partial W + 3W\partial \mu_{2} - \frac{\beta^{2}}{24}\mu_{3}\left(\partial^{3}T+\frac{16}{k}T\partial T\right) - \frac{\beta^{2}}{48}\partial \mu_{3}\left(9\partial^{2}T+\frac{32}{k}T^{2} \right)
\nonumber\\
&\hphantom{=}
 - \frac{5\beta^{2}}{16}\partial^{2}\mu_{3}\partial T - \frac{5\beta^{2}}{24}T\partial^{3}\mu_{3} - \frac{\beta^{2}}{96}k \partial^{5}\mu_{3}\,.
\end{align}

In agreement with \eqref{gauge parameter from azbar}, the generator $\Lambda$ of a general infinitesimal gauge transformation that preserves the form of $a_{z}$ is obtained by replacing  $\mu_{2} \to \epsilon$ and $\mu_{3} \to \chi$ in \eqref{azbar DS spin 3}: 
\begin{equation}
\Lambda =\left. a_{\bar{z}}\right|_{\mu_{2}\to \epsilon ,\, \mu_{3} \to \chi}\,.
\end{equation}

\noindent Under such gauge transformations, the currents transform as
\begin{align}\label{delta T spin 3}
\delta T &=
 \epsilon \partial T + 2T \partial \epsilon + \frac{k}{2}\partial^{3}\epsilon + 2\chi \partial W + 3W\partial \chi\,.
 \\
 \delta W &=
  \epsilon \partial W + 3W \partial \epsilon -\frac{\beta^{2}}{24}\chi \left( \partial^{3}T+\frac{16}{k}T\partial T\right) - \frac{\beta^{2}}{48}\partial \chi \left(9 \partial^{2}T+\frac{32}{k}T^{2}\right) 
  \nonumber\\
  &\hphantom{=}
  - \frac{5\beta^{2}}{16}\partial^{2}\chi \partial T-\frac{5\beta^{2}}{24}T\partial^{3}\chi  - \frac{\beta^{2}}{96}k \partial^{5}\chi\,.
  \label{delta W spin 3}
\end{align}

\noindent Similarly, from
\begin{equation}
\delta a_{\bar{z}} = \delta \mu_{2}L_1 -\frac{\beta}{4}\delta \mu_{3}W_{2} + \text{higher weights}
\end{equation}

\noindent we find the transformation of the sources
\begin{align}
\delta \mu_{2} 
&=
 \bar{\partial}\epsilon + \epsilon \partial \mu_{2}- \mu_{2}\partial \epsilon  -\frac{\beta^{2}}{24k}\chi\left(k \partial^{3}\mu_{3} + 16 T\partial \mu_{3}\right)
 \nonumber\\
 &\hphantom{=}
 +\frac{\beta^{2}}{48k}\partial \chi \left(32T\mu_{3} + 3k\partial^{2}\mu_{3}\right)-\frac{\beta^{2}}{16}\partial^{2}\chi \partial \mu_{3} + \frac{\beta^{2}}{24}\partial^{3}\chi \, \mu_{3}
\\
\delta \mu_{3}
 &=
 \bar{\partial}\chi + 2\chi\partial \mu_{2} - \mu_{2}\partial\chi +\epsilon \partial \mu_{3} - 2\mu_{3}\partial \epsilon\,.
\end{align}

\noindent Comparing the variations \eqref{delta T spin 3}-\eqref{delta W spin 3} with Noether's theorem \eqref{Noether theorem} we can read off the large-$c$ $\mathcal{W}_{3}$ OPEs. The standard normalization \eqref{OPE normalization} requires
\begin{equation}\label{normalization spin 3}
k = \frac{c}{6}=2k_{cs}\text{Tr}\left[L_0L_0\right]= 4k_{cs}\,,\qquad \beta^{2} = -\frac{8}{5}
\end{equation}

\noindent and we obtain
\begin{align}
\label{TW OPE}
T(z)W(w)
&\sim
 \frac{3W(w)}{(z-w)^{2}} + \frac{\partial W(w)}{z-w}
\\
W(z)W(w)
&\sim
\frac{c/3}{(z-w)^{6}} + \frac{2T(w)}{(z-w)^{4}} + \frac{\partial T(w)}{(z-w)^{3}} + \frac{1}{10}\frac{3\partial^{2}T(w)+\frac{64}{c}T^{2}(w)}{(z-w)^{2}}
\nonumber\\
&\phantom{\sim}
+\frac{1}{15}\frac{\partial^{3}T(w) + \frac{96}{c}T(w)\partial T(w)}{z-w}
\end{align}

\noindent with $TT$ as in \eqref{TT OPE}. 

With the normalization \eqref{normalization spin 3} we find that the flat connection in Drinfeld-Sokolov form satisfies
\begin{align}
-k_{cs}\text{Tr}\left[a_{\bar{z}}^{2}\right] 
={}&
 2\mu_{2}^{2}T  + 6\mu_{2}\mu_{3}W + \frac{1}{10}\mu_{3}^{2}\left(3\partial^{2}T + \frac{64}{c}T^{2}\right) 
\nonumber\\
& 
+\mu_{2}\,\partial \left(\frac{c}{4}\partial \mu_{2}\right)  + \mu_{3}\,\partial\left(T\partial \mu_{3} + \frac{c}{72}\partial^{3}\mu_{3}\right)
\nonumber\\
&
 -\partial^{2}\left(\frac{1}{6}\mu_{3}^{2} T+\frac{c}{24}\mu_{2}^{2} -\frac{c}{180}\left(\partial \mu_{3}\right)^{2} + \frac{c}{120}\mu_{3}\partial^{2}\mu_{3}\right).
\end{align}

\noindent On the other hand, in the $N=3$ case we have $\Delta L_{3} = \mu_{2}T + \mu_{3}W$ and
\begin{align}
\text{Res}_{z\to w}\Bigl[(z-w)\Delta L_{3}(z)\Delta L_{3}(w)\Bigr]
={}&
 2\mu_{2}^{2}T  + 6\mu_{2}\mu_{3}W + \frac{1}{10}\mu_{3}^{2}\left(3\partial^{2}T + \frac{64}{c}T^{2}\right) 
\nonumber\\
& 
+\mu_{2}\,\partial \left(\frac{c}{4}\partial \mu_{2}\right)  + \mu_{3}\,\partial\left(T\partial \mu_{3} + \frac{c}{72}\partial^{3}\mu_{3}\right). 
\end{align}

\noindent  Hence, $\text{Tr}\left[a_{\bar{z}}^{2}\right]$ verifies \eqref{Trazbar sq} with 
\begin{equation}
P_{3} = -\frac{1}{6}\mu_{3}^{2} T -\frac{c}{24}\mu_{2}^{2}  +\frac{c}{180}\left(\partial \mu_{3}\right)^{2} -\frac{c}{120}\mu_{3}\partial^{2}\mu_{3}\,.
\end{equation}

\subsection{$N=4$}
We employ the matrix realization of the $sl(4,\mathds{R})$ generators given in \cite{Tan:2011tj}. The Drinfeld-Sokolov connection is of the form
\begin{align}\label{az DS spin 4}
a_{z} &=
 L_{1} + \frac{1}{k}T(z,\bar{z})L_{-1}+ \frac{1}{k\, \beta}W(z,\bar{z})W_{-2} + \frac{1}{k\,\gamma}U(z,\bar{z})U_{-3}
 \\
a_{\bar{z}} &=
 \mu_{2}(z,\bar{z}) L_{1} -\frac{5\beta}{12}\mu_{3}(z,\bar{z})W_{2} +\frac{5\gamma}{18}\mu_{4}(z,\bar{z})U_{3}
 \\
 &\hphantom{=}
  +\sum_{j=-1}^{0}f_{j}(z,\bar{z})L_{j} + \sum_{m=-2}^{1}g_{m}(z,\bar{z})W_{m} + \sum_{n=-3}^{2}h_{n}(z,\bar{z})U_{n}
\label{azbar DS spin 4}
\end{align}

\noindent where the constants $k$, $\beta$ and $\gamma$ will be fixed by demanding the OPEs to have the standard normalization. The flatness conditions yields
\begin{align}
f_{0} 
&=  -\partial \mu_{2}
\\
f_{-1}
&=
\frac{1}{k}\left(T\mu_{2}+ 2\mu_{3}W + 3\mu_{4}U\right) +\frac{1}{2}\partial^{2}\mu_{2}
\\
g_{1}
&=
\frac{5\beta}{12}\partial\mu_{3}
\\
g_{0}
&=
-\frac{5\beta}{3k}\left(\frac{1}{2}\mu_{3}T + \frac{\gamma}{\beta^{2}}\mu_{4}W+\frac{k}{8}\partial^{2}\mu_{3}\right)
\\
g_{-1}
&=
\frac{5\beta}{36k}\left(2\mu_{3}\partial T + 5T\partial \mu_{3}  +\frac{4\gamma}{\beta^{2}}\left(\mu_{4}\partial W + 2W\partial \mu_{4}\right)+ \frac{k}{2}\partial^{3}\mu_{3}\right)
\\
g_{-2}
&=
\frac{1}{\beta k}\mu_{2}W-\frac{5\beta}{36k}\biggl[\mu_{3}\left(\frac{1}{2}\partial^{2}T + \frac{3}{k}T^{2} - \frac{9U}{\gamma}\right) + \frac{7}{4}\partial \mu_{3}\partial T + 2T\partial^{2}\mu_{3} + \frac{k}{8}\partial^{4}\mu_{3}
\nonumber\\
&\hphantom{=\frac{1}{\beta k}\mu_{2}W-\frac{5\beta}{36k}\biggl[}
 +\frac{\gamma}{\beta^{2}}\mu_{4}\left(\partial^{2}W + \frac{48 }{5k}TW\right) + \frac{3\gamma}{\beta^{2}}\partial \mu_{4}\partial W + \frac{13\gamma}{5\beta^{2}}W\partial^{2}\mu_{4}
  \biggr]
\end{align}
\begin{align}
  h_{2} 
  &=
  -\frac{5\gamma}{18}\partial\mu_{4}
  \\
  h_{1}
  &=
  \frac{5\gamma}{36}\left(\partial^{2}\mu_{4} + \frac{6}{k}\mu_{4}T\right)
  \\
  h_{0}
  &=
  -\frac{10\gamma}{27k}\left(\frac{3}{4}\mu_{4}\partial T + 2T\partial \mu_{4} + \frac{k}{8}\partial^{3}\mu_{4}\right)
  \\
  h_{-1} 
  &=
 -\frac{5}{6k}\mu_{3}W+  \frac{5\gamma}{108 k}\left[ 3\mu_{4}\left(\frac{1}{2}\partial^{2}T + \frac{6}{k}T^{2} + \frac{6}{\gamma}U\right) + \frac{11}{2}\partial T \partial\mu_{4} + 7T\partial^{2}\mu_{4} +\frac{k}{4}\partial^{4}\mu_{4}\right]
   \\
  h_{-2}
  &=
 \frac{1}{6k}\left(\mu_{3}\partial W + 4W\partial \mu_{3}\right) -\frac{\gamma}{18k}\biggl[ \mu_{4}\left(\frac{1}{4}\partial^{3}T + \frac{9}{k}T\partial T + \frac{3}{\gamma}\partial U\right) 
  \nonumber\\
  &\hphantom{=-\frac{\gamma}{18k}\biggl[}
  + \partial \mu_{4}\left(\frac{7}{6}\partial^{2}T + \frac{11}{k}T^{2} + \frac{9}{\gamma}U\right)
  +\frac{25}{12}\partial^{2}\mu_{4}\partial T + \frac{5}{3}T\partial^{3}\mu_{4} + \frac{k}{24}\partial^{5}\mu_{4}\biggr]
\end{align}
\begin{align}
h_{-3}
={}&
\frac{1}{36k}\biggl[
\frac{36}{\gamma} \mu_{2} U -\mu_{3} \left(\partial^{2}W+\frac{30}{k}T W\right)-5\partial W \partial \mu_{3}-9 W \partial^{2}\mu_{3} 
\nonumber\\
&
+\mu_{4} \left(\frac{23 \gamma}{6 k} T\partial^{2}T +\frac{3 \gamma}{k} \left(\partial T\right)^2  +\frac{\gamma}{12} \partial^{4}T+\frac{22}{k }T U+\frac{10 \gamma }{k^2}T^3 +\partial^{2}U-\frac{40 \gamma }{k  \beta ^2}W^2\right)
\nonumber\\
&
+\partial \mu_{4} \left(\frac{241 \gamma }{18 k }T \partial T+\frac{17\gamma}{36}   \partial^{3}T+4 \partial U\right)
+ \partial^{2}\mu_{4}\left(\frac{13\gamma}{12}\partial^{2}T + \frac{68\gamma}{9k}T^{2} + 5U\right)
\nonumber\\
&
+\frac{5\gamma}{4}\left( \partial T \partial^{3}\mu_{4}+\frac{5}{9}  T \partial^{4}\mu_{4}
+\frac{k}{90}\partial^{6}\mu_{4}
\right)
\biggr]
\end{align}

\noindent plus the Ward identities
\begin{align}
\overline{\partial}T
 ={}&
 \mu_{2}\partial T + 2T\partial \mu_{2} + 2\mu_{3}\partial W + 3W\partial \mu_{3} + 3\mu_{4}\partial U + 4U\partial \mu_{4} + \frac{k}{2}\partial^{3}\mu_{2}
\label{Ward N4 T} 
 \\
 \overline{\partial}W 
 ={}&
\mu_{2} \partial W+3 W \partial \mu_{2}-5\beta^2 \mu_{3} \left(\frac{2}{9 k}T \partial T+\frac{1}{72} \partial^{3} T - \frac{1}{4\gamma}\partial U\right)
 \nonumber\\
 &
 -5\beta^{2} \partial \mu_{3} \left(\frac{1}{16} \partial^{2}T+\frac{2}{9 k} T^2-\frac{1}{2\gamma}U\right) -\frac{25 \beta^{2} }{48}\partial T \partial^{2}\mu_{3}-\frac{25\beta^{2}}{72}  T\partial^{3}\mu_{3} -\frac{5k  \beta^{2}}{288}  \partial^{5}\mu_{3}
 \nonumber\\
 &
 -\frac{\gamma}{k} \mu_{4} \left(\frac{3}{2}W \partial T +\frac{17}{9} T  \partial W + \frac{5k}{36} \partial^{3}W\right)-\frac{\gamma}{9}  \partial \mu_{4} \left(\frac{26}{k} T W+5 \partial^{2}W\right)
 \nonumber\\
 &
  -\frac{7 \gamma}{9}  \partial W \partial^{2}\mu_{4}-\frac{7\gamma }{18}  W \partial^{3}\mu_{4}
\end{align}
\begin{align}
 \overline{\partial}U
  ={}&  
  \mu_{2}\partial U+4U\partial \mu_{2}
   -\frac{\gamma}{18k}\mu_{3} \left(25W\partial T +18 T \partial W + \frac{k}{2}\partial^{3}W\right)
  \nonumber\\
  &
   -\frac{\gamma}{6}\partial \mu_{3} \left( \partial^{2}W+\frac{52 }{3 k}T W\right)  
  -\frac{7\gamma}{18}\partial W \partial^{2}\mu_{3}   -\frac{7\gamma}{18}W \partial^{3}\mu_{3}
\nonumber\\  
&
+  \frac{\gamma^{2}}{3k} \mu_{4}\left[\partial T\left(\frac{59}{72} \partial^{2} T+\frac{4}{k}T^2+\frac{7}{3 \gamma}U\right)+\frac{13}{36}T \partial^{3} T+\frac{k}{144} \partial^{5}T+\frac{7}{3 \gamma}T \partial U+\frac{k}{12 \gamma} \partial^{3}U-\frac{10}{\beta^{2}}W \partial W\right]
  \nonumber\\
  &
  +\frac{\gamma^{2}}{9k}\partial \mu_{4}\left(\frac{44}{9}T\partial^{2} T+\frac{295}{72} (\partial T)^2+\frac{5k}{36} \partial^{4}T+\frac{14}{\gamma}T U+\frac{8}{ k}T^3+\frac{5k}{4 \gamma} \partial^{2}U-\frac{30}{\beta^{2}}W^2\right)
  \nonumber\\
  &
  +\gamma^{2}\partial^{2}\mu_{4} \left(\frac{49}{54 k } T \partial T+\frac{7}{162} \partial^{3} T+\frac{1}{4 \gamma }\partial U\right)+\frac{\gamma^{2}}{6}\partial^{3}\mu_{4} \left(\frac{7}{18} \partial^{2} T+\frac{49}{27 k}T^2+\frac{1}{\gamma }U\right)
  \nonumber\\
  &
 +\frac{\gamma^{2}}{648}\left(35 \partial T\partial^{4}\mu_{4} +14 T \partial^{5}\mu_{4}+\frac{k}{4}  \partial^{7}\mu_{4}\right)
 \label{Ward N4 U} 
\end{align}

In agreement with \eqref{gauge parameter from azbar}, the generator $\Lambda$ of the most general infinitesimal gauge transformation that preserves the form of $a_{z}$ is obtained as
\begin{equation}
\Lambda =\left. a_{\bar{z}}\right|_{\mu_{2}\to \epsilon ,\, \mu_{3} \to \chi,\, \mu_{4} \to \xi}\,.
\end{equation}

\noindent Under such gauge transformations, the transformation of the currents is obtained from the right hand side of the Ward identities \eqref{Ward N4 T}-\eqref{Ward N4 U} by replacing $\mu_{2}\to \epsilon\,$, $\mu_{3}\to \chi$ and $\mu_{4} \to \xi\,$ (c.f. \eqref{Ward identities from variations}). Comparing these variations with Noether's theorem \eqref{Noether theorem} we read off the large-$c$ $\mathcal{W}_{4}$ OPEs. The standard normalization \eqref{OPE normalization} requires
\begin{equation}\label{normalization spin 4}
k = \frac{c}{6}=2k_{cs}\text{Tr}\left[L_0L_0\right]= 10k_{cs}\,,\qquad \beta^{2} = -\frac{24}{25}\,,\qquad \gamma^{2} = \frac{27}{35}\,.
\end{equation}

\noindent In addition to \eqref{TT OPE} and \eqref{TW OPE} we then obtain the following OPEs
\begin{align}
T(z)U(w)
&\sim
 \frac{4U(w)}{(z-w)^{2}} + \frac{\partial U(w)}{z-w}
\\
W(z)W(w)
&\sim
\frac{c/3}{(z-w)^{6}} + \frac{2T(w)}{(z-w)^{4}} + \frac{\partial T(w)}{(z-w)^{3}} + \frac{1}{10}\frac{3\partial^{2}T(w)+\frac{64}{c}T^{2}(w) - \frac{24}{\gamma}U(w)}{(z-w)^{2}}
\nonumber\\
&\phantom{\sim}
+\frac{1}{15}\frac{\partial^{3}T(w) + \frac{96}{c}T(w)\partial T(w) - \frac{18}{\gamma}\partial U(w)}{z-w}
\\
\frac{W(z)U(w)}{-\gamma}
 &\sim
  \frac{7}{3}\frac{W(w)}{(z-w)^{4}} +  \frac{7}{9}\frac{\partial W(w)}{(z-w)^{3}} +\frac{1}{6}\frac{ \partial^{2}W(w)+\frac{104}{c}T(w)W(w)}{(z-w)^{2}}
\nonumber\\
&\phantom{\sim}
+\frac{1}{36}\frac{\partial^{3}W(w) + \frac{300}{c}W(w)\partial T(w) + \frac{216}{c}T(w)\partial W(w)}{z-w}
\end{align}

\begin{align}
U(z)U(w)
 \sim{}&
 \frac{c/4}{(z-w)^{8}} + \frac{2T(w)}{(z-w)^{6}} + \frac{\partial T(w)}{(z-w)^{5}} +\frac{3}{10} \frac{\partial^{2}T(w) + \frac{28}{c}T(w)^{2} + \frac{18}{7\gamma}U(w)}{(z-w)^{4}}
 \nonumber\\
 &
  + \frac{1}{5}\frac{\frac{1}{3}\partial^{3}T(w) + \frac{42}{c}T(w)\partial T(w) + \frac{27}{14\gamma}\partial U(w)}{(z-w)^{3}}
  +\frac{1}{84} \frac{\partial^{4}T(w) + \frac{9}{\gamma}\partial^{2}U(w)}{(z-w)^{2}}
\\
& 
+ \frac{1}{c}\frac{\frac{225}{14}W(w)^{2} + \frac{36}{5\gamma}T(w)U(w)+\frac{59}{28}\left(\partial T(w)\right)^{2}+ \frac{88}{35}T(w)\partial^{2}T(w)  +\frac{864}{35c}T(w)^{3}}{(z-w)^{2}}
\nonumber\\
&
+\frac{1}{c}\frac{\frac{3c}{140\gamma}\partial^{3}U(w)+\frac{18}{5\gamma}\left(U(w)\partial T(w)+ T(w)\partial U(w)\right) + \frac{225}{14}W(w)\partial W(w) }{z-w}
\nonumber
\nonumber\\
&
+\frac{1}{7c}\frac{\frac{c}{80}\partial^{5}T(w)+ \frac{177}{20}\partial T(w)\partial^{2}T(w) + \frac{39}{10}T(w)\partial^{3}T(w)+ \frac{1296}{5c}T(w)^{2}\partial T(w)}{z-w}
\nonumber
\end{align}

\noindent We have left explicit factors of $\gamma$ in the OPEs in order to have the freedom to choose the overall sign in the normalization of $U$ (c.f. \eqref{normalization spin 4}).

With the normalization \eqref{normalization spin 3} the flat connection in Drinfeld-Sokolov form satisfies
\begin{align}\label{trazbarsq spin 4}
-k_{cs}\text{Tr}\left[a_{\bar{z}}^{2}\right] 
={}&
2  \mu_{2}^2 T +6 \mu_{2} \mu_{3}W+8 \mu_{2} \mu_{4} U
\nonumber\\
&
 +\mu_{3}^2 \left(\frac{32}{5 c}T^{2}+\frac{3}{10}\partial^{2}T-\frac{12}{5 \gamma}U\right)
-\frac{13 \gamma }{18}\mu_{3} \mu_{4} \left(\partial^{2}W+\frac{48}{c} T W\right)
\nonumber\\
&
+\mu_{4}^2 \left(\frac{864}{35 c^2}T^{3}+\frac{88}{35 c}T\partial^{2}T+\frac{59}{28 c}\left(\partial T\right)^2+\frac{36}{5 c \gamma}TU +\frac{225}{14 c} W^{2}+\frac{1}{84} \partial^{4}T+\frac{3}{28 \gamma} \partial^{2}U\right)
\nonumber\\
&
+ \mu_{2} \left(\frac{c}{4}\partial^{2}\mu_{2}\right)
+\mu_{3} \left(\frac{c}{72}\partial^{4}\mu_{3}+ \partial T  \partial \mu_{3}+T \partial^{2}\mu_{3}-\frac{14 \gamma}{9}  \partial W  \partial \mu_{4}-\frac{7 \gamma}{6} W \partial^{2}\mu_{4}\right)
\nonumber\\
&
+\mu_{4} \left(\frac{1}{12}T \partial^{4}\mu_{4}+\frac{1}{15} \partial^{3}T  \partial\mu_{4} +\frac{3}{20} \partial^{2}T\partial^{2} \mu_{4}+\frac{42}{5 c}T\partial T  \partial\mu_{4}+\frac{21}{5 c}T^2\partial^{2} \mu_{4}
\right.
\nonumber\\
&\left.
+\frac{1}{6}  \partial T \partial^{3}\mu_{4}+\frac{27}{70 \gamma }\partial \left(U  \partial \mu_{4}\right)
-\frac{7\gamma}{9} \partial W  \partial \mu_{3} - \frac{7 \gamma}{6}W \partial^{2}\mu_{3} +\frac{c }{2880}\partial^{6}\mu_{4}\right)
\\
&
 +\partial^{2}\left(P_{4}\right)
\nonumber
\end{align}

\noindent where the quantity $P_{4}$ in the last line is defined as
\begin{align}\label{P4}
P_{4}
 ={}&
-\frac{c}{24}\mu_{2}^2-\frac{1}{6} T\mu_{3}^2-\frac{1}{120} \mu_{4}^2 \left(\partial^{2}T+\frac{84}{c}T^{2} +10\gamma   U\right)
+\frac{7\gamma}{18}  W\mu_{3}  \mu_{4}
\nonumber\\
&
-\mu_{4} \left(\frac{c}{4032}\partial^{4}\mu_{4}+\frac{1}{60} \partial T \partial \mu_{4}+\frac{1}{20} T \partial^{2}\mu_{4}\right)+\frac{1}{30} T \left(\partial \mu_{4}\right)^2
\nonumber\\
&
-\frac{c}{120}  \mu_{3} \partial^{2}\mu_{3}+\frac{c}{180} \left(\partial \mu_{3}\right)^2+\frac{c}{2520} \partial \mu_{4}\partial^{3} \mu_{4}-\frac{c }{4480}\left(\partial^{2}\mu_{4}\right)^2\,.
\end{align}

\noindent Denoting $\Delta L_{4} = \mu_{2}T + \mu_{3}W + \mu_{4} U$ one finds that $\text{Res}_{z\to w}\Bigl[(z-w)\Delta L_{4}(z)\Delta L_{4}(w)\Bigr]$ is given precisely by the first six lines of \eqref{trazbarsq spin 4}. Therefore, $\text{Tr}\left[a_{\bar{z}}^{2}\right]$ verifies \eqref{Trazbar sq} with $P_{4}$ given by \eqref{P4}.



\begin{thebibliography}{10}

\bibitem{Klebanov:2002ja}
I.~Klebanov and A.~Polyakov, ``{AdS dual of the critical O(N) vector model},''
  \href{http://dx.doi.org/10.1016/S0370-2693(02)02980-5}{{\em Phys.Lett.} {\bf
  B550} (2002)  213--219},
\href{http://arxiv.org/abs/hep-th/0210114}{{\tt arXiv:hep-th/0210114
  [hep-th]}}.

\bibitem{Fradkin:1986qy}
E.~Fradkin and M.~A. Vasiliev, ``{Cubic Interaction in Extended Theories of
  Massless Higher Spin Fields},''
\href{http://dx.doi.org/10.1016/0550-3213(87)90469-X}{{\em Nucl.Phys.} {\bf
  B291} (1987)  141}.

\bibitem{Fradkin:1987ks}
E.~Fradkin and M.~A. Vasiliev, ``{On the Gravitational Interaction of Massless
  Higher Spin Fields},''
\href{http://dx.doi.org/10.1016/0370-2693(87)91275-5}{{\em Phys.Lett.} {\bf
  B189} (1987)  89--95}.

\bibitem{Vasiliev:1995dn}
M.~A. Vasiliev, ``{Higher spin gauge theories in four-dimensions,
  three-dimensions, and two-dimensions},''
  \href{http://dx.doi.org/10.1142/S0218271896000473}{{\em Int.J.Mod.Phys.} {\bf
  D5} (1996)  763--797},
\href{http://arxiv.org/abs/hep-th/9611024}{{\tt arXiv:hep-th/9611024
  [hep-th]}}.

\bibitem{Giombi:2012ms}
S.~Giombi and X.~Yin, ``{The Higher Spin/Vector Model Duality},''
  \href{http://dx.doi.org/10.1088/1751-8113/46/21/214003}{{\em J.Phys.} {\bf
  A46} (2013)  214003},
\href{http://arxiv.org/abs/1208.4036}{{\tt arXiv:1208.4036 [hep-th]}}.

\bibitem{Gaberdiel:2010pz}
M.~R. Gaberdiel and R.~Gopakumar, ``{An AdS$_3$ Dual for Minimal Model CFTs},''
  \href{http://dx.doi.org/10.1103/PhysRevD.83.066007}{{\em Phys.Rev.} {\bf D83}
  (2011)  066007},
\href{http://arxiv.org/abs/1011.2986}{{\tt arXiv:1011.2986 [hep-th]}}.

\bibitem{Gaberdiel:2012uj}
M.~R. Gaberdiel and R.~Gopakumar, ``{Minimal Model Holography},''
  \href{http://dx.doi.org/10.1088/1751-8113/46/21/214002}{{\em J.Phys.} {\bf
  A46} (2013)  214002},
\href{http://arxiv.org/abs/1207.6697}{{\tt arXiv:1207.6697 [hep-th]}}.

\bibitem{Prokushkin:1998bq}
S.~Prokushkin and M.~A. Vasiliev, ``{Higher spin gauge interactions for massive
  matter fields in 3-D AdS space-time},''
  \href{http://dx.doi.org/10.1016/S0550-3213(98)00839-6}{{\em Nucl.Phys.} {\bf
  B545} (1999)  385},
\href{http://arxiv.org/abs/hep-th/9806236}{{\tt arXiv:hep-th/9806236
  [hep-th]}}.

\bibitem{Prokushkin:1998vn}
S.~Prokushkin and M.~A. Vasiliev, ``{3-d higher spin gauge theories with
  matter},''
\href{http://arxiv.org/abs/hep-th/9812242}{{\tt arXiv:hep-th/9812242
  [hep-th]}}.

\bibitem{Bouwknegt:1992wg}
P.~Bouwknegt and K.~Schoutens, ``{W symmetry in conformal field theory},''
  \href{http://dx.doi.org/10.1016/0370-1573(93)90111-P}{{\em Phys.Rept.} {\bf
  223} (1993)  183--276},
\href{http://arxiv.org/abs/hep-th/9210010}{{\tt arXiv:hep-th/9210010
  [hep-th]}}.

\bibitem{Castro:2011iw}
A.~Castro, R.~Gopakumar, M.~Gutperle, and J.~Raeymaekers, ``{Conical Defects in
  Higher Spin Theories},''
  \href{http://dx.doi.org/10.1007/JHEP02(2012)096}{{\em JHEP} {\bf 1202} (2012)
   096},
\href{http://arxiv.org/abs/1111.3381}{{\tt arXiv:1111.3381 [hep-th]}}.

\bibitem{Perlmutter:2012ds}
E.~Perlmutter, T.~Prochazka, and J.~Raeymaekers, ``{The semiclassical limit of
  $W_N$ CFTs and Vasiliev theory},''
  \href{http://dx.doi.org/10.1007/JHEP05(2013)007}{{\em JHEP} {\bf 1305} (2013)
   007},
\href{http://arxiv.org/abs/1210.8452}{{\tt arXiv:1210.8452 [hep-th]}}.

\bibitem{Campoleoni:2013lma}
A.~Campoleoni, T.~Prochazka, and J.~Raeymaekers, ``{A note on conical solutions
  in 3D Vasiliev theory},''
  \href{http://dx.doi.org/10.1007/JHEP05(2013)052}{{\em JHEP} {\bf 1305} (2013)
   052},
\href{http://arxiv.org/abs/1303.0880}{{\tt arXiv:1303.0880 [hep-th]}}.

\bibitem{Campoleoni:2013iha}
A.~Campoleoni and S.~Fredenhagen, ``{On the higher-spin charges of conical
  defects},'' \href{http://dx.doi.org/10.1016/j.physletb.2013.08.012}{{\em
  Phys.Lett.} {\bf B726} (2013)  387--389},
\href{http://arxiv.org/abs/1307.3745}{{\tt arXiv:1307.3745}}.

\bibitem{Gaberdiel:2013cca}
M.~R. Gaberdiel, R.~Gopakumar, and M.~Rangamani, ``{The Spectrum of Light
  States in Large N Minimal Models},''
\href{http://arxiv.org/abs/1310.1744}{{\tt arXiv:1310.1744 [hep-th]}}.

\bibitem{Gaberdiel:2011zw}
M.~R. Gaberdiel, R.~Gopakumar, T.~Hartman, and S.~Raju, ``{Partition Functions
  of Holographic Minimal Models},''
  \href{http://dx.doi.org/10.1007/JHEP08(2011)077}{{\em JHEP} {\bf 1108} (2011)
   077},
\href{http://arxiv.org/abs/1106.1897}{{\tt arXiv:1106.1897 [hep-th]}}.

\bibitem{Kraus:2011ds}
P.~Kraus and E.~Perlmutter, ``{Partition functions of higher spin black holes
  and their CFT duals},'' \href{http://dx.doi.org/10.1007/JHEP11(2011)061}{{\em
  JHEP} {\bf 1111} (2011)  061},
\href{http://arxiv.org/abs/1108.2567}{{\tt arXiv:1108.2567 [hep-th]}}.

\bibitem{Gaberdiel:2012yb}
M.~R. Gaberdiel, T.~Hartman, and K.~Jin, ``{Higher Spin Black Holes from
  CFT},'' \href{http://dx.doi.org/10.1007/JHEP04(2012)103}{{\em JHEP} {\bf
  1204} (2012)  103},
\href{http://arxiv.org/abs/1203.0015}{{\tt arXiv:1203.0015 [hep-th]}}.

\bibitem{Hijano:2013fja}
E.~Hijano, P.~Kraus, and E.~Perlmutter, ``{Matching four-point functions in
  higher spin $AdS_3/CFT_2$},''
  \href{http://dx.doi.org/10.1007/JHEP05(2013)163}{{\em JHEP} {\bf 1305} (2013)
   163},
\href{http://arxiv.org/abs/1302.6113}{{\tt arXiv:1302.6113 [hep-th]}}.

\bibitem{Gaberdiel:2013jca}
M.~R. Gaberdiel, K.~Jin, and E.~Perlmutter, ``{Probing higher spin black holes
  from CFT},'' \href{http://dx.doi.org/10.1007/JHEP10(2013)045}{{\em JHEP} {\bf
  1310} (2013)  045},
\href{http://arxiv.org/abs/1307.2221}{{\tt arXiv:1307.2221 [hep-th]}}.

\bibitem{deBoer:2013vca}
J.~de~Boer and J.~I. Jottar, ``{Entanglement Entropy and Higher Spin Holography
  in AdS$_3$},'' \href{http://dx.doi.org/10.1007/JHEP04(2014)089}{{\em JHEP}
  {\bf 1404} (2013)  },
\href{http://arxiv.org/abs/1306.4347}{{\tt arXiv:1306.4347 [hep-th]}}.

\bibitem{Ammon:2013hba}
M.~Ammon, A.~Castro, and N.~Iqbal, ``{Wilson Lines and Entanglement Entropy in
  Higher Spin Gravity},'' \href{http://dx.doi.org/10.1007/JHEP10(2013)110}{{\em
  JHEP} {\bf 1310} (2013)  110},
\href{http://arxiv.org/abs/1306.4338}{{\tt arXiv:1306.4338 [hep-th]}}.

\bibitem{Datta:2014uxa}
S.~Datta, J.~R. David, M.~Ferlaino, and S.~P. Kumar, ``{A universal correction
  to higher spin entanglement entropy},''
\href{http://arxiv.org/abs/1405.0015}{{\tt arXiv:1405.0015 [hep-th]}}.

\bibitem{Drinfeld:1984qv}
V.~Drinfeld and V.~Sokolov, ``{Lie algebras and equations of Korteweg-de Vries
  type},''
\href{http://dx.doi.org/10.1007/BF02105860}{{\em J.Sov.Math.} {\bf 30} (1984)
  1975--2036}.

\bibitem{deBoer:2013gz}
J.~de~Boer and J.~I. Jottar, ``{Thermodynamics of higher spin black holes in
  $AdS_3$},'' \href{http://dx.doi.org/10.1007/JHEP01(2014)023}{{\em JHEP} {\bf
  1401} (2014)  023},
\href{http://arxiv.org/abs/1302.0816}{{\tt arXiv:1302.0816 [hep-th]}}.

\bibitem{Kraus:2013esi}
P.~Kraus and T.~Ugajin, ``{An Entropy Formula for Higher Spin Black Holes via
  Conical Singularities},''
  \href{http://dx.doi.org/10.1007/JHEP05(2013)160}{{\em JHEP} {\bf 1305} (2013)
   160},
\href{http://arxiv.org/abs/1302.1583}{{\tt arXiv:1302.1583 [hep-th]}}.

\bibitem{Dijkgraaf:1996iy}
R.~Dijkgraaf, ``{Chiral deformations of conformal field theories},''
  \href{http://dx.doi.org/10.1016/S0550-3213(97)00153-3}{{\em Nucl.Phys.} {\bf
  B493} (1997)  588--612},
\href{http://arxiv.org/abs/hep-th/9609022}{{\tt arXiv:hep-th/9609022
  [hep-th]}}.

\bibitem{Caux:2012nk}
J.-S. Caux and R.~M. Konik, ``{Constructing the generalized Gibbs ensemble
  after a quantum quench},''
  \href{http://dx.doi.org/10.1103/PhysRevLett.109.175301}{{\em Phys.Rev.Lett.}
  {\bf 109} (2012)  175301},
\href{http://arxiv.org/abs/1203.0901}{{\tt arXiv:1203.0901
  [cond-mat.quant-gas]}}.

\bibitem{Gopakumar:2012gd}
R.~Gopakumar, A.~Hashimoto, I.~R. Klebanov, S.~Sachdev, and K.~Schoutens,
  ``{Strange Metals in One Spatial Dimension},''
  \href{http://dx.doi.org/10.1103/PhysRevD.86.066003}{{\em Phys.Rev.} {\bf D86}
  (2012)  066003},
\href{http://arxiv.org/abs/1206.4719}{{\tt arXiv:1206.4719 [hep-th]}}.

\bibitem{Datta:2014ska}
S.~Datta, J.~R. David, M.~Ferlaino, and S.~P. Kumar, ``{Higher spin
  entanglement entropy from CFT},''
\href{http://arxiv.org/abs/1402.0007}{{\tt arXiv:1402.0007 [hep-th]}}.

\bibitem{Kapusta:1981aa}
J.~I. Kapusta, ``{Bose-Einstein Condensation, Spontaneous Symmetry Breaking,
  and Gauge Theories},''
\href{http://dx.doi.org/10.1103/PhysRevD.24.426}{{\em Phys.Rev.} {\bf D24}
  (1981)  426--439}.

\bibitem{Landsman:1986uw}
N.~Landsman and C.~van Weert, ``{Real and Imaginary Time Field Theory at Finite
  Temperature and Density},''
\href{http://dx.doi.org/10.1016/0370-1573(87)90121-9}{{\em Phys.Rept.} {\bf
  145} (1987)  141}.

\bibitem{Hull:1989wu}
C.~Hull, ``{Gauging the Zamolodchikov $W$ Algebra},''
\href{http://dx.doi.org/10.1016/0370-2693(90)90417-5}{{\em Phys.Lett.} {\bf
  B240} (1990)  110}.

\bibitem{Hull:1990pg}
C.~Hull, ``{Higher spin extended conformal algebras and W gravities},''
\href{http://dx.doi.org/10.1016/0550-3213(91)90324-Q}{{\em Nucl.Phys.} {\bf
  B353} (1991)  707--756}.

\bibitem{Schoutens:1990ja}
K.~Schoutens, A.~Sevrin, and P.~van Nieuwenhuizen, ``{A New Gauge Theory for
  $W$ Type Algebras},''
\href{http://dx.doi.org/10.1016/0370-2693(90)90846-X}{{\em Phys.Lett.} {\bf
  B243} (1990)  245--249}.

\bibitem{Zamolodchikov:1985wn}
A.~Zamolodchikov, ``{Infinite Additional Symmetries in Two-Dimensional
  Conformal Quantum Field Theory},''
\href{http://dx.doi.org/10.1007/BF01036128}{{\em Theor.Math.Phys.} {\bf 65}
  (1985)  1205--1213}.

\bibitem{Fateev:1987vh}
V.~Fateev and A.~Zamolodchikov, ``{Conformal Quantum Field Theory Models in
  Two-Dimensions Having Z(3) Symmetry},''
\href{http://dx.doi.org/10.1016/0550-3213(87)90166-0}{{\em Nucl.Phys.} {\bf
  B280} (1987)  644--660}.

\bibitem{Mikovic:1991rf}
A.~R. Mikovic, ``{Hamiltonian construction of W gravity actions},''
  \href{http://dx.doi.org/10.1016/0370-2693(92)90710-L}{{\em Phys.Lett.} {\bf
  B278} (1992)  51--55},
\href{http://arxiv.org/abs/hep-th/9108002}{{\tt arXiv:hep-th/9108002
  [hep-th]}}.

\bibitem{Romans:1990ag}
L.~Romans, ``{Realizations of classical and quantum W(3) symmetry},''
\href{http://dx.doi.org/10.1016/0550-3213(91)90108-A}{{\em Nucl.Phys.} {\bf
  B352} (1991)  829--848}.

\bibitem{Hull:1990mu}
C.~Hull, ``{Chiral W gravities for general extended conformal algebras},''
\href{http://dx.doi.org/10.1016/0370-2693(91)90135-D}{{\em Phys.Lett.} {\bf
  B259} (1991)  68--72}.

\bibitem{Witten:1983ar}
E.~Witten, ``{Nonabelian Bosonization in Two-Dimensions},''
\href{http://dx.doi.org/10.1007/BF01215276}{{\em Commun.Math.Phys.} {\bf 92}
  (1984)  455--472}.

\bibitem{Mikovic:1990ha}
A.~R. Mikovic, ``{Gauging the extended conformal algebras},''
\href{http://dx.doi.org/10.1016/0370-2693(91)90972-S}{{\em Phys.Lett.} {\bf
  B260} (1991)  75--80}.

\bibitem{Gutperle:2011kf}
M.~Gutperle and P.~Kraus, ``{Higher Spin Black Holes},''
  \href{http://dx.doi.org/10.1007/JHEP05(2011)022}{{\em JHEP} {\bf 1105} (2011)
   022},
\href{http://arxiv.org/abs/1103.4304}{{\tt arXiv:1103.4304 [hep-th]}}.

\bibitem{Kraus:2006wn}
P.~Kraus, ``{Lectures on black holes and the AdS(3) / CFT(2) correspondence},''
  {\em Lect.Notes Phys.} {\bf 755} (2008)  193--247,
\href{http://arxiv.org/abs/hep-th/0609074}{{\tt arXiv:hep-th/0609074
  [hep-th]}}.

\bibitem{Gaberdiel:2010ar}
M.~R. Gaberdiel, R.~Gopakumar, and A.~Saha, ``{Quantum $W$-symmetry in
  $AdS_3$},'' \href{http://dx.doi.org/10.1007/JHEP02(2011)004}{{\em JHEP} {\bf
  1102} (2011)  004},
\href{http://arxiv.org/abs/1009.6087}{{\tt arXiv:1009.6087 [hep-th]}}.

\bibitem{Campoleoni:2010zq}
A.~Campoleoni, S.~Fredenhagen, S.~Pfenninger, and S.~Theisen, ``{Asymptotic
  symmetries of three-dimensional gravity coupled to higher-spin fields},''
  \href{http://dx.doi.org/10.1007/JHEP11(2010)007}{{\em JHEP} {\bf 1011} (2010)
   007},
\href{http://arxiv.org/abs/1008.4744}{{\tt arXiv:1008.4744 [hep-th]}}.

\bibitem{Henneaux:2010xg}
M.~Henneaux and S.-J. Rey, ``{Nonlinear $W_{infinity}$ as Asymptotic Symmetry
  of Three-Dimensional Higher Spin Anti-de Sitter Gravity},''
  \href{http://dx.doi.org/10.1007/JHEP12(2010)007}{{\em JHEP} {\bf 1012} (2010)
   007},
\href{http://arxiv.org/abs/1008.4579}{{\tt arXiv:1008.4579 [hep-th]}}.

\bibitem{Campoleoni:2011hg}
A.~Campoleoni, S.~Fredenhagen, and S.~Pfenninger, ``{Asymptotic W-symmetries in
  three-dimensional higher-spin gauge theories},''
  \href{http://dx.doi.org/10.1007/JHEP09(2011)113}{{\em JHEP} {\bf 1109} (2011)
   113},
\href{http://arxiv.org/abs/1107.0290}{{\tt arXiv:1107.0290 [hep-th]}}.

\bibitem{Gaberdiel:2011wb}
M.~R. Gaberdiel and T.~Hartman, ``{Symmetries of Holographic Minimal Models},''
  \href{http://dx.doi.org/10.1007/JHEP05(2011)031}{{\em JHEP} {\bf 1105} (2011)
   031},
\href{http://arxiv.org/abs/1101.2910}{{\tt arXiv:1101.2910 [hep-th]}}.

\bibitem{Ammon:2011nk}
M.~Ammon, M.~Gutperle, P.~Kraus, and E.~Perlmutter, ``{Spacetime Geometry in
  Higher Spin Gravity},'' \href{http://dx.doi.org/10.1007/JHEP10(2011)053}{{\em
  JHEP} {\bf 1110} (2011)  053},
\href{http://arxiv.org/abs/1106.4788}{{\tt arXiv:1106.4788 [hep-th]}}.

\bibitem{Verlinde:1989ua}
H.~L. Verlinde, ``{Conformal Field Theory, 2-$D$ Quantum Gravity and
  Quantization of Teichmuller Space},''
\href{http://dx.doi.org/10.1016/0550-3213(90)90510-K}{{\em Nucl.Phys.} {\bf
  B337} (1990)  652}.

\bibitem{Bais:1990bs}
F.~Bais, T.~Tjin, and P.~van Driel, ``{Covariantly coupled chiral algebras},''
\href{http://dx.doi.org/10.1016/0550-3213(91)90484-F}{{\em Nucl.Phys.} {\bf
  B357} (1991)  632--654}.

\bibitem{Bilal:1991cf}
A.~Bilal, ``{W algebras from Chern-Simons theory},''
\href{http://dx.doi.org/10.1016/0370-2693(91)90898-Z}{{\em Phys.Lett.} {\bf
  B267} (1991)  487--496}.

\bibitem{deBoer:1991jc}
J.~de~Boer and J.~Goeree, ``{W gravity from Chern-Simons theory},''
  \href{http://dx.doi.org/10.1016/0550-3213(92)90650-Z}{{\em Nucl.Phys.} {\bf
  B381} (1992)  329--359},
\href{http://arxiv.org/abs/hep-th/9112060}{{\tt arXiv:hep-th/9112060
  [hep-th]}}.

\bibitem{DeBoer:1992vm}
J.~De~Boer and J.~Goeree, ``{Covariant W gravity and its moduli space from
  gauge theory},'' \href{http://dx.doi.org/10.1016/0550-3213(93)90308-C}{{\em
  Nucl.Phys.} {\bf B401} (1993)  369--412},
\href{http://arxiv.org/abs/hep-th/9206098}{{\tt arXiv:hep-th/9206098
  [hep-th]}}.

\bibitem{deBoer:1993iz}
J.~de~Boer and T.~Tjin, ``{The Relation between quantum W algebras and Lie
  algebras},'' \href{http://dx.doi.org/10.1007/BF02103279}{{\em
  Commun.Math.Phys.} {\bf 160} (1994)  317--332},
\href{http://arxiv.org/abs/hep-th/9302006}{{\tt arXiv:hep-th/9302006
  [hep-th]}}.

\bibitem{deBoer:1998ip}
J.~de~Boer, ``{Six-dimensional supergravity on S**3 x AdS(3) and 2-D conformal
  field theory},'' \href{http://dx.doi.org/10.1016/S0550-3213(99)00160-1}{{\em
  Nucl.Phys.} {\bf B548} (1999)  139--166},
\href{http://arxiv.org/abs/hep-th/9806104}{{\tt arXiv:hep-th/9806104
  [hep-th]}}.

\bibitem{Ammon:2012wc}
M.~Ammon, M.~Gutperle, P.~Kraus, and E.~Perlmutter, ``{Black holes in three
  dimensional higher spin gravity: A review},''
  \href{http://dx.doi.org/10.1088/1751-8113/46/21/214001}{{\em J.Phys.} {\bf
  A46} (2013)  214001},
\href{http://arxiv.org/abs/1208.5182}{{\tt arXiv:1208.5182 [hep-th]}}.

\bibitem{Banados:2012ue}
M.~Banados, R.~Canto, and S.~Theisen, ``{The Action for higher spin black holes
  in three dimensions},'' \href{http://dx.doi.org/10.1007/JHEP07(2012)147}{{\em
  JHEP} {\bf 1207} (2012)  147},
\href{http://arxiv.org/abs/1204.5105}{{\tt arXiv:1204.5105 [hep-th]}}.

\bibitem{Perez:2012cf}
A.~Perez, D.~Tempo, and R.~Troncoso, ``{Higher spin gravity in 3D: Black holes,
  global charges and thermodynamics},''
  \href{http://dx.doi.org/10.1016/j.physletb.2013.08.038}{{\em Phys.Lett.} {\bf
  B726} (2013)  444--449},
\href{http://arxiv.org/abs/1207.2844}{{\tt arXiv:1207.2844 [hep-th]}}.

\bibitem{Perez:2013xi}
A.~Perez, D.~Tempo, and R.~Troncoso, ``{Higher spin black hole entropy in three
  dimensions},'' \href{http://dx.doi.org/10.1007/JHEP04(2013)143}{{\em JHEP}
  {\bf 1304} (2013)  143},
\href{http://arxiv.org/abs/1301.0847}{{\tt arXiv:1301.0847 [hep-th]}}.

\bibitem{Henneaux:2013dra}
M.~Henneaux, A.~Perez, D.~Tempo, and R.~Troncoso, ``{Chemical potentials in
  three-dimensional higher spin anti-de Sitter gravity},''
  \href{http://dx.doi.org/10.1007/JHEP12(2013)048}{{\em JHEP} {\bf 1312} (2013)
   048},
\href{http://arxiv.org/abs/1309.4362}{{\tt arXiv:1309.4362 [hep-th]}}.

\bibitem{Compere:2013nba}
G.~Comp{\`e}re, J.~I. Jottar, and W.~Song, ``{Observables and Microscopic
  Entropy of Higher Spin Black Holes},''
  \href{http://dx.doi.org/10.1007/JHEP11(2013)054}{{\em JHEP} {\bf 1311} (2013)
   054},
\href{http://arxiv.org/abs/1308.2175}{{\tt arXiv:1308.2175 [hep-th]}}.

\bibitem{Compere:2013gja}
G.~Comp{\`e}re and W.~Song, ``{$\mathcal{W}$ symmetry and integrability of
  higher spin black holes},''
  \href{http://dx.doi.org/10.1007/JHEP09(2013)144}{{\em JHEP} {\bf 1309} (2013)
   144},
\href{http://arxiv.org/abs/1306.0014}{{\tt arXiv:1306.0014 [hep-th]}}.

\bibitem{Bunster:2014mua}
C.~Bunster, M.~Henneaux, A.~Perez, D.~Tempo, and R.~Troncoso, ``{Generalized
  Black Holes in Three-dimensional Spacetime},''
\href{http://arxiv.org/abs/1404.3305}{{\tt arXiv:1404.3305 [hep-th]}}.

\bibitem{Campoleoni:2012hp}
A.~Campoleoni, S.~Fredenhagen, S.~Pfenninger, and S.~Theisen, ``{Towards
  metric-like higher-spin gauge theories in three dimensions},''
  \href{http://dx.doi.org/10.1088/1751-8113/46/21/214017}{{\em J.Phys.} {\bf
  A46} (2013)  214017},
\href{http://arxiv.org/abs/1208.1851}{{\tt arXiv:1208.1851 [hep-th]}}.

\bibitem{Hijano:2014sqa}
E.~Hijano and P.~Kraus, ``{A new spin on entanglement entropy},''
\href{http://arxiv.org/abs/1406.1804}{{\tt arXiv:1406.1804 [hep-th]}}.

\bibitem{Li:2013rsa}
W.~Li, F.-L. Lin, and C.-W. Wang, ``{Modular Properties of 3D Higher Spin
  Theory},'' \href{http://dx.doi.org/10.1007/JHEP12(2013)094}{{\em JHEP} {\bf
  1312} (2013)  094},
\href{http://arxiv.org/abs/1308.2959}{{\tt arXiv:1308.2959 [hep-th]}}.

\bibitem{Afshar:2012nk}
H.~Afshar, M.~Gary, D.~Grumiller, R.~Rashkov, and M.~Riegler, ``{Non-AdS
  holography in 3-dimensional higher spin gravity - General recipe and
  example},'' \href{http://dx.doi.org/10.1007/JHEP11(2012)099}{{\em JHEP} {\bf
  1211} (2012)  099},
\href{http://arxiv.org/abs/1209.2860}{{\tt arXiv:1209.2860 [hep-th]}}.

\bibitem{Gary:2012ms}
M.~Gary, D.~Grumiller, and R.~Rashkov, ``{Towards non-AdS holography in
  3-dimensional higher spin gravity},''
  \href{http://dx.doi.org/10.1007/JHEP03(2012)022}{{\em JHEP} {\bf 1203} (2012)
   022},
\href{http://arxiv.org/abs/1201.0013}{{\tt arXiv:1201.0013 [hep-th]}}.

\bibitem{Kraus:2006nb}
P.~Kraus and F.~Larsen, ``{Partition functions and elliptic genera from
  supergravity},'' \href{http://dx.doi.org/10.1088/1126-6708/2007/01/002}{{\em
  JHEP} {\bf 0701} (2007)  002},
\href{http://arxiv.org/abs/hep-th/0607138}{{\tt arXiv:hep-th/0607138
  [hep-th]}}.

\bibitem{Hull:1993kf}
C.~Hull, ``{Lectures on W gravity, W geometry and W strings},''
\href{http://arxiv.org/abs/hep-th/9302110}{{\tt arXiv:hep-th/9302110
  [hep-th]}}.

\bibitem{Banados:2004nr}
M.~Banados and R.~Caro, ``{Holographic ward identities: Examples from 2+1
  gravity},'' \href{http://dx.doi.org/10.1088/1126-6708/2004/12/036}{{\em JHEP}
  {\bf 0412} (2004)  036},
\href{http://arxiv.org/abs/hep-th/0411060}{{\tt arXiv:hep-th/0411060
  [hep-th]}}.

\bibitem{Tan:2011tj}
H.-S. Tan, ``{Aspects of Three-dimensional Spin-4 Gravity},''
  \href{http://dx.doi.org/10.1007/JHEP02(2012)035}{{\em JHEP} {\bf 1202} (2012)
   035},
\href{http://arxiv.org/abs/1111.2834}{{\tt arXiv:1111.2834 [hep-th]}}.

\end{thebibliography}
\providecommand{\href}[2]{#2}\begingroup\raggedright\endgroup

\end{document}